\documentclass[hyper,letterpaper]{JHEP3}
\usepackage{amsmath,amssymb,amsfonts,bm}
\usepackage{graphicx}
\usepackage{multirow}
\usepackage{verbatim}
\usepackage{appendix}
\usepackage{rotating}
\usepackage{multirow}

\numberwithin{equation}{section}

\newcommand{\bea}{\begin{eqnarray}}
\newcommand{\eea}{\end{eqnarray}}
\newcommand{\be}{\begin{equation}}
\newcommand{\ee}{\end{equation}}

\newcommand{\wwp}{{\omega_{\raisebox{-.02in}{\scriptsize WP}}}}

\newcommand{\Teich}{{\rm Teich}}
\newcommand{\wt}{\widetilde}
\newcommand{\wh}{\widehat}
\newcommand{\verteq}{{\begin{turn}{90}=\end{turn}}}
\newcommand{\adj}[1]{{{#1}_\flat}}
\newcommand{\ol}{\overline}
\newcommand{\ds}{\displaystyle}
\newcommand{\eqdef}{=\hspace{-.1cm}\raisebox{.02cm}{:}}
\newcommand{\defeq}{\raisebox{.02cm}{:}\hspace{-.1cm}=}

\newcommand{\eg}{\emph{e.g.}}
\newcommand{\ie}{\emph{i.e.}}
\newcommand{\cf}{\emph{cf.}}

\newcommand{\Z}{{\mathbb Z}}
\newcommand{\R}{{\mathbb R}}
\newcommand{\C}{{\mathbb C}}

\newcommand{\Tr}{{\rm Tr \,}}
\renewcommand{\Re}{{\rm Re}}
\renewcommand{\Im}{{\rm Im}}
\newcommand{\bs}{\backslash}
\newcommand{\pd}{\partial}

\newcommand{\CA}{\mathcal{A}}
\newcommand{\CB}{\mathcal{B}}
\newcommand{\CC}{\mathcal{C}}
\newcommand{\CD}{\mathcal{D}}

\newcommand{\CH}{\mathcal{H}}

\newcommand{\CJ}{\mathcal{J}}

\newcommand{\CL}{\mathcal{L}}
\newcommand{\CM}{\mathcal{M}}
\newcommand{\CN}{\mathcal{N}}
\newcommand{\CO}{\mathcal{O}}
\newcommand{\CP}{\mathcal{P}}

\newcommand{\CT}{\mathcal{T}}

\newcommand{\CW}{\mathcal{W}}

\newcommand{\CY}{\mathcal{Y}}

\title{Chern-Simons Theory and S-duality}

\author{Tudor Dimofte$^1$ and Sergei Gukov$^{2,3}$
\\ ~
\\
$^1$ Institute for Advanced Study, Einstein Dr., Princeton, NJ 08540, USA\\
$^2$ California Institute of Technology, Pasadena, CA 91125, USA \\
$^3$ Max-Planck-Institut f\"ur Mathematik, Vivatsgasse 7, D-53111 Bonn, Germany}

\abstract{
We study $S$-dualities in analytically continued $SL(2)$ Chern-Simons theory on a 3-manifold $M$. By realizing Chern-Simons theory via a compactification of a 6d five-brane theory on $M$, various objects and symmetries in Chern-Simons theory become related to objects and operations in dual 2d, 3d, and 4d theories. For example, the space of flat $SL(2,\C)$ connections on $M$ is identified with the space of supersymmetric vacua in a dual 3d gauge theory. The hidden symmetry $\hbar \to - \frac{4\pi^2}{\hbar}$ of $SL(2)$ Chern-Simons theory can be identified as the $S$-duality transformation of $\CN=4$ super-Yang-Mills theory (obtained by compactifying the five-brane theory on a torus); whereas the mapping class group action in Chern-Simons theory on a three-manifold $M$ with boundary $C$ is realized as $S$-duality in 4d $\CN=2$ super-Yang-Mills theory associated with the Riemann surface $C$. We illustrate these symmetries by considering simple examples of $3$-manifolds
that include knot complements and punctured torus bundles, on the one hand, and mapping cylinders associated with mapping class group transformations, on the other. A generalization of mapping class group actions further allows us to study the transformations between several distinguished coordinate systems on the phase space of Chern-Simons theory, the $SL(2)$ Hitchin moduli space.
\\
\\
\\
\\
\\
\\
{\tt CALT-68-2841}
}

\begin{document}


\section{Introduction}

In the past year, several closely related proposals emerged  \cite{DGH,Rtwisting,Wfiveknots}
on how to realize analytic continuation of Chern-Simons theory
on the world-volume of a fivebrane system.
In all of these proposals, the Hilbert space $\CH$ of Chern-Simons theory on $\R \times C$
is obtained by quantizing the space of classical solutions, realized as a real slice
inside the Hitchin moduli space, $\CM_H (C)$, of the Riemann surface $C$.

In this paper we continue studying the relation between
analytically continued Chern-Simons theory --- sometimes called Chern-Simons theory with complex gauge group --- and
the three-dimensional $\CN=2$ effective field theory obtained by compactifying
the six-dimensional fivebrane theory on a 3-manifold $M$
(and subject to the $\Omega$-deformation in the remaining three dimensions).
This relation has two important implications:

\begin{itemize}

\item {\bf ``$\CN=2$ S-duality'':}
when a 3-manifold $M$ is a mapping torus of a Riemann surface $C$, the mapping class groupoid of $C$ acts
on ``holomorphic blocks'' of Chern-Simons theory as S-duality of the four-dimensional $\CN=2$ gauge theory;

\item {\bf ``$\CN=4$ S-duality'':}
Chern-Simons theory has hidden symmetry that acts on the coupling constant as
\be \hbar \to {}^L\hbar = - \frac{4\pi^2}{\hbar} \label{Lh} \ee
and changes the gauge group $G$ to the Langlands or GNO dual group ${}^LG$.

\end{itemize}

The first statement is essentially due to the AGT correspondence \cite{AGT}, whereas the second claim is fairly new and more mysterious.
First hints of a hidden symmetry $\hbar \to {}^L\hbar$ in Chern-Simons theory come from the work
of Lawrence and Zagier \cite{LawrenceZag} and its generalizations (see {\it e.g.} \cite{Hikami-Kirillov,Hikami-Brieskorn})
where non-hyperbolic 3-manifolds provide a ``laboratory'' for experiments.
Closer to the current developments is a series of observations \cite{GukovZagier,QMF}
that analytically continued partition function on hyperbolic 3-manifolds exhibits even more delicate
modular behavior which, in the context of state integral models for Chern-Simons theory \cite{DGLZ, Dimofte-QRS},
can be traced to the corresponding behavior of the quantum dilogarithm function.
The transformation of the coupling constant \eqref{Lh} and the exchange of the gauge group $G$
with the dual group ${}^LG$ suggests that this hidden symmetry should have a physical
explanation in the construction of Chern-Simons Hilbert spaces and wavefunctions in terms
of four-dimensional $\CN=4$ super-Yang-Mills theory \cite{GW-branes,Witten-path,Wfiveknots}
or, equivalently, in terms of six-dimensional $(2,0)$ theory compactified on a torus.
By understanding the relation between Chern-Simons theory
and the five-brane theory in the spirit of \cite{DGH,Rtwisting,Wfiveknots},
we will be able to understand both of the above-mentioned S-duality symmetries,
and even see a connection between them.

\begin{figure}[htb]
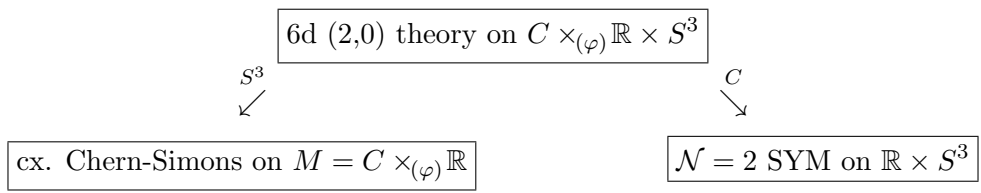

\centering
\be
\begin{array}{ccc}
&\boxed{\text{6d (2,0) theory on $C\times_{\!(\varphi)\!}\R\times S^3$}}& \\
\overset{S^3}{\swarrow} && \overset{C}{\searrow}
\end{array}\notag
\ee
\vspace{-.3cm}
\be \boxed{\text{cx. Chern-Simons on $M=C\times_{\!(\varphi)\!} \R$}}\hspace{1in} \boxed{\text{$\CN=2$ SYM on $\R\times S^3$}} \notag\ee
\caption{Two compactifications of the fivebrane theory. Eventually we will twist the product $C\times\R$ by an action ``$\varphi$'' of
$\CN=2$ S-duality group ({\it i.e.} mapping class group of $C$) to form more interesting manifolds~$M$.}
\label{fig:fivebrane}
\end{figure}

As a starting point, let us consider putting $N$ M5 branes on a (Euclidean) spacetime of the form $M\times S^3 = C\times \R\times S^3$,
where $M$ is the product of a punctured Riemann surface $C$ and ``time'' $\R$ (Figure \ref{fig:fivebrane}). Compactification on $C$ leads
to an $\CN=2$ theory on $\R\times S^3$, with gauge group and matter content determined by the surface $C$ \cite{Gaiotto-dualities}.
(The gauge group is a product of $G=SU(N)$ factors.) We moreover impose an $\Omega$-deformation on $S^3$ with equivariant
parameters $\epsilon_1$ and $\epsilon_2$, as in \cite{AGT, NekWitten}. To this 4-dimensional theory, one naturally associates
a Hilbert space $\CH^{\CN=2}_{\epsilon_{1,2}}(S^3)$ of states that depends on $\epsilon_1$ and $\epsilon_2$
only through the ratio $\epsilon_1/\epsilon_2$.

Going in the opposite direction, we could try to compactify the fivebrane theory on~$S^3$. As observed in several contexts \cite{DGH, Wfiveknots}, compactification of a fivebrane theory to three dimensions in the presence of an $\Omega$-deformation produces a complexified, or analytically continued, version of Chern-Simons theory. Here, we find analytically continued Chern-Simons theory on $M=C\times \R$. The classical solutions of this theory are flat $G_\C=SL(N,\C)$ connections, and upon quantization one obtains (analytic continuations of wavefunctions in%
\footnote{\label{foot:CSanal}In analytically continued Chern-Simons theory one typically does not find an honest Hilbert space of wavefunctions associated to a spatial slice or boundary $C$, but rather analytic continuations of a subset of the wavefunctions in a standard Hilbert space. In the present case, we can think of $\CH_\hbar^{CS}(C)$ as an analytic continuation of the space of wavefunctions of real $G_\R=SL(2,\R)$ Chern-Simons theory.}%
) a Hilbert space~$\CH_\hbar^{CS}(C)$.

{}From the duality diagram in Figure \ref{fig:fivebrane} we expect, of course, that $\CH^{CS}_\hbar(C) = \CH^{\CN=2}_{\epsilon_{1,2}}(S^3)$.
The Chern-Simons coupling constant $\hbar$ is related to the ratio of $\Omega$-deformation parameters $\epsilon_1,\epsilon_2$ as
\be \hbar  = 2\pi i\, \frac{\epsilon_1}{\epsilon_2}\,. \label{he} \ee
Thus, the ``$\CN=4$ S-duality'' of Chern-Simons theory, $\hbar\leftrightarrow {}^L\hbar$ becomes an exchange of
deformation parameters $\epsilon_1\leftrightarrow \epsilon_2$ in the $\CN=2$ theory on $\R\times S^3$.
Note that it is crucial for this correspondence that the Chern-Simons theory be analytically continued.
If this were Chern-Simons theory with compact gauge group $G=SU(N)$, then the coupling $\hbar$ would be related to the quantized
level $k\in \Z$ as $\hbar = 2\pi i/k$, and a duality $\hbar\leftrightarrow {}^L\hbar$ would make no sense. Moreover, the Hilbert space in compact Chern-Simons theory would be finite dimensional, with no hope of being equal to $\CH^{\CN=2}_{\epsilon_{1,2}}(S^3)$. In analytically continued Chern-Simons theory \cite{gukov-2003} (see also \cite{Wit-anal} and discussions in \cite{DGLZ, Dimofte-QRS}), $\hbar$ can be taken to be an arbitrary nonzero complex number, and $\CH^{CS}_\hbar(C)$ is, appropriately, infinite dimensional.

In the case where we wrap $N=2$ M5 branes on $C\times \R\times S^3$, it is actually easy to see that
$\CH^{CS}_\hbar(C) = \CH^{\CN=2}_{\epsilon_{1,2}}(S^3)$ from a different --- though obviously related --- point of view.
The Hilbert space $\CH^{CS}_\hbar(C)$ can be related to (an analytic continuation of wavefunctions in)
the so-called quantum Teichm\"uller space $\CH^{Teich}_\hbar(C)$ \cite{FockChekhov, Kash-Teich},
essentially a quantization of part of the moduli space of flat $SL(2,\R)$ connections on $C$.
Moreover, quantum Teichm\"uller space is equivalent to the space of conformal blocks of Liouville theory on $C$,
$\CH^{Teich}_\hbar(C) = \CH^{Liouv}_b(C)$ labeled by a parameter $b$ that determines the central charge
of the theory \cite{Teschner-TeichMod, Teschner-TeichLiouv, Verlinde-TeichLiouv}.
Finally, from the AGT conjecture \cite{AGT}, we know that $\CH^{Liouv}_b(C) = \CH^{\CN=2}_{\epsilon_{1,2}}(S^3)$.
Thus, there really exists a square of equivalences,
\be \begin{array}{ccc}
 \CH^{CS}_\hbar(C) & = & \CH^{\CN=2}_{\epsilon_{1,2}}(S^3) \\
 \verteq & & \verteq \\
 \CH^{Teich}_\hbar(C) & = & \CH^{Liouv}_b(C)\,.
 \end{array}
 \label{Hilbsquare}
\ee
The Liouville coupling constant $b$ is related to $\hbar$ as
\be \hbar = 2\pi i b^2\,, \label{Liouvhbar} \ee
and transforms as $b\leftrightarrow b^{-1}$ under ``$\CN=4$ S-duality,'' as observed in \cite{ZZ-3pt, Tesch-Liouv}.

Although the square of Hilbert space equivalences \eqref{Hilbsquare} is written for $N=2$ M5 branes on $C\times \R\times S^3$, it can be extended to any number of branes --- or in fact to any ADE $(2,0)$ theory on $C\times\R\times S^3$. One should replace Teichm\"uller theory with ``higher'' Teichm\"uller theory \cite{FG-Teich}, and Liouville theory with an appropriate Toda CFT \cite{AGT, Wyllard-An}.

Each of the Hilbert spaces $\CH$ in \eqref{Hilbsquare} has an algebra $\hat\CA$ of operators acting on it, which must also transform under ``$\CN=4$ S-duality.''
Thus:
\be (\CH^{\,\bullet}_\hbar,\,\hat\CA^{\,\bullet}_\hbar) \sim (\CH^{\,\bullet}_{{}^L\hbar},\,\hat\CA^{\,\bullet}_{{}^L\hbar})\,,\ee
where the placeholder ``$\bullet$'' can be either ``$CS$,'' ``$\CN=2$,'' ``$Teich$,'' or ``$Liouv$.''
For example, in $\CN=2$ theory, $\hat\CA^{\CN=2}_{\epsilon_{1,2}}(S^3)$ is an algebra of line operators $\wh W_\gamma$,
labeled by the electric and magnetic charges $\gamma$ of the theory \cite{AGGTV, Tesch-loop}.
In Liouville theory, these become Verlinde loop operators $\wh \Lambda_\gamma$ \cite{Verlinde-loop}
that wrap various closed cycles $\gamma$ on $C$. In both Teichm\"uller and Chern-Simons theory,
the operators correspond to holonomies of flat $SL(2)$ connections around cycles $\gamma$ of $C$.
While S-duality of the Teichm\"uller and Liouville algebras was noticed%
\footnote{The standard statement in the literature is that both Hilbert spaces $\CH^{Teich}_{\hbar}$, $\CH^{Liouv}_b$, etc. and algebras $\hat \CA^{Teich}_{\hbar}$, $\hat \CA^{Liouv}_b$, etc. are invariant under S-duality. The dualization of the gauge group $G\leftrightarrow {}^LG$ (\ie\ $SU(2)\leftrightarrow SO(3)$, or $SL(2,\R)\leftrightarrow PSL(2,\R)$, or $SL(2,\C)\leftrightarrow PSL(2,\C)$) is largely invisible in the construction of these objects; thus they simply appear S-invariant. Similarly, in the study of 4d gauge theory, it is well known that instanton partition functions of Vafa-Witten type \cite{VafaWitten} are invariant under $\tau\to-1/\tau$ even though the gauge group secretly changes from $G$ to ${}^LG$. In this paper, we will see the subtle effect of dualizing the group $G$ only in Section \ref{sec:mirrors}.} %
some time ago \cite{FockChekhov, Tesch-Liouv}, S-duality for $\hat\CA^{CS}_\hbar(C)$ in Chern-Simons theory appeared only recently, for the case $C=T^2$, in \cite{Dimofte-QRS}. \bigskip

\FIGURE{
\includegraphics[width=2in]{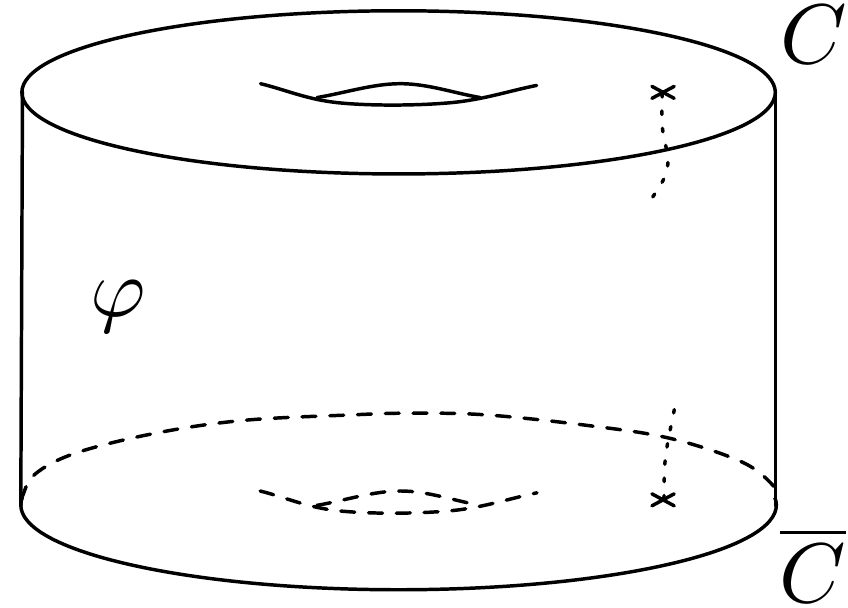}
\caption{A mapping cylinder for an element $\varphi\in\bm\Gamma(C)$ acting on the punctured torus $C=T^2\bs\{p\}$.}
\label{fig:MCgeneric}
}

What about ``$\CN=2$ S-duality''? In $\CN=2$ gauge theory coming from compactification of a fivebrane theory on $C$, the S-duality groupoid can be identified with the mapping class groupoid $\bm\Gamma(C)$ of the surface $C$ \cite{Gaiotto-dualities}. In particular, the duality transformation corresponding to an element $\varphi\in\bm\Gamma(C)$ acts as an operator $\wh\CO_\varphi\,:\,\CH\to \CH$, where by $\CH$ we mean the Hilbert space $\CH^{\CN=2}_{\epsilon_{1,2}}(S^3)$ in \eqref{Hilbsquare}. The action of this operator can be represented by an integral kernel $K^{\CN=2}_\varphi(\epsilon_1,\epsilon_2)\in \CH\otimes\CH^*$. Then, in terms of Chern-Simons theory, the natural object to identify with $K^{\CN=2}_\varphi(\epsilon_1,\epsilon_2)$ would be the partition function (or wavefunction) on a \emph{mapping cylinder} $M=C\!\times_\varphi I$ (Figure \ref{fig:MCgeneric}). In the mapping cylinder, the copy of $C$ at the top of the interval $I$ is related to the copy at the bottom by the mapping class group twist $\varphi$. We expect that
\be Z^{CS}(C\!\times_\varphi\! I;\hbar) = K^{\CN=2}_\varphi(\epsilon_1,\epsilon_2)\qquad \in\quad \CH\otimes\CH^*\,. \label{ZCSgauge} \ee

Both sides of \eqref{ZCSgauge} depend on additional parameters, not explicitly written here. The Chern-Simons partition function $Z^{CS}(C\!\times_\varphi\! I;\hbar)$ is a function of boundary conditions on $\pd M = C\sqcup \ol{C}$. These boundary conditions describe flat $G_\C=SL(N,\C)$ connections on $C\sqcup \ol{C}$, and if $C$ has genus $g$ and $s$ punctures, consist of $(2g-2+s)\dim G$ complexified parameters. On the 4d gauge theory side, the kernel $K^{\CN=2}_\varphi(\epsilon_1,\epsilon_2)$ depends on Coulomb vevs and hypermultiplet masses, of which (again) there should be $(2g-2+s)\dim G$. We will review the precise identification of parameters starting in Section \ref{sec:class}. Although the manifold $M=C\times_\varphi I$ is topologically equivalent to the product $C\times I$, it is the choice of relative boundary conditions or parameters on $\pd M = C\sqcup \ol{C}$ --- determined by $\varphi$ --- that makes kernels \eqref{ZCSgauge} nontrivial.

The kernel $K^{\CN=2}_\varphi(\epsilon_1,\epsilon_2)$ also has an interesting interpretation as the partition function $Z^{3d}_\varphi(\epsilon_1/\epsilon_2)$ of 3-dimensional $\CN=2$ gauge theory ${\bf T}_{M}$ living on $S^3$. This is the theory of an S-duality domain wall at a fixed ``time'' in $\R\times S^3$, which implements the S-duality action $\varphi$ \cite{DGG-defects}; it is a generalization of the $T[SU(2)]$ theory of \cite{GW-Sduality}, which was further studied in \cite{HLP-wall, Yamazaki-3d}. The extra parameters that $K^{\CN=2}_\varphi(\epsilon_1,\epsilon_2)$ depends on --- Coulomb vevs and masses --- become identified with twisted masses and FI parameters in three dimensions.

To complete a square of equivalences for ``$\CN=2$ S-dualities,'' we should note
that both Liouville and Teichm\"uller theory have analogues of $K^{\CN=2}_\varphi(\epsilon_1,\epsilon_2)$.
In Liouville theory, this is the Moore-Seiberg kernel $K^{Liouv}_\varphi(b)$ \cite{MS-groupoid}
that implements a mapping class group action on the conformal blocks associated to $C$.
It should equal $K^{\CN=2}_\varphi(\epsilon_1,\epsilon_2)=Z^{3d}_\varphi(\epsilon_1/\epsilon_2)$ by the AGT conjecture,
and this equivalence was checked by careful calculations in \cite{HLP-wall} (see also \cite{Yamazaki-3d}).
In quantum Teichm\"uller theory, we have a kernel $K^{Teich}_\varphi(\hbar)$ \cite{FockChekhov, Kash-Teich}
(also \cf\ \cite{FG-qdl-cluster}), which intertwines a mapping class group action in the algebra of quantum operators.
The equality $K^{Teich}_\varphi(\hbar)=K^{Liouv}_\varphi(b)$ follows by a rather nontrivial change of basis in the Hilbert space $\CH$,
and formed the main thrust of \cite{Teschner-TeichLiouv, Tesch-LtoQG, Teschner-TeichMod}. Thus:
\be \hspace{.5in} \begin{array}{cccc}
 Z^{CS}_\varphi(\hbar) & \simeq & K^{\CN=2}_\varphi(\epsilon_1,\epsilon_2) & = Z^{3d}_\varphi(\epsilon_1/\epsilon_2) \\
 \verteq & & \verteq & \\
 K^{Teich}_\varphi(\hbar) & \simeq & K^{Liouv}_\varphi(b)\,. &
 \end{array}
 \label{Ksquare}
\ee
The final equality $K^{CS}_\varphi(\hbar)=K^{Teich}_\varphi(\hbar)$ follows directly from definitions and analytic continuation.

Our present interest in the mapping-cylinder kernels/wavefunctions \eqref{Ksquare} centers on two important properties that they share. First, just like the Hilbert spaces \eqref{Hilbsquare}, they all enjoy ``$\CN=4$ S-duality'' in the sense that
\begin{align*}
Z^{CS}_\varphi(\hbar) &= Z^{CS}_\varphi({}^L\hbar) \\
K^{Teich}_\varphi(\hbar) &= K^{Teich}_\varphi({}^L\hbar) \\
K^{Liouv}_\varphi(b) &= K^{Liouv}_\varphi(b^{-1}) \\
K^{\CN=2}_\varphi(\epsilon_1,\epsilon_2) &= K^{\CN=2}_\varphi(\epsilon_2,\epsilon_1)\,.
\end{align*}
Whereas the last three of these dualities have been understood for some time, the first is an example of S-duality for wavefunctions in analytically continued Chern-Simons theory --- one of our main new proposals. It quickly follows from the equivalences \eqref{Ksquare}.

Second, all the wavefunctions in \eqref{Ksquare} are annihilated by a system of difference equations
\be \wh\Delta_i\,K^\bullet_\varphi(...) = 0\,,\qquad \wh\Delta_i\in \hat\CA(C)\otimes \hat\CA(\ol{C})\,. \label{DeltaK} \ee
This statement is immediate for Chern-Simons theory, since the wavefunction of any three-manifold with boundary is annihilated by a system of difference equations \cite{gukov-2003, DGLZ}. The Chern-Simons statement then translates in an interesting way to all the other mapping-class kernels. In particular, we find that Moore-Seiberg kernels in Liouville theory are annihilated by operators composed form Verlinde loops, and that S-duality kernels in $\CN=2$ gauge theory are annihilated by combinations of line operators.

The number of operators $\wh\Delta_i$ in \eqref{DeltaK} is equal to $(2g-2+s)\dim G$. In gauge theory language, they are composed of multiplication and differentiation in the $(2g-2+s)\dim G$ Coulomb vevs $a_j$ and masses $m_k$ that $K^{\CN=2}_\varphi(\epsilon_1,\epsilon_2)$ depends on. We will be particularly interested in a subset of $(2g-2+s)\dim G-s\,{\rm rank}\,G = (N^2-1)(2g-2)+N(N-1)s$ operators that differentiate only Coulomb vevs, while leaving the masses fixed. In this case, the classical $\hbar\to 0$ limit of the operators $\wh\Delta_i$ has a nice geometric interpretation. One obtains a system of classical equations $\Delta_i=0$ that cut out a complex Lagrangian submanifold
\be \CL_\varphi = \{\Delta_i=0\}\quad\subset\quad Y\times \ol{Y}\,, \label{Lagintro} \ee
where $Y=\CM_H(G;C)$ is the Hitchin moduli space associated to $C$ \cite{Hitchin-SD} (and $\ol{Y}$ is just $Y$ with opposite symplectic structure).

The space $Y$ can be thought of intuitively as a complexification of the Coulomb branch of 4-dimensional $\CN=2$ theory --- it is parameterized by complexified Coulomb vevs. The Lagrangian $\CL_\varphi$ is then a graph that encodes how Coulomb vevs map to one another in the classical limit of an $(\CN=2)$ S-duality transformation $\varphi$. More interestingly, we will be able to see $\CL_\varphi$ arising directly from the low-energy superpotential $\CW_{\rm eff}$ of a 3-dimensional ${\bf T}_{M_\varphi}$ domain wall theory corresponding to $\varphi$. In particular, $\CW_{\rm eff}$ is the ``potential function'' that determines the graph $\CL_\varphi$. \\

Having made a loop through the circle of S-dualities related to Chern-Simons theory, we come now summarize the content and organization of the remainder of the paper. The Hitchin moduli space $Y=\CM_H(G,C)$ turns out to be a central ingredient in understanding both $\CN=4$ and $\CN=2$ S-duality. Its quantization (or rather quantization of one of its real slices) produces the Hilbert spaces in \eqref{Hilbsquare}, while mapping class group actions on $Y$ correspond to the wavefunctions/kernels in \eqref{Ksquare}. Thus, we will begin in Section \ref{sec:class} by studying the classical geometry on $Y$ and three useful systems of coordinates for parametrizing it. We illustrate general ideas with the concrete example of $G_{\C} = SL(2,\C)$ and a punctured torus $C=T^2\bs\{p\}$, which, in the context of AGT correspondence, is known as the four-dimensional $\CN=2^*$ theory with gauge group $G=SU(2)$.

We proceed with the semi-classical limit of the relations \eqref{Ksquare} described by Lagrangians (correspondences) in $Y\times \ol{Y}$ and then in Sections \ref{sec:tori}--\ref{sec:3dgauge} close up mapping cylinders into mapping tori. In the case of mapping tori, the Lagrangians \eqref{Lagintro} become classical A-polynomials \cite{cooper-1994} for knot and link complements. We thus find two new interpretations of the classical $A$-polynomial (and its generalizations): as the spectrum of eigenbranes on the Hitchin moduli space and as a space of SUSY moduli in the ``effective'' 3d $\CN=2$ gauge theory ${\bf T}_M$ associated with $M$.

In Section \ref{sec:quant} we quantize $Y$ and construct quantum analogs of all objects considered in Section \ref{sec:class}. Specifically, in Section \ref{sec:qalg} we quantize the algebra of functions on $Y \cong \CM_{{\rm flat}} (G_{\C},M)$ in all three coordinate systems of Section \ref{sec:MHit}. As in Section \ref{sec:class}, we illustrate general ideas with the example $G_{\C} = SL(2,\C)$ and $C=T^2\bs\{p\}$. The partition functions of mapping tori \eqref{Ksquare}, corresponding to duality walls in $\CN=2^*$ 4d gauge theory theory, are computed in Section \ref{sec:qcyl} along with the operators \eqref{DeltaK} that annihilate them. Working with the punctured torus as the main example, our discussion would be incomplete without two special duality walls that correspond to $T$ and $S$ elements of the $SL(2,\Z)$ duality group of $\CN=2^*$ theory. These special elements correspond to the $T$-move and $S$-move in Liouville theory and are analyzed in detail in Section \ref{sec:eg}. Another interesting example considered in Section \ref{sec:eg} is a quantum version of the coordinate transformation between ``shear'' and ``Fenchel-Nielsen'' coordinates on $Y$. It would be interesting to understand the corresponding duality wall in the four-dimensional $\CN=2^*$ theory.

Throughout our discussion in Sections \ref{sec:class} and \ref{sec:quant} we verify that all formulas are manifestly invariant under the $\CN=4$ S-duality \eqref{Lh}, which then becomes the central theme of Section~\ref{sec:mirrors}. Following \cite{BJSV,Kapustin-Witten}, we interpret this duality as mirror symmetry between moduli spaces $Y=\CM_H(G,C)$ and $\widetilde{Y}=\CM_H(^LG,C)$. Using this approach, we study how flat $G_{\C}$ connections on knot complements transform under Langlands or GNO duality. In particular, we obtain the $SO(3)$ version of the $A$-polynomial from mirror symmetry and discuss the corresponding $\CD$-modules. Finally, in appendices we summarize various technical details used in the main text.


\section{Classical theory}
\label{sec:class}

As a starting point in the study of both ``$\CN=2$'' and ``$\CN=4$'' S-dualities, we can try to understand their semiclassical limits. In physical gauge theory, an S-duality typically inverts the strength of quantum fluctuations, mapping a weakly coupled theory to a strongly coupled one and vice versa; thus a ``semiclassical limit'' of S-duality may not immediately make sense. What we mean by semiclassical, however, is the limit $\hbar \to 0$, where
\be \hbar  = 2\pi i b^2  = 2\pi i \frac{\epsilon_1}{\epsilon_2} \ee
is the ``quantum'' parameter of the introduction. Therefore, if we are thinking of $\CN=2$ gauge theory, this semiclassical limit has nothing to do with the physical coupling constant. We can safely ask how objects like expectation values of Wilson and 't Hooft loops or (equivariant) instanton partition functions  transform ``semiclassically'' under $\CN=2$ S-duality. This is our present goal.

On the other hand, $\CN=4$ S-duality does invert $\hbar \to {}^L\hbar = -4\pi^2/\hbar$ as in \eqref{Lh}. We will start tackling $\CN=4$ S-duality in Section \ref{sec:quant}, once we have moved away from $\hbar \approx 0$. By studying certain protected objects, we will also be able to make sense of a semiclassical limit of $\CN=4$ S-duality, but that will have to wait until Section \ref{sec:mirrors}.

As discussed in the introduction, a central construct in the study of S-dualities is the Hitchin moduli space \cite{Hitchin-SD}
\be Y = \CM_H(G,C)\,, \ee
where $G$ is a compact gauge group and $C$ is a punctured Riemann surface. $\CN=2$ S-duality acts on $Y$ via the mapping class group of $C$, whereas $\CN=4$ S-duality acts via mirror symmetry. The semiclassical limit $\hbar\to 0$ of $\CN=2$ S-duality can be understood in terms of the classical geometry of $Y$. Therefore, we begin in Section \ref{sec:MHit} by reviewing the various geometric structures and coordinates on $Y$, and their relation to quantities in gauge theory, Liouville theory, Teichm\"uller theory, and Chern-Simons theory. In Section \ref{sec:cyl}, we consider semiclassical ($\CN=2$) S-duality transformations, again relating them to the geometry of $Y$. Finally, in Sections \ref{sec:tori}--\ref{sec:3dgauge}, we use classical S-duality to construct more interesting 3-manifolds $M$ (mapping tori for mapping class group actions), and to relate Chern-Simons theory on these spaces with SUSY gauge theory in three and four dimensions. In particular, we find new physical interpretations of the $A$-polynomial and, more generally, of the classical moduli space $\CM_{{\rm flat}} (G_{\C},M)$ as
\begin{itemize}

\item[$a)$] the spectrum of eigenbranes on $Y$,

\item[$b)$] the space of supersymmetric vacua in the 3d ``effective'' $\CN=2$ gauge theory.

\end{itemize}
\noindent
Throughout this section and the next, we focus on the case where $C$ is a punctured torus
\be C = T^2\bs\{p\}\,, \ee
and $G=SU(2)$ (or $G_\C = SL(2,\C)$). In other words, we wrap $N=2$ M5 branes on spaces $(T^2\bs\{p\})\times_{\!(\varphi)\!}\R\times S^3$. This is purely for clarity of exposition: all concepts should generalize to arbitrary $C$ and (in principle) to higher rank.

\subsection{The Hitchin moduli space}
\label{sec:MHit}

We can begin by recalling that the Hitchin moduli space $Y = \CM_H (G,C)$ is hyper-K\"ahler. In particular, it has a $\mathbb{CP}^1$ of global complex structures all compatible with its hyper-K\"ahler metric $ g$, generated by a triplet of complex structures $(I,J,K)$ that obey the quaternion relations $IJ=K$, and so forth. Likewise, to each complex structure $\CJ = I,J,K,\text{etc.}$ there is associated a real symplectic form $\omega_\CJ =  g \CJ$. Namely, $\omega_\CJ$ is the K\"ahler form in complex structure $\CJ$. Moreover, in every complex structure $\CJ$ there is a \emph{holomorphic} symplectic form $\Omega_\CJ$, which is made up from the ``complementary'' real symplectic forms. For example, $i\Omega_I = \omega_J+i\omega_K$, $i\Omega_J = \omega_K + i\omega_I$, and $i\Omega_K = \omega_I+i\omega_J$.

In one of the complex structures, which, following \cite{Hitchin-SD}, we will call $I$, the space $\CM_H (G,C)$ can be identified with the moduli space of stable Higgs pairs $(A,\Phi)$
on $C$ with the structure group $G$.
(Hence the notation $\CM_H (G,C)$, where ``H'' stands for ``Hitchin'' or ``Higgs.'')
For a connection with Chern-Simons theory, however, perhaps one of the most important facts is
that in complex structure $J$ the space $Y$ can be identified with the space of flat $G_\C$ connections on $C$,
\be
Y \cong \CM_{{\rm flat}} (G_{\C},C)\,. \label{Mflat}
\ee
This is the classical phase space for analytically continued $(G_\C)$ Chern-Simons theory on any 3-manifold whose boundary is $C$, \cf\ \cite{gukov-2003}.

In complex structure $J$, there are several holomorphic coordinate systems that are often used in the literature:
\begin{itemize}
\item loop (or Fricke-Klein) coordinates , \eg\ $(x,y,z)$\,,
\item shear (or Fock) coordinates, \eg\ $(t,t',t'') = (e^{T}, e^{T'}, e^{T''})$\,,
\item complexified Fenchel-Nielsen coordinates, \eg\ $(\tau,\lambda) = (e^\CT,e^\Lambda)$\,,
\end{itemize}
Each coordinate system --- which we will proceed momentarily to describe in detail --- has its inherent advantages and disadvantages, and combining all three gives a powerful tool for analyzing S-dualities. For example, we will see that loop coordinates are very natural in 4d gauge theory (where they correspond to classical expectation values of circular line operators), whereas shear coordinates are best for quantization in both Teichm\"uller and Chern-Simons theory. Fenchel-Nielsen coordinates bridge the gap between the other two systems.

Specializing now to $G=SU(2)$, we can define the Teichm\"uller space $\Teich(C)$ as the component of the moduli space of flat $G_\R=SL(2,\R)$ connections on $C$ all of whose holonomies are hyperbolic%
\footnote{This means that the eigenvalues of all holonomy matrices are real, as opposed to being on the unit circle.} %
:
\be \Teich(C) := \CM_{\rm flat}(G_\R,C)\big|_{\rm hyp} \quad \subset\quad
 Y\,. \ee
This is clearly a real slice of $Y$ with respect to complex structure $J$, and can be parametrized by restricting any of the $J$-holomorphic coordinates above (loop, shear, etc.) to be real.
$\Teich(C)$ must be Lagrangian in $Y$ with respect to the symplectic form $\omega_J$.

Moreover, it is easy to see that $\Teich(C)\subset Y$ is
a component of the fixed point set of the involution $(A,\Phi) \mapsto (A, - \Phi)$ (modulo gauge transformations).
This involution is holomorphic in complex structure $I$ and anti-holomorphic in complex structures $J$ and $K$.
Together, these properties force $\Teich(C)$ to also be analytic with respect
to the complex structure $I$ and Lagrangian with respect
to the third symplectic form $\omega_K$, making it a brane of ``type $(B,A,A)$.''
(Another such $(B,A,A)$ brane in $Y$ is the base ${\bf B}$ of the Hitchin fibration \cite{Hitchin-SD}
that will be discussed later in Section \ref{sec:params} and also in Section \ref{sec:mirrors}.)
Among different types of half-BPS branes in $Y$, the branes of type $(B,A,A)$
that we encounter here are perhaps the most delicate ones \cite{mirrorbranes}.
In particular, under ``$\CN=4$ S-duality'' (mirror symmetry on $Y$) they are transformed
to branes of type $(B,B,B)$ holomorphic in all complex structures.

Since $\Teich(C) \subset Y$ is Lagrangian with respect to $\omega_J$ and $\omega_K$, we have $\omega_J\big|_{\Teich(C)}=0$ and $\omega_K\big|_{\Teich(C)}=0$. On the other hand, $\omega_I$ restricts to a symplectic form on $\Teich(C)$. As shown in the original paper by Hitchin \cite{Hitchin-SD}, the symplectic form $\omega_I \big|_{\Teich(C)}$ is identical to the standard Weil-Petersson symplectic form $\wwp$ on the Teichm\"uller space,
\be  \wwp = \omega_I\big|_{\Teich(C)} = \Re\, \Omega_J\big|_{\Teich(C)} = \Omega_J\big|_{\Teich(C)}\,, \label{wps} \ee
where the last equality follows from the Lagrangian condition.
Due to \eqref{wps}, we can think of $\Omega_J = \omega_I -i\omega_K$ as a natural complexification of the Weil-Petersson symplectic form on $Y$. Note that, in terms of flat $G_{\C}$ connections $\CA$ on $C$, we have
\be \Omega_J \; = \; \frac{1}{4 \pi^2 \hbar} \int_C  \Tr\big(\delta\CA\wedge \delta \CA\big)\,.\ee
This is easily recognized as the symplectic structure induced on $Y = \CM_{\rm flat}(G_\C,C)$ by the holomorphic, analytically continued Chern-Simons action $S_{CS} = \frac{1}{4 \pi \hbar} \int\Tr\big(\CA\,d\CA+\frac23\CA^3\big)$. The space $\CH^{CS}_\hbar(C)$ (and the three other ``Hilbert'' spaces mentioned in the introduction, \cf\ \eqref{Hilbsquare}) can be obtained most rigorously by quantizing $\Teich(C)\subset Y$ with respect to $\wwp$ and then analytically continuing the resulting space of wavefunctions, allowing them to be locally holomorphic functions of complex parameters. Being imprecise, however, we could just say that $\CH^{CS}_\hbar(C)$ ``is a holomorphic quantization'' of $(Y,\Omega_J)$.

We now proceed to discuss the three $J$-holomorphic coordinate systems on $Y = \CM_{\rm flat}(G_\C,C)$ more explicitly for the case that $C = T^2\bs \{p\}$ is a punctured torus and $G_\C = SL(2,\C)$. Thus $\Teich(C)$ is the Teichm\"uller space of the punctured torus, and the relevant four-dimensional gauge theory is the $\CN=2^*$ theory with gauge group $G=SU(2)$. In order to have a well-defined, non-degenerate holomorphic 2-form $\Omega_J$ on $Y$ when $C$ has punctures, the conjugacy class of the $G_\C$ holonomy surrounding each puncture must be fixed. For the punctured torus, we will call the puncture holonomy matrix $g_v$ and set%
\footnote{Fixing the eigenvalues is sufficient to fix the conjugacy class as long as $\ell\neq \pm 1$. At the points $\ell =\pm1$, $Y$ actually becomes singular, and one must additionally choose $g_v$ to be either diagonal or parabolic.}
\be \text{eigenvalues of $g_v$} \equiv \{\ell,\ell^{-1}\} =: \{e^v,e^{-v}\} \ee
for fixed $\ell \in \C^*/\Z_2$ (or $v\in (\C/2\pi i\Z)/\Z_2$), or
\be \Tr g_v \equiv \ell+\ell^{-1} = 2\cosh v\,. \label{gvvholonomy} \ee
(The choice of symbols $\ell$ and $v$ has to do with $A$-polynomial of knot complements in Chern-Simons theory, and will become clear in Section \ref{sec:tori}.) In terms of $\CN=2^*$ gauge theory, fixing $v$ amounts to fixing the mass of the adjoint hypermultiplet.

As a preliminary exercise, we can count the dimension of $Y$. For general $G_\C$ and $C$ of genus $g$ with $s$ punctures, we expect
\be  Y \approx {\rm Hom}(\pi_1(C),G_\C\,|\, \text{puncture evals fixed})/G_\C\,, \ee
so
\begin{align} \dim_\C Y &= (2g+s)\dim_\C G_\C - s\,{\rm rank}_\C\,G_\C - 2\dim_\C G_\C \notag \\
 &= (2g-2)\dim_\C G_\C + s(\dim_\C G_\C - {\rm rank}_\C\, G_\C)\,, \label{dimY}
\end{align}
where we begin with $(2g+s)$ generators of $\pi_1(C)$, impose the one relation among these generators in $\pi_1$ (subtracting $\dim_\C G_\C$), fix the puncture holonomy eigenvalues (subtracting $s\,{\rm rank}_\C\,G_\C$), and divide out by $G_\C$ conjugation (subtracting $\dim_\C G_\C$ again). The dimension \eqref{dimY} is always even, consistent with the fact that $Y$ is hyper-K\"ahler. For $G_\C = SL(2,\C)$ and $C = T^2\bs\{p\}$, we immediately find
\be \dim_{\C} \CM_{\rm flat}(SL(2,\C),T^2\bs\{p\}) = 2 \,. \ee

\subsubsection{Loop coordinates}
\label{sec:loop}

\begin{figure}[htb]
\centering
\includegraphics[width=4.5in]{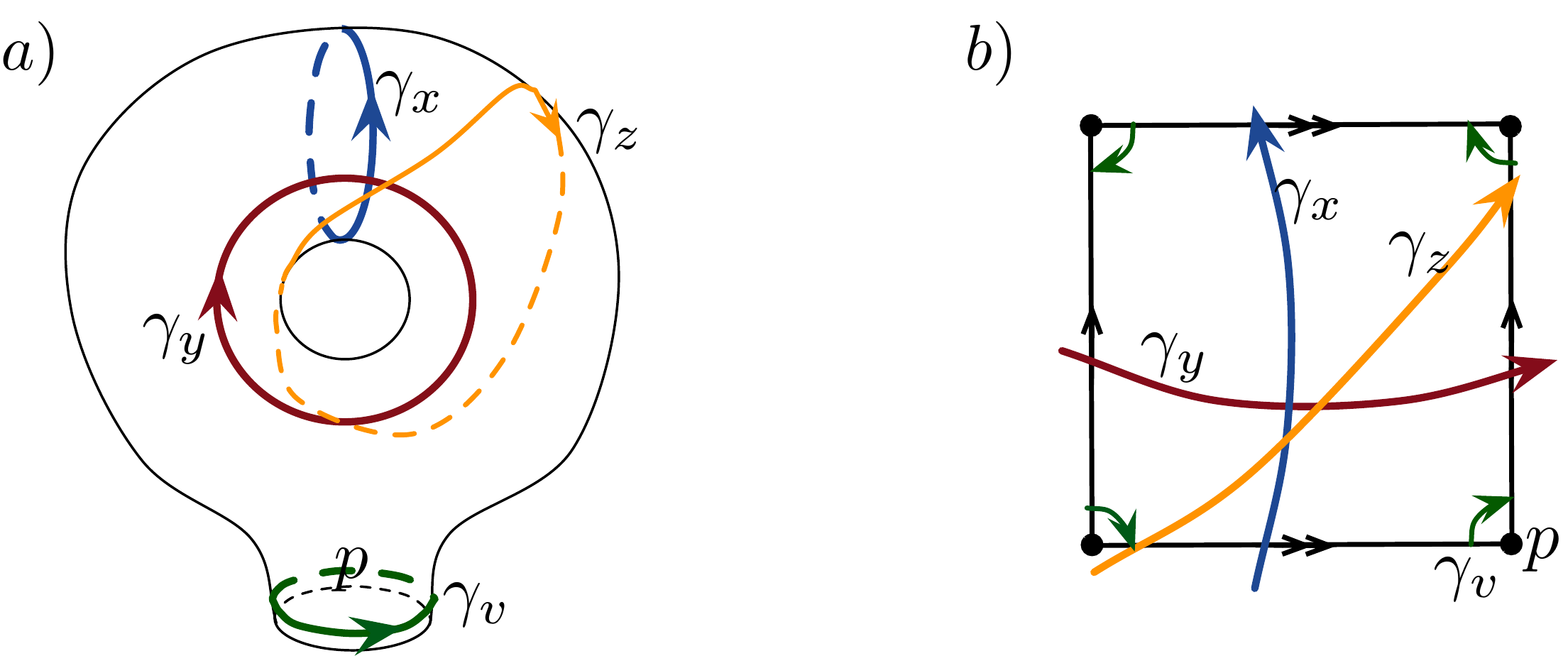}
\caption{Loops on the punctured torus.}
\label{fig:loops}
\end{figure}

Loop coordinates are given by the traces of holonomy matrices around the nontrivial cycles of $C = T^2\bs\{p\}$.
We call them ``loop coordinates'' to emphasize their relationship with loop operators in gauge and Liouville theory, though another standard name is ``Fricke-Klein coordinates,'' after the work of Fricke and Klein in the the late 1800's \cite{FrickeKlein}.
By construction, they form a global, but overdetermined, coordinate system for $Y$. Specifically, let us draw the punctured torus as in Figure \ref{fig:loops}(a), and consider loops $\gamma_x,\,\gamma_y,\,\gamma_v \in \pi_1(C)$, as well as the product (a ``diagonal'' loop) $\gamma_z = \gamma_x \gamma_y \in \pi_1(C)$. Another way to draw the punctured torus is shown in Figure \ref{fig:loops}(b), with corresponding cycles $\gamma_i$. Letting $g_i$ denote the $SL(2,\C)$ holonomy matrix around $\gamma_i$, we then define
\be -x := \Tr g_x\,,\qquad -y  := \Tr g_y\,,\qquad -z := \Tr g_z\,.
\label{xyz:def}
\ee
The triplet $(x,y,z)$ parametrizes a three-complex-dimensional space, which is one dimension too many. However, the loops $\gamma_x,\gamma_y,\gamma_z$ are not all independent; indeed,
\be \pi_1(C) = \{\gamma_x,\gamma_y\,|\,\gamma_x\gamma_y\gamma_x^{-1}\gamma_y^{-1} = \gamma_v\}\,, \ee
and by repeatedly using the well-known identity $\Tr(AB)+\Tr(AB^{-1})=\Tr(A)\Tr(B)$ for $A,B\in SL(2,\C)$, one easily finds that the relation in $\pi_1(C)$ translates to the condition
\be x^2+y^2+z^2+xyz = \Tr g_v + 2 = \ell+\ell^{-1}+2\,, \label{toruscubic} \ee
so we have
\be Y = \{x,y,z\,|\,x^2+y^2+z^2+xyz -\ell-\ell^{-1}-2=0\} \quad\subset \;\C^3\,. \ee
For $\ell \neq \pm1$, the surface $Y$ is nonsingular. Moreover, if we restrict to $g_x,g_y,g_z,g_v$ with real eigenvalues we obtain the Teichm\"uller space,
\be \Teich(C) = Y\cap\{x,y,z \leq -2\}\,. \ee
Equation \eqref{toruscubic}, which defines $Y$ as a cubic surface, is very well known in the literature
on dynamical systems and Markov processes for its large group of automorphisms, a property to which
we return in Section \ref{sec:cyl}.

Any surface that is defined as the zero-locus of a polynomial $f(x,y,z)=0$ in $\C^3$ has a natural holomorphic symplectic form given by $\Omega = dx\wedge dy\big/ (\pd f/\pd z)\,.$ In the present case, with proper normalization, we therefore find
\be \Omega_J = -\frac{2}{\hbar}\,\frac{dx\wedge dy}{xy+2z}\,. \label{Omegaloop}\ee
The fact that \eqref{Omegaloop} is nontrivial (in particular, non-diagonal) makes quantization in loop coordinates rather difficult, and ultimately is related to the fact that loop operators in $\CN=2^*$ gauge theory, or in Liouville theory on the torus, satisfy nontrivial commutation relations. The precise relation between $(x,y,z)$ and classical expectation values of Wilson and 't Hooft loops will be explained in Section \ref{sec:params}.

\subsubsection{Shear coordinates}
\label{sec:shear}

A choice of holomorphic coordinates that produces a diagonal symplectic structure and is much more suitable for quantization of $Y = \CM_{\rm flat}(G_\C,C)$ is ``shear'' (or ``Fock'') coordinates. Shear coordinates were originally introduced by Thurston \cite{Thurston-shear}, and studied systematically by Penner \cite{Penner-decorated, Penner-volumes}, Fock \cite{Fock-WP, Fock-Teich}, and others. They formed the basis for the original quantization of Teichm\"uller space \cite{FockChekhov, Kash-Teich}. Shear coordinates also enter very naturally in state integral models of Chern-Simons theory \cite{DGLZ, Dimofte-QRS}. In particular, if we consider Chern-Simons theory on a three-manifold $M$ with boundary $\pd M=C$, the shape parameters of a 3-dimensional ideal triangulation of $M$ automatically induce shear coordinates for $C$ \cite{DGGV-hybrid}. The \emph{dis}advantage of shear coordinates is that they are not very simply related to Liouville or gauge theory quantities --- like the expectation values of loop operators. A further distinction from loop coordinates $(x,y,z)$ is that they are not quite global, and only cover (algebraically) open patches of $Y$.%
\footnote{The details of combining these open patches in the case of complex shear coordinates can be found in (\eg) \cite{FG-Teich, Bonahon-shearing, GMNII}, though they will not be very important for us here.}

\FIGURE{
\;\;\includegraphics[width=1.4in]{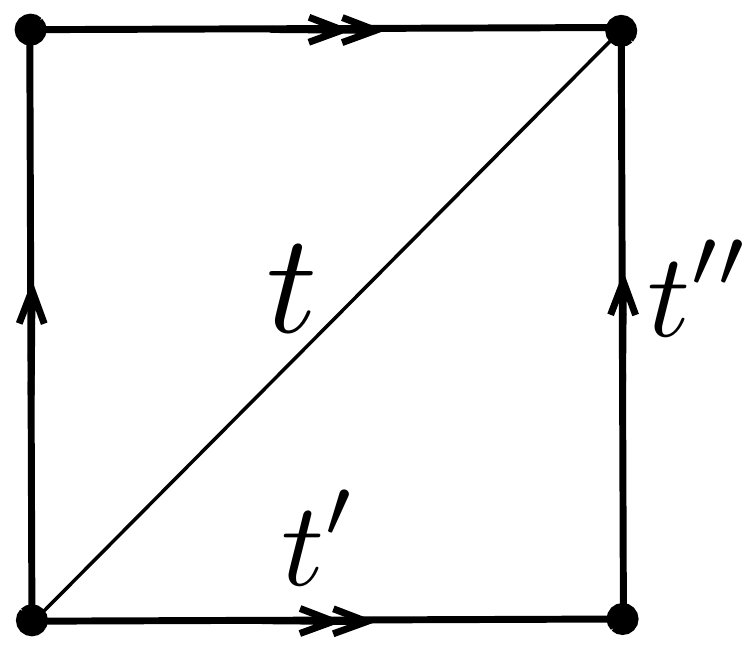}
\caption{A triangulated punctured torus $C$.\\\\}
\label{fig:triang}
}

A system of shear coordinates for an open patch of $Y$ is associated to each topological ideal triangulation of the surface $C$. An ideal triangulation is one where all edges begin and end on punctures. In the case $C=T^2\bs\{p\}$, every triangulation can be mapped to the one shown in Figure \ref{fig:triang}. To each edge, we then assign a complex shear coordinate $t,\,t',$ and $t''$, subject to the single constraint
\be t t' t'' = -\ell = e^{v+i\pi}\,, \label{shearcond} \ee
where, say, ${\rm abs}(v)\geq 0$ (otherwise, we set $tt't''=e^{-v-i\pi}$). Altogether, the three coordinates $(t,t',t'')$ parametrize a $\C^*\times \C^*$ patch of $Y$. We could also choose logarithms $(T,T',T'')$ such that $t=e^T,\,t'=e^{T'},$ and $t''=e^{T''}$; then the constraint \eqref{shearcond} becomes
\be T + T' + T'' = v+i\pi\,. \label{logvshear} \ee

Conceptually, the shear coordinate of an edge $e$ can roughly be thought of as a the square of a partial holonomy eigenvalue along a path $\gamma \subset C$ that crosses $e$. In particular, this idea leads to the correct constraint \eqref{shearcond} for the path $\gamma_v$ that surrounds the puncture $p$. Unfortunately, when considering non-boundary cycles like $\gamma_x$ or $\gamma_y$ (Figure \ref{fig:triangpath}), one must be a little more careful. Following the complete rules for constructing holonomy matrices from shear coordinates (\cf\ \cite{FockChekhov}), we find a dictionary between loop and shear coordinates:
\FIGURE{
\includegraphics[width=1.4in]{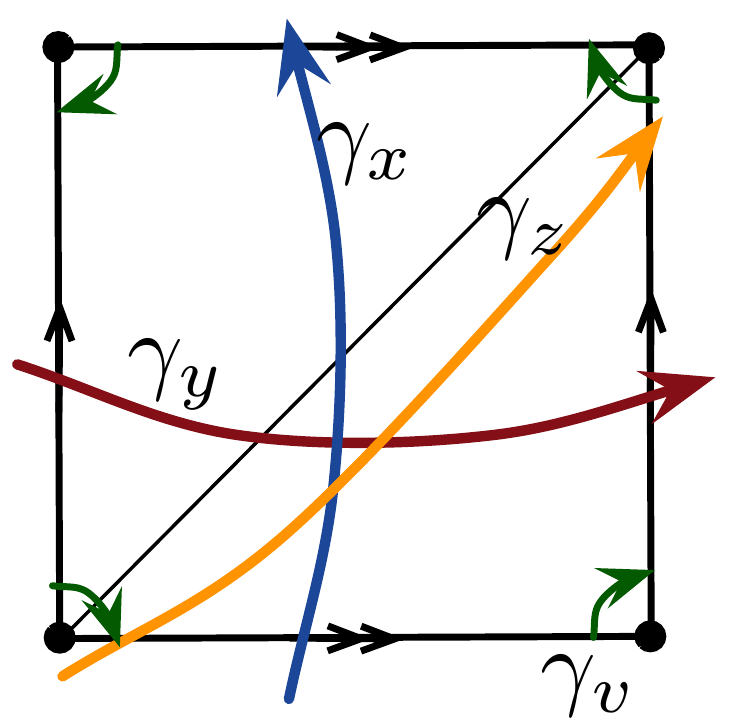}
\caption{Paths on the triangulated torus.}
\label{fig:triangpath}
}
\begin{subequations} \label{loopshear}
\begin{align}
 -x &= \Tr g_x = \sqrt{tt'}+\frac{1}{\sqrt{tt'}}+\sqrt{\frac{t'}{t}}\,, \\
 -y &= \Tr g_y = \sqrt{tt''}+\frac{1}{\sqrt{tt''}}+\sqrt{\frac{t}{t''}}\,, \\
 -z &= \Tr g_z = \sqrt{t't''}+\frac{1}{\sqrt{t't''}}+\sqrt{\frac{t'}{t''}}\,.
\end{align}
\end{subequations}

Since the $(t,t',t'')$ patch of the complex surface $Y$ is given as the zero-locus of a polynomial \eqref{shearcond}, the holomorphic symplectic form $\Omega_J$ should be proportional to $dt\wedge dt'/\pd_{t''}(tt't'') = \frac{dt}{t}\wedge \frac{dt'}{t''} = dT\wedge dT'$. In fact, the Poisson structure in shear coordinates is always given by $\frac\hbar{2}\sum_{e,e'} n_{ee'} (\pd/\pd T_e)\wedge (\pd/\pd T_{e'})$, where the sum is over edges and $n_{e,e'}$ is the (oriented) number of triangles shared by edges $e$ and $e'$
. In the present case, this quickly leads to
\be \Omega_J = \frac{1}{2\hbar}\,\frac{dt}{t}\wedge\frac{dt'}{t'} = \frac{1}{2\hbar}\,dT\wedge dT' = \frac{1}{2\hbar}\,dT'\wedge dT'' = \frac{1}{2\hbar}\,dT''\wedge dT\,,
\label{shearholom}\ee
since any two edges share $n=2$ triangles.

\subsubsection{Fenchel-Nielsen coordinates}
\label{sec:FN}

Fenchel-Nielsen coordinates on $Y$ \cite{Fenchel-Nielsen, Wolpert-deformation, Wolpert-symplectic} provide a sort of compromise between having a canonical, diagonalized symplectic form $\Omega_J$ (\ie\ having Darboux coordinates) while maintaining a simple relationship to holonomies around \emph{some} of the curves $\gamma$ on $C$. To define Fenchel-Nielsen coordinates, one begins by choosing a maximal set $\CC$ of nonintersecting closed curves $\gamma_i$ on $C$ that are not homotopic to the boundary. This is equivalent to choosing a pants decomposition for $C$. In general, for $C$ of genus $g$ with $s$ punctures, there will be $3g-3+s = \frac12\dim_CY$ curves $\gamma_i$ in $\CC$. It turns out that eigenvalues $\lambda_i^{\pm 1}$ of the corresponding holonomy matrices $g_i$, viewed as functions on $Y$, all Poisson-commute. In other words, $\Omega_J^{-1}(d\lambda_i,d\lambda_j)=0$ for any two such eigenvalues. Then, one simply has to choose $3g-3+s$ canonical duals $\tau_i$ to the $\lambda_i$ to obtain a complete set of Darboux coordinates on $Y$.

For the punctured torus $C=T^2\bs\{p\}$, the set $\CC$ of nonintersecting closed curves can only contain a single element, either $\gamma_x$, or $\gamma_y$, or any nontrivial concatenation of $\gamma_x$ and $\gamma_y$, such as $\gamma_z$. Indeed, it is easy to see that the choices for $\CC$ are in one-to-one correspondence with possible ``pants decompositions'' of $C$ (\cf\ Figure \ref{fig:appFN} in Appendix \ref{app:FNMarkov}). Let us choose $\CC = \{\gamma_x\}$, and set $\lambda_x^{\pm 1} = \exp(\pm \Lambda_x)$ to be the eigenvalues of the holonomy matrix $g_x$. Then,%
\footnote{Note that classically $\Lambda_x$ is only defined modulo $2\pi i\Z$, and up to multiplication by $\pm 1$.}
\be -x = \Tr g_x = \lambda_x+\lambda_x^{-1}= 2\cosh \Lambda_x\,. \ee
If we restrict to $\Teich(C) \subset Y$, viewed as the space of hyperbolic structures on $C$, it is well known that $\Lambda_x \in \R$ is half the hyperbolic \emph{length} of a geodesic homotopic to $\gamma_x$ (\cf\ \cite{Fenchel-Nielsen}),
\be 2|\Lambda_x| = \text{length}(\gamma_x)\,. \ee
Therefore, $\Lambda_x$ is often called a complexified length coordinate.

The coordinate $\CT_x$ that is canonically dual to $\Lambda_x$, in the sense that
\be \Omega_J = \frac{1}{\hbar}\,d\CT_x \wedge d \Lambda_x\,, \label{FNholom} \ee
is only well-defined up to the addition of any function $f(\Lambda_x)$. One standard choice for $\CT_x$ is the so-called Fenchel-Nielsen \emph{twist} $\CT_x^{(FN)}$ \cite{Fenchel-Nielsen, Wolpert-symplectic}, which in terms of $\Teich(C)$ and hyperbolic structures literally describes how far one has to twist two legs of a hyperbolic pair of pants before gluing them together to form our punctured torus $C$.
This is described in further detail in Appendix \ref{app:FNMarkov}. The loop coordinates $y$ and $z$ are related to the Fenchel-Nielsen twist $\CT_x^{(FN)}$ and its exponential $\tau_x^{(FN)} = \exp(\CT_x^{(FN)})$ in a fairly complicated way, as (\cf\ \cite{Okai-pants})
\begin{subequations} \label{loopFNgeom}
\begin{align}
-y &= \Tr g_y = \big(\tau_x^{(FN)}{}^{\frac12}+\tau_x^{(FN)}{}^{-\frac12}\big)\frac{\sqrt{\lambda_x^{\,2}+\lambda_x^{-2}-\ell-\ell^{-1}}}{\lambda_x-\lambda_x^{-1}}\,,\\
-z &=\Tr g_z = \big( \lambda_x\tau_x^{(FN)}{}^{\frac12}+\lambda_x^{-1}\tau_x^{(FN)}{}^{-\frac12}\big) \frac{\sqrt{\lambda_x^{\,2}+\lambda_x^{-2}-\ell-\ell^{-1}}}{\lambda_x-\lambda_x^{-1}}\,.
\end{align}
\end{subequations}
These standard complex Fenchel-Nielsen coordinates $(\Lambda_x,\CT_x^{(FN)})$ are identical to the Darboux coordinates $(\alpha,\beta)$ used recently by \cite{NRS} (not just for the punctured torus, but for any punctured Riemann surface).

In the relation to Liouville theory and gauge theory, it is a bit more natural to choose a different twist coordinate $\CT_x$, related to $\CT_x^{(FN)}$ as
\be \tau_x := \frac{\ell^{\frac12} \lambda-\ell^{-\frac12}\lambda^{-1}}{\ell^{-\frac12}\lambda-\ell^{\frac12}\lambda^{-1}}\tau_x^{(FN)}
= \frac{\sinh(\Lambda_x+v/2)}{\sinh(\Lambda_x-v/2)}\exp({\CT_x^{(FN)}})\,, \qquad \tau_x := \exp(\CT_x)\,. \label{FNgeom} \ee
This choice of twist reflects a natural choice of polarization for Liouville conformal blocks and Nekrasov partition functions \cite{Tesch-loop}, as well as Chern-Simons partition functions \cite{DGGV-hybrid}. We will say more about this in Section \ref{sec:eg}.
Dropping the subscripts ``$x$'' from $(\Lambda_x,\CT_x)$, we must still have
\be \Omega_J = \frac{1}{\hbar}\,d\CT\wedge dL = \frac{1}{\hbar}\,\frac{d\tau}{\tau}\wedge \frac{d\lambda}{\lambda}\,, \ee
and now
\begin{subequations} \label{loopFN}
\begin{align}
-x = \Tr g_x &= \lambda+\lambda^{-1}\,,\\
-y = \Tr g_y &= \frac{\ell^{-\frac 12}\lambda-\ell^{\frac 12}\lambda^{-1}}{\lambda-\lambda^{-1}}\tau^{\frac 12}+\frac{\ell^{\frac12}\lambda-\ell^{-\frac 12}\lambda^{-1}}{\lambda-\lambda^{-1}}\tau^{-\frac 12}\,, \\
-z = \Tr g_z &= \frac{\ell^{-\frac 12}\lambda-\ell^{\frac 12}\lambda^{-1}}{\lambda-\lambda^{-1}}\lambda^{-1}\tau^{\frac 12}+\frac{\ell^{\frac12}\lambda-\ell^{-\frac 12}\lambda^{-1}}{\lambda-\lambda^{-1}}\lambda\tau^{-\frac 12}\,.
\end{align}
\end{subequations}
The explicit relation to shear coordinates can also be written down. After taking square roots of $t,t',t''$, we find
\be \label{shearFN}
 \sqrt{t} = -i\frac{\tau^{\frac12}-\tau^{-\frac12}}{\lambda-\lambda^{-1}}\,,\qquad \sqrt{t'}=i\frac{\lambda-\lambda^{-1}}{\tau^{\frac12}/\lambda-\lambda/\tau^{\frac12}}\,,\qquad\sqrt{t''} = -i\sqrt{\ell}\frac{\tau^{\frac12}/\lambda-\lambda/\tau^{\frac12}}{\tau^{\frac12}-\tau^{-\frac12}}\,.\ee


\subsection{Coordinates and moduli of $\CN=2$ gauge theories}
\label{sec:params}

As we presented loop, shear, and Fenchel-Nielsen coordinates in Section \ref{sec:MHit}, we mentioned how each was more or less related to Chern-Simons theory, Liouville theory, and gauge theory. Here we make these relations a bit more precise, reviewing the (generally well-established) dictionary between coordinates and semiclassical physical parameters.

\FIGURE{
\includegraphics[width=2in]{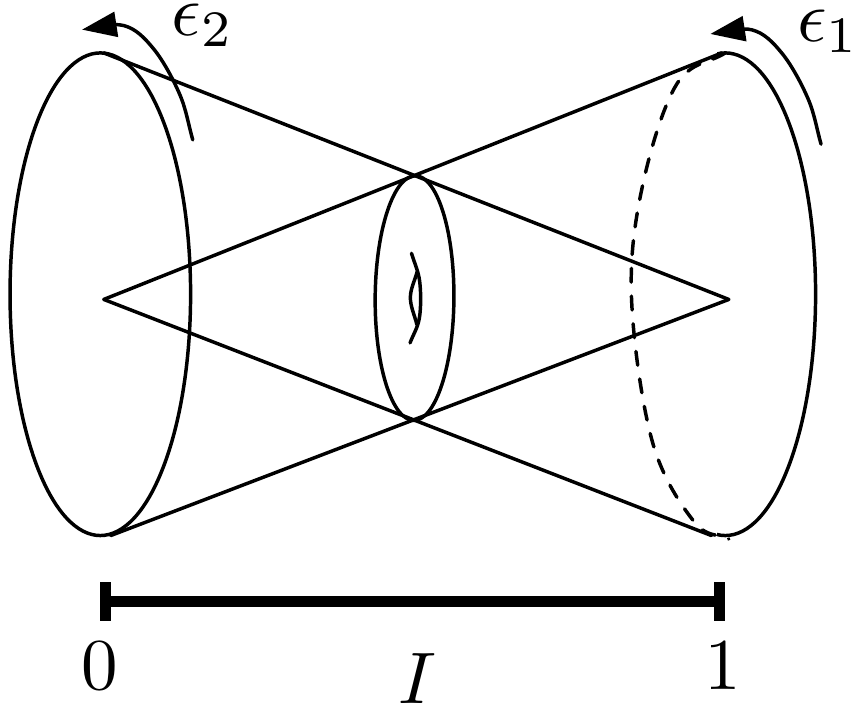}
\caption{$\Omega$-deformed $S^3$}
\label{fig:S3}
}

Let us consider $\Omega$-deformed $\CN=2^*$ gauge theory on $S^3\times \R$, obtained by compactifying the theory of two M5 branes on $S^3\times\R\times C$ for $C=T^2\bs \{p\}$. As described very carefully in \cite{NekWitten}, the $\Omega$-deformation is defined by the equivariant action of two $U(1)$ rotations inside $S^3$, with parameters $\epsilon_1$ and $\epsilon_2$. It is clear that two such rotations are possible, since the isometry group of $S^3$ is $SO(4)\approx SU(2)\times SU(2)$, which contains a subgroup $U(1)_{\epsilon_1}\times U(1)_{\epsilon_2}$. More explicitly, one can view $S^3$ as a fibration over an interval $I = [0,1]$ whose generic fibers are $T^2 = S^1_{\epsilon_1}\times S^1_{\epsilon_2}$, on which the group $U(1)_{\epsilon_1}\times U(1)_{\epsilon_2}$ acts by respective rotations. The two cycles $S^1_{\epsilon_1}$ and $S^1_{\epsilon_2}$ degenerate to points at the endpoints $\{0\}$ and $\{1\}$ of $I$, respectively (Figure \ref{fig:S3}). 

After further reduction of the six-dimensional theory on $C \times T^2$, one obtains a two-dimensional sigma-model on $\R \times I$, with suitable boundary conditions $\CB_0$ and $\CB_1$ at the end-points of $I$, dictated by the geometry of the fibration $T^2 \to I$.
These boundary conditions can be conveniently described as ``branes'' in the target space $Y$ of the resulting sigma-model \cite{NekWitten}, and for the problem at hand one finds that both boundary conditions are related to the so-called ``brane of opers'' $\CB_{{\rm opers}}$, which is holomorphic in complex structure $J$ and Lagrangian with respect to $i \Omega_J = \omega_K + i \omega_I$.
In other words, $\CB_{{\rm opers}}$ is a brane of type $(A,B,A)$. It is supported on a middle-dimensional submanifold of $Y$, known as the space of opers.
 In the complex Fenchel-Nielsen coordinates, the space of opers can be defined by the conditions:
\be
\CT_i = \frac{\partial \CY (\Lambda)}{\partial \Lambda_i} \,,
\label{opers}
\ee
where $\CY (\Lambda)$ is the so-called Yang-Yang function \cite{NRS} which, in the limit of large $\Lambda_i$, coincides with the prepotential of the corresponding four-dimensional $\CN=2$ gauge theory. Clearly, these equations define a subvariety of $Y$ which is Lagrangian with respect to the holomorphic symplectic form \eqref{FNholom}.
As a complex manifold, the space of opers is naturally isomorphic to the space of quadratic differentials
on $C$, {\it i.e.} the base of the Hitchin fibration \cite{Hitchin-SD}:
\be
{\bf B} :=  H^0 (C, K_C^2) \,.
\ee
Note, however, that the brane of opers $\CB_{{\rm opers}}$ is a brane of type $(A,B,A)$, whereas the brane $\CB_{{\bf B}}$ supported on the base of the Hitchin fibration is a brane of type $(B,A,A)$, much like the brane $\CB_{\Teich}$ supported on $\Teich(C)$:
\be \begin{array}{ccc}
\underline{\text{\emph{brane}}} && \underline{\text{\emph{type}}} \vspace{.2cm} \\
\CB_{\Teich} &:~~  & (B,A,A) \\
\CB_{{\rm opers}} &:~~ & (A,B,A) \\
\CB_{{\bf B}} &:~~  & (B,A,A) \\
\end{array} \label{branetypes} \ee
Indeed, the base ${\bf B}$ is parametrized by the Coulomb branch parameters of $\CN=2$ gauge theory, which are holomorphic
functions on $Y$ in complex structure $I$. However, when restricted to the brane of opers they appear to coincide with the $J$-holomorphic coordinates $\Lambda_i$.%
\footnote{We thank A. Neitzke and D. Gaiotto for helpful discussions on this point.} %
Various aspects of the distinguished branes \eqref{branetypes} were discussed in \cite{NekWitten,GMNII,NRS,Teschner-opers,mirrorbranes}.

In the two-dimensional sigma-model on $\R \times I$, the Hilbert space $\CH^{\CN=2}_{\epsilon_{1,2}} (S^3)$
is simply the space of open strings between branes $\CB_0$ and $\CB_1$ on $Y$,
\be
\CH^{\CN=2}_{\epsilon_{1,2}} (S^3) \; = \; {\rm space~of~} (\CB_0,\CB_1) {\rm ~strings} \,.
\ee
In the present case, it leads to the general setup of ``brane quantization'' \cite{GW-branes},
so that $\CH^{\CN=2}_{\epsilon_{1,2}} (S^3)$ can be identified with the quantization of the space of opers.
Conjecturally, this problem is equivalent to the quantization of the Teichm\"uller space \cite{NekWitten,Teschner-opers},
thus, justifying one of the key relations $\CH^{Liouv}_b(C) = \CH^{\CN=2}_{\epsilon_{1,2}}(S^3)$ in the AGT correspondence,
{\it cf.} \eqref{Hilbsquare}. Moreover, in this approach the ``$\CN=4$ S-duality'' \eqref{Lh}
has an elegant geometric interpretation as a symmetry that exchanges the two $S^1$'s in the fibration of Figure \ref{fig:S3}
or, equivalently, as a modular transformation of $T^2 = S^1_{\epsilon_1}\times S^1_{\epsilon_2}$,
\be
\hbar \to {}^L\hbar = - \frac{4\pi^2}{\hbar} \,.
\ee
Indeed, according to \eqref{he} a symmetry that exchanges $\epsilon_1 \leftrightarrow \epsilon_2$
transforms $\hbar  = 2\pi i \frac{\epsilon_1}{\epsilon_2}$ into ${}^L\hbar = 2\pi i \frac{\epsilon_2}{\epsilon_1} = - \frac{4\pi^2}{\hbar}$.
This symmetry can be also viewed as the electric-magnetic duality of the $\CN=4$ super-Yang-Mills
with gauge group $G$ --- obtained by compactifying the six-dimensional theory
on $T^2 = S^1_{\epsilon_1}\times S^1_{\epsilon_2}$ --- thus, finally justifying our choice of terminology.

\FIGURE{
\;\;\includegraphics[width=1.5in]{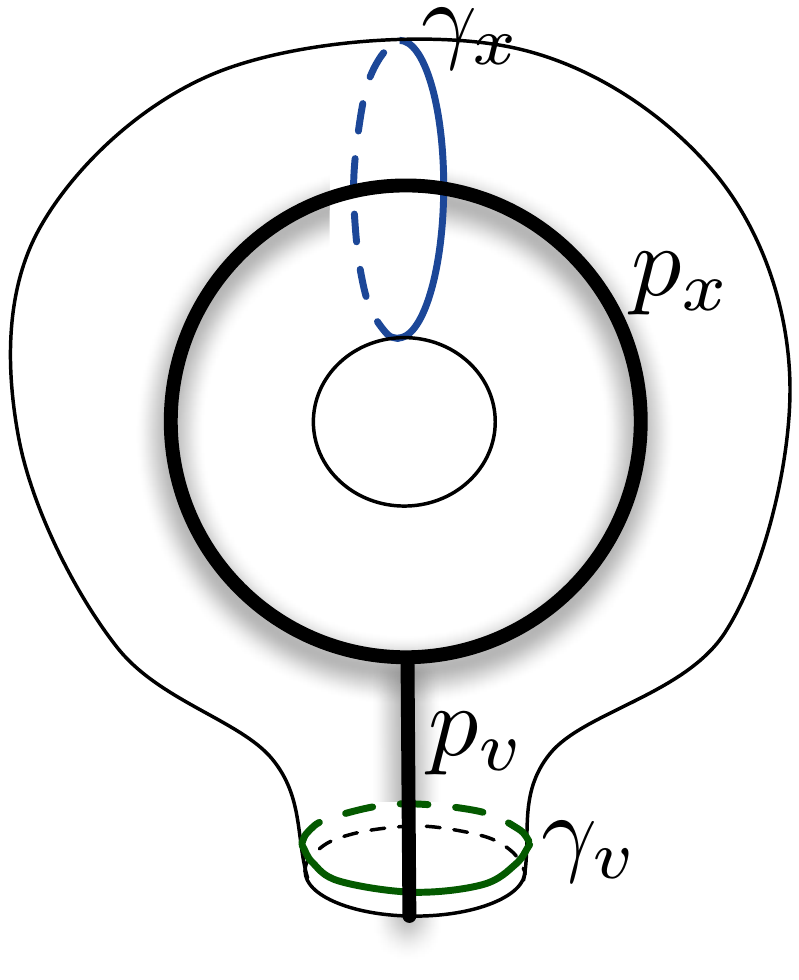}
\caption{Moore-Seiberg graph as the skeleton of $C$. \\}\label{fig:MSgraph}
}

Another important part of the AGT dictionary \cite{AGT} is the relation of Coulomb vevs $a_j$ appearing in the Nekrasov partition function with the Liouville momenta $p_j$, or with the corresponding length coordinates $\Lambda_j$,
\be
\frac{a_j}{\sqrt{\epsilon_1\epsilon_2}} = i p_j = - \frac{\Lambda_j}{2 \pi i b} \,,
\label{lprelation}
\ee
where $b = \sqrt{\epsilon_1/\epsilon_2}$, or $\hbar = 2\pi ib^2$.
In the Liouville theory, physical values of $p_j \in \R$ correspond to the primary fields
$e^{2 \alpha_j \phi}$ with
\be
\alpha_j = \frac{Q}{2} + i p_j = \frac{Q}{2} + \frac{a_j}{\sqrt{\epsilon_1\epsilon_2}}
\ee
and with conformal dimension $\Delta_j = \alpha_j (Q - \alpha_j) = \frac{Q^2}{4} + p_j^2$, where $Q = b+b^{-1}$.
In the four-dimensional $\CN=2$ gauge theory, they correspond to pure imaginary values
of the vevs $a_j$.

The dictionary \eqref{lprelation} between Coulomb moduli and classical coordinates
also extends to mass parameters of the $\CN=2$ gauge theory \cite{AGT}.
In our prime example of $C=T^2\bs \{p\}$, the eigenvalue $v$ of the holonomy \eqref{gvvholonomy}
at $p$ corresponds to the mass parameter of the adjoint matter multiplet
of the $\CN=2^*$ theory \cite{OkudaPestun},
\be
\frac{m_{\rm adj}}{\sqrt{\epsilon_1\epsilon_2}} = i p_v = - \frac{v}{2 \pi i b}-\frac{Q}{2} \,,
\label{madjvrelation}
\ee
where the subscript of $p_v$ refers to the ``external'' Liouville momentum in the Moore-Seiberg graph shown in Figure \ref{fig:MSgraph}.

More generally, given any path $\gamma$ on $C$ one can associate to it an $SL(2,\C)$ holonomy matrix $g_{\gamma}$, as we did in Section \ref{sec:loop}:
\be
g_{\gamma} \; = \; P \exp \oint_{\gamma} \CA \,,
\ee
where $\CA$ is the gauge connection\footnote{To be more precise, depending on whether one is interested in a unitary theory or its analytic continuation, the gauge connection $\CA$ is either $\mathfrak{g}_{\R}$- or $\mathfrak{g}_{\C}$-valued.} of complex Chern-Simons theory on $M= \R \times C$. In particular, the coordinates $(x,y,z)$ introduced in \eqref{xyz:def} are special examples of the operators $W_{\gamma} \equiv \Tr g_{\gamma}$ associated to cycles shown on Figure \ref{fig:loops}. In classical Chern-Simons theory, the operators $W_{\gamma} = \Tr g_{\gamma}$ are ($J$-holomorphic) functions on the moduli space \eqref{Mflat}. Upon quantization, they become elements of the non-commutative algebra $\CA^{CS}_{\hbar}(C)$.

\FIGURE{
\includegraphics[width=4in]{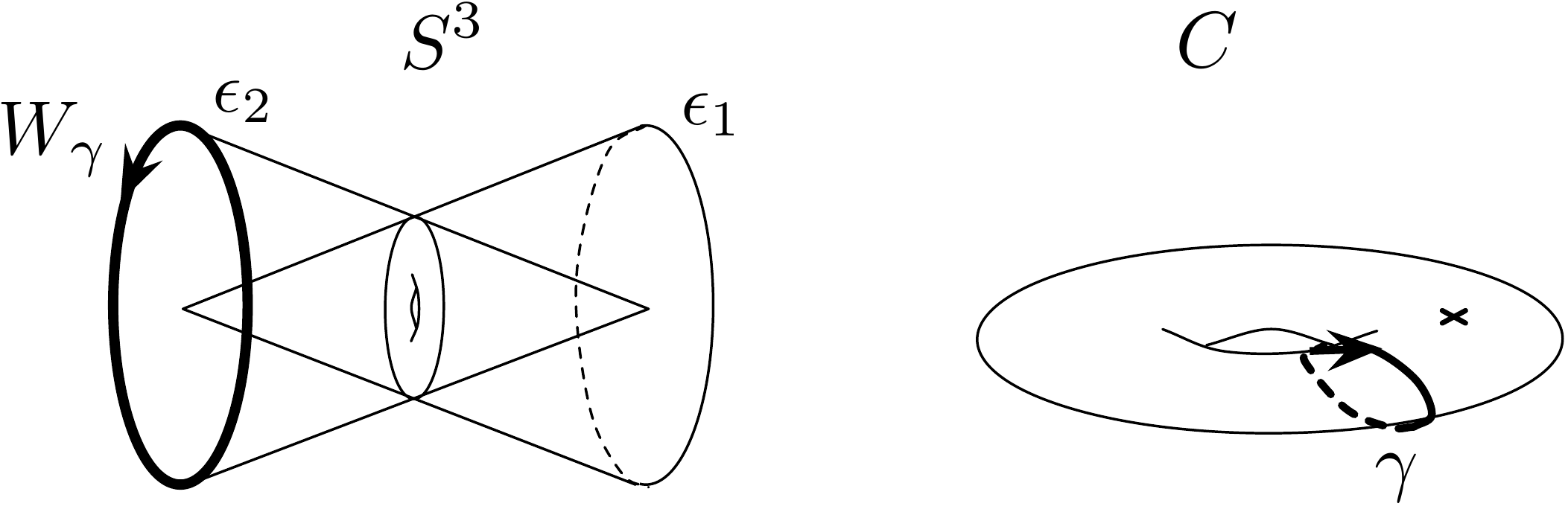}
\caption{Loop operator of charge $\gamma$ coming from a surface operator in $S^3\times\R\times C$.}
\label{fig:S3loops}
}

From the viewpoint of the four-dimensional $\CN=2$ gauge theory --- constructed by compactifying the 6d five-brane theory on $C$ --- the operators $W_{\gamma}$ can be identified either as half-BPS UV line operators \cite{DMO} or, via passing to a double cover of $C$, as BPS charges in the four-dimensional $\CN=2$ supersymmetry algebra \cite{SW-I,Lerche-noncrit}. (For example, a Wilson loop can be equivalently viewed as a static quark, a 't Hooft loop as a monopole source, {\it etc}.) The BPS charge of every such operator is determined by the choice of $\gamma \in \pi_1 (C)$ which, for a given ``pants decomposition'' of $C$, can be written as a word in the A-cycles and B-cycles, regarded as generators of $\pi_1 (C)$. The resulting line operators include familiar Wilson operators $W_j$ and 't Hooft operators $H_j$, labeled by a half-integer spin $j$ of $G=SU(2)$ and associated to holonomies around A- and B-cycles on $C$, respectively. For example, in the case of the $\CN=2^*$ theory, the primitive cycles $\gamma_x$, $\gamma_y$, and $\gamma_z$ shown on Figure  \ref{fig:loops} correspond, respectively, to Wilson, 't Hooft, and Wilson-'t Hooft operators of lowest charge (\ie\ of spin $j=\frac{1}{2}$), so that semiclassically
\be
W_{1/2} = 2 \cosh \Lambda_x = -x \qquad , \qquad H_{1/2} = 2 \cosh \Lambda_y = -y \,.
\ee
In particular, this clarifies why we refer to the holomorphic coordinates of Section \ref{sec:loop} as the ``loop coordinates.''

When $\CN=2$ four-dimensional gauge theory is subject to the $\Omega$-deformation, supersymmetry requires line operators to be invariant under the $U(1)_{\epsilon_1}\times U(1)_{\epsilon_2}$ symmetry action. This includes interesting line operators supported on the great circles $S^1 \subset S^3$, \ie\ on the singular fibers at the two endpoints of $I$ in Figure \ref{fig:S3}. Semiclassically, only the operators supported on $S^1_{\epsilon_2}$ above $\{0\}\in I$ are visible (Figure \ref{fig:S3loops}). For example, according to the pants decomposition of Figure \ref{fig:MSgraph} and the dictionary \eqref{lprelation}, the Wilson loop at $\{0\}\in I$ has semiclassical expectation value
\be W_{1/2} = 2\cosh\frac{2\pi ia}{\epsilon_2}\,,\ee
which remains finite as $\epsilon_1/\epsilon_2\to 0$. The loop operators on the other circle $S^1_{\epsilon_1}$ will appear once we quantize, in Section \ref{sec:quant}.


\subsection{Mapping cylinders and Lagrangians}
\label{sec:cyl}

Having reviewed the classical geometry of the Hitchin moduli space $Y=\CM_H(G,C) = \CM_{\rm flat}(G_\C,C)$ and its relation to Chern-Simons theory, Liouville theory, and gauge theory, we now begin to look at the classical limit of ``$\CN=2$ S-duality'' actions. Geometrically, an S-duality of $\CN=2$ gauge theory --- coming from compactification on $S^3\times \R\times C$ --- corresponds to a mapping class transformation on $C$ \cite{Gaiotto-dualities},
\be \varphi\,:\;C\to C\,. \ee
As explained in the introduction, it is very natural to visualize such an action as encoded by a 3-dimensional mapping cylinder $M_\varphi = C\times_\varphi I$ (Figure \ref{fig:MCgeneric}). Topologically, this space is homeomorphic to the trivial product $C\times I$. However, the top and bottom boundaries $C$ and $\ol{C}$ are thought of (and, in particular, coordinatized) as if they were twisted relative to one another by $\varphi$,
\be C = \varphi(\ol C)\,. \ee
By putting a bar `\,$\raisebox{.07cm}{--}$\,' on $\ol C$, we indicate that, as a boundary of $M_\varphi$, its orientation is reversed relative to that of the top boundary $C$.

In Chern-Simons theory, one associates a semi-classical phase space $\CP_{\pd M} = \CM_{\rm flat}(G_\C,\pd M)$ to the boundary of any 3-manifold $M$. In the case of a mapping cylinder $M_\varphi = C\times I$, this is
\be \CP_{\pd M_\varphi} = \CP_C\times \CP_{\ol C} = Y\times \ol Y\,, \label{Pcyl} \ee
with holomorphic symplectic form
\be \Omega_{\pd M_\varphi} = \Omega_J - \ol\Omega_J\,. \ee
The opposite orientation of $\ol C$ inverts the sign of the symplectic structure on $\ol Y$. Thus, for example, if $C=T^2\bs\{p\}$ is a punctured torus and we parametrize $Y$ and $\ol Y$ in shear coordinates $(t,t',t'')$ and $(\adj t,\adj t', \adj t'')$, respectively, then
\be \Omega_{\pd M_\varphi} = \frac1{2\hbar}\frac{dt}{t}\wedge \frac{dt'}{t'}-\frac1{2\hbar}\frac{d\adj t}{\adj t}\wedge \frac{d\adj t'}{\adj t'}\,. \ee
We will always use flats `$\flat$' to distinguish coordinates on $\ol Y$.

While the boundary of any 3-manifold $M$ determines the phase space $\CP_{\pd M}$, the actual internal structure of $M$ defines a Lagrangian submanifold $\Delta_M \subset \CP_{\pd M}$. This Lagrangian is the set of flat $G_\C$ connections on $\pd M$ that can extend to be flat connections throughout all of $M$,
\be \Delta_M = \{\text{flat connections in bulk $M$}\}\quad\subset \quad
\CP_{\pd M} = \{\text{flat conn$^{\rm s}$ on bdy $\pd M$}\}\,. \label{defDelta}
\ee
In the particular case of a mapping cylinder, $\Delta_\varphi \equiv \Delta_{M_\varphi}$ is very easy to characterize: it is the graph of the coordinate transformation on $Y$ induced from the mapping class group action of $\varphi$. Such a graph is called a ``correspondence'' in $Y\times \ol{Y}$,
\be
\Delta_{\varphi} \; \subset \; Y \times \ol{Y} \,. \label{correspondence}
\ee
For example, if we are working in shear coordinates for a punctured torus $C$ (Figure \ref{fig:MCshear}), and the element $\varphi$ corresponds to a map $t = f_1(\adj t,\adj t')$, $t'= f_2(\adj t,\adj t')$ (with the action on $\adj t''$ given implicitly, since $tt't''=\adj t\adj t'\adj t''=-\ell$), then the Lagrangian is just
\be \Delta_\varphi = \{t-f_1(\adj t,\adj t')=0,\,t'-f_2(\adj t,\adj t')=0\}\,. \label{Deltaeg} \ee
The fact that the induced mapping class group actions on $Y$ always preserve the symplectic structure $\Omega_J$ guarantees that their graphs define Lagrangian submanifolds of $\CP_{\pd M_\varphi}$.

\FIGURE{
\includegraphics[width=1.4in]{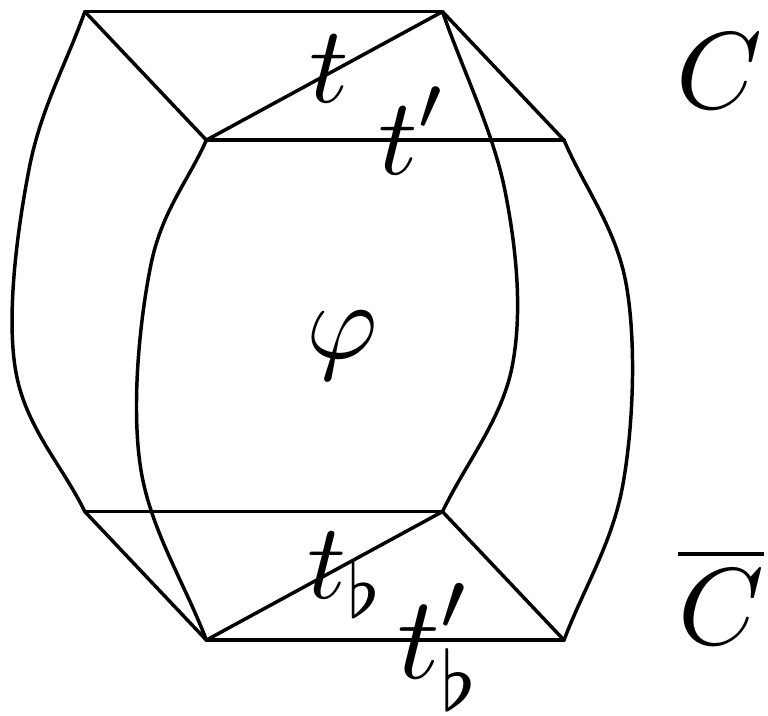}
\caption{A mapping cylinder in shear coordinates for $C=T^2\bs\{p\}$}
\label{fig:MCshear}
}

The notion of extending a flat connection to $M$ in \eqref{defDelta} must be treated with some care when the boundary of $M$ has punctures. In particular, the definition of the 3-manifold $M$ in this case should specify a network of codimension-one holonomy defects that end at the punctures of $\pd M$. (In Chern-Simons theory, holonomy defects are equivalent to Wilson loops \cite{Wit-Jones, Rozansky-triv1}.) For a mapping cylinder $M_\varphi = C\times_\varphi I$, punctures can be extended naturally as ``timelike'' line defects, whose braiding in the vertical time direction is fully specified by the mapping class group element $\varphi$. In terms of branes, $\varphi$ acts as an autoequivalence of the derived category of branes on $Y$, {\it i.e.} as a functor that maps
\be
\varphi \,:\, \adj \CB \mapsto \CB \,.
\label{actiononbranes}
\ee
For example, if $\CB$ is an $(A,B,A)$ brane supported on $A(t,t')=0$ and $\varphi$ is described by \eqref{Deltaeg} then $\adj \CB$ is also a brane of type $(A,B,A)$ defined by the zero locus of $A(f_1(\adj t,\adj t'), f_2(\adj t,\adj t'))$. In Section \ref{sec:tori} we will illustrate how mapping class group/braid group actions on branes give rise to familiar knot invariants, such as the $A$-polynomial.

Given a mapping cylinder $M_\varphi$, quantum Chern-Simons theory computes a partition function or wavefunction $Z_\varphi$. This is a wavefunction in the sense that
\be Z_\varphi \in \CH_\hbar(C)\otimes \CH_\hbar(C)^*\,, \label{ZDelta} \ee
where $\CH_\hbar(C)$ is the Hilbert space obtained from quantization of $Y$ with respect to $\Omega_J$ and $\CH_\hbar(C)^*$ is its dual, obtained from quantization of $\ol{Y}$ with respect to $-\ol \Omega_J$. More precisely, this wavefunction can be chosen to depend on a maximal commuting set of coordinates on $Y\times \ol{Y}$. Thus, for $C=T^2\bs\{p\}$, we could have $Z_\varphi = Z_\varphi(T,\adj T)$ in (logarithmic%
\footnote{Analytically continued Chern-Simons wavefunctions always depend on the logarithms of coordinates on $Y$ rather than  the coordinates themselves. This subtle fact manifested itself in \cite{gukov-2003, DGLZ} and was further discussed in \cite{Wit-anal, Dimofte-QRS}.}%
) shear coordinates or $Z_\varphi= Z_\varphi(\Lambda,\adj \Lambda)$ in (logarithmic) Fenchel-Nielsen coordinates. As explained in the introduction, a wavefunction $Z_\varphi$ should also be identified with a Moore-Seiberg kernel for Liouville theory or an S-duality kernel for $\CN=2$ gauge theory; then the classical parameters (such as $(T,\adj T)$) that $Z_\varphi$ depends on are identified with Liouville momenta or Coulomb vevs according to the dictionary in Section \ref{sec:params} above. We postpone further details of the quantum wavefunctions $Z_\varphi$ and the operators that annihilate them --- quantized versions $\wh\Delta_\varphi$ of the Lagrangians $\Delta_\varphi$ --- until Section \ref{sec:qcyl}.

\begin{figure}[htb]
\centering
\medskip
\includegraphics[width=5in]{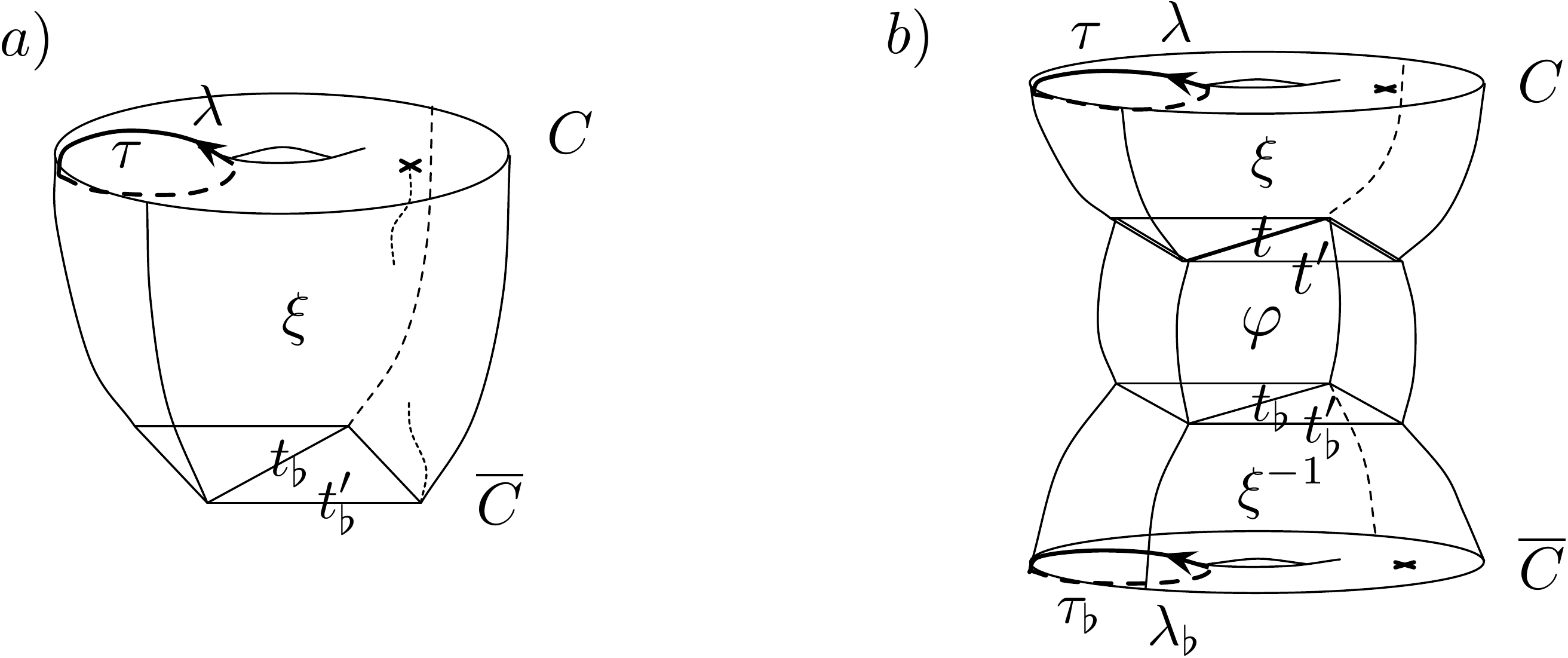}
\caption{Sandwiching of coordinate-transformation cylinders}
\label{fig:sFNcyl}
\end{figure}

While we have described mapping cylinders $M_\varphi$ and Lagrangians $\Delta_\varphi$ for actual mapping class group elements $\varphi$, it is easy to extend our picture to include any coordinate transformation $\xi\,:\;Y\to Y$. For example, we could construct a mapping cylinder $M_\xi$ that interpolates between shear and Fenchel-Nielsen coordinates on $Y$ (Figure \ref{fig:sFNcyl}(a)). Its Lagrangian $\Delta_\xi$ is given by
\be \Delta_\xi\,:\quad i\frac{\tau^{\frac12}-\tau^{-\frac12}}{\lambda-\lambda^{-1}} +\sqrt{\adj t}=0\,,\qquad i\frac{\lambda-\lambda^{-1}}{\tau^{\frac12}/\lambda-\lambda/\tau^{\frac12}} -\sqrt{\adj t'}=0
\label{Deltaxi}
\ee
when $C=T^2\bs\{p\}$, with $Y\times \ol{Y}$ parametrized by $(\lambda,\tau,\adj t,\adj t')$.
The Chern-Simons wavefunction $Z_\xi(\Lambda,\adj T)$ for such a cylinder is simply the kernel that transforms wavefunctions from one coordinate system to another. It is familiar from the study of quantum Teichm\"uller \cite{Kash-kernel} and Liouville theory \cite{Teschner-TeichMod}. Having $M_\xi$ and its wavefunction, one could then obtain a mapping class kernel $Z_\varphi(\Lambda,\adj \Lambda)$ in Fenchen-Nielsen coordinates by sandwiching a shear-coordinate mapping cylinder $M_\varphi$ in between $M_\xi$ and its inverse $M_{\xi^{-1}}$ (Figure \ref{fig:sFNcyl}(b)). In terms of wavefunctions,
\be Z_\varphi(\Lambda,\adj \Lambda) = \int dTd\adj T\, Z_\xi(\Lambda,T)\,Z_\varphi(T,\adj T)\,Z_{\xi^{-1}}(\adj T,\adj \Lambda)\,.\ee

Such constructions of mapping cylinders (and concatenations of mapping cylinders) are quite useful for understanding various operations in Chern-Simons, Liouville, and gauge theory. Moreover, they are \emph{not} merely academic tools. For example, in gauge theory, any of the mapping cylinders just described can be interpreted as an interesting duality domain wall. In Chern-Simons theory, both mapping-class and coordinate-transformation cylinders can be given natural 3d triangulations. The Chern-Simons wavefunctions $Z_\varphi(T,\adj T)$, $Z_\xi(\Lambda,\adj T)$, etc. can then be computed via the methods of \cite{Dimofte-QRS}.%
\footnote{We will discuss this further in \cite{DGGV-hybrid}. See also \cite{Yamazaki-layered} for a recent example of mapping class group kernels computed via triangulations related to Teichm\"uller theory.}

We now describe in detail the induced mapping class group actions on the three main coordinate systems of $Y = \CM_{\rm flat}(SL(2,\C),C)$ for a punctured torus $C=T^2\bs\{p\}$.

\subsubsection{Loop coordinates}
\label{sec:loopclassMCG}

We begin with loop (or Fricke-Klein) coordinates, where the action of the mapping class group is most intuitive and simplest to describe. We follow the detailed discussion in \cite{GoldmanNeumann}. The automorphism group that acts on the complex surface $Y$, viewed as the zero-locus of the Markov cubic \eqref{toruscubic}, is a semidirect product
\be {\rm Aut}(Y) = PGL(2,\Z) \ltimes \Xi\,. \label{AutY} \ee
Here $PGL(2,\Z) = GL(2,\Z)/\{\pm I\}$ is a double cover of the mapping class group of the (oriented) punctured torus $C$,
\be \bm \Gamma(C) = PSL(2,\Z) = PGL(2,\Z)\big/\big\langle I,\,\left(\begin{smallmatrix}1&0\\0 &-1\end{smallmatrix}\right)\big\rangle\,. \ee
The group $PSL(2,\Z)$ acts on the cycles $(\gamma_x,\gamma_y)$ discussed in Section \ref{sec:MHit} by standard matrix multiplication. The element $\kappa = \left(\begin{smallmatrix} 1&0\\0&-1 \end{smallmatrix}\right)$ that extends $\Gamma(C) = PSL(2,\Z)$ to $PGL(2,\Z)$ also acts by matrix multiplication, but reverses the orientation of $C$ --- and hence the sign of the symplectic form $\Omega_J$ on $Y$. The extra factor $\Xi \cong \Z_2\times \Z_2$ in \eqref{AutY} is the Klein 4-group, and acts on loop coordinates as $(x,y,z)\mapsto (\pm x,\pm y,\pm z)$ with an even number of sign changes.

To understand how loop coordinates transform under $PGL(2,\Z)$, we can first consider the action of $S = \left(\begin{smallmatrix} 0&-1\\1&0 \end{smallmatrix}\right)$. Since $S:\,(\gamma_x,\gamma_y)\mapsto (\gamma_y^{-1},\gamma_x)$ on $C$, the loop coordinates $(x,y)$ must map to $(y,x)$. In terms of gauge theory, this is a version of the statement that
\be
S\,:\,(a,a^D)\mapsto (-a^D,a)\,.
\ee
The transformation of the third coordinate $z$ follows by imposing the cubic equation \eqref{toruscubic} and requiring that the full transformation preserves the symplectic form $\Omega_J$ \eqref{Omegaloop}, including its sign. The simple result is that $z\mapsto -z-xy$.

Similarly, one can write down the induced action of other generators of $PGL(2,\Z)$. Altogether, we find
\be \begin{array}{c|c@{\quad}c@{\quad}c}  \varphi & x\mapsto & y\mapsto & z\mapsto  \\[.1cm] \hline
 S = {\small \begin{pmatrix} 0 & -1 \\ 1 & 0\end{pmatrix}} & y & x & -z-xy \\[.4cm]
R = T = {\small \begin{pmatrix} 1 & 1 \\ 0 & 1\end{pmatrix}} & z & y & -x-yz \\[.4cm]
R^{-1} = {\small \begin{pmatrix} 1 & -1 \\ 0 & 1\end{pmatrix}} & -z-xy & y & x\\[.4cm]
L = T^t = {\small \begin{pmatrix} 1 & 0 \\ 1 & 1\end{pmatrix}} & x & z & -y-xz \\[.4cm]
L^{-1} = {\small \begin{pmatrix} 1 & 0 \\ -1 & 1\end{pmatrix}} & x &-z-xy & y \\[.4cm]
\kappa = {\small \begin{pmatrix} 1 & 0 \\ 0 & -1\end{pmatrix}} & x & y & -z - x y
\end{array}
\label{MCGloop}
\ee
The action of a general element of the mapping class group $\bm\Gamma(C)=PSL(2,\Z)$ can be obtained by either composing the actions of generators $S$ and $T^{\pm1}$, or of generators $R^{\pm1}$ and $L^{\pm1}$.

\subsubsection{Fenchel-Nielsen coordinates}
\label{sec:FNclassMCG}

The mapping class group action for Fenchel-Nielsen coordinates can be obtained in a crude form by combining \eqref{loopFN} and the loop coordinate transformations \eqref{MCGloop}. In some cases, the Fenchel-Nielsen transformation is remarkably simple. For example, corresponding to $T^t\,:\;(\gamma_x,\gamma_y)\mapsto (\gamma_x,\gamma_z)$ on $C$ we have
\be T^t\;:\quad (\Lambda,\CT)\;\mapsto\; (\Lambda,\CT-2L)\,. \label{Tclass} \ee
The corresponding Lagrangian $\Delta_{T^t}$ in $Y\times \ol{Y}$, parametrized by $(\Lambda,\CT,\adj \Lambda,\adj \CT)$, is
\be \Delta_{T^t}\;=\;\{ L-\adj \Lambda = 0\,,\; \CT + 2L-\adj \CT = 0\}\,. \label{DTclass} \ee
Geometrically, $\CT\mapsto \CT-2L$ means that we have cut open the Fenchel-Nielsen pants forming the punctured torus $C$, and glued it back together with one additional (negative) full twist.%
\footnote{Although the twist $\CT$ that we are using is not exactly the geometric Fenchel-Nielsen twist $\CT^{(FN)}$, \cf\ \eqref{FNgeom}, it is closely related and the geometric argument still works here.} %

In contrast, the $S$ transformation is remarkably complicated in Fenchel-Nielsen coordinates. The Lagrangian $\Delta_S$ is given by
\begin{align} \label{DSclass}
\Delta_S\;=\;\bigg\{&\lambda+\lambda^{-1}-\frac{\ell^{-\frac 12}\adj \lambda-\ell^{\frac 12}\adj \lambda^{-1}}{\adj \lambda-\adj \lambda^{-1}}\adj \tau^{\frac 12}-\frac{\ell^{\frac12}\adj \lambda-\ell^{-\frac 12}\adj \lambda^{-1}}{\adj \lambda-\adj \lambda^{-1}} \adj \tau^{-\frac 12}=0  \\
 &\quad \adj \lambda+\adj \lambda^{-1}-\frac{\ell^{-\frac 12}\lambda-\ell^{\frac 12}\lambda^{-1}}{\lambda-\lambda^{-1}}\tau^{\frac 12}-\frac{\ell^{\frac12}\lambda-\ell^{-\frac 12}\lambda^{-1}}{\lambda-\lambda^{-1}}\tau^{-\frac 12} =0 \bigg\}\,. \notag
\end{align}
One interesting simplification of the $S$ transformation happens when we send the puncture parameter $v\to 0$ (or $\ell\to 1$), effectively removing the puncture from $C$. In this limit, $\tau^{\frac12}$ can be identified with the holonomy eigenvalue $\lambda_y$ for the cycle $\gamma_y$ dual to $\gamma_x$ on a (now) smooth torus. In logarithmic coordinates, we have $\Lambda_y=\CT/{2}$. Then \eqref{DSclass} reduces to a union of two Lagrangians
\be\Delta_S^{v\to 0}:\quad \{\Lambda_x = \Lambda_y{}_\flat\,,\;\Lambda_y = - \Lambda_x{}_\flat\}\quad\text{or}\quad  \{\Lambda_x = - \Lambda_y{}_\flat\,,\;\Lambda_y =  \Lambda_x{}_\flat\}\, \label{classSlimit} \ee
in $Y\times \ol Y$, now parametrized by symplectic coordinates $(\Lambda_x,\Lambda_y, \Lambda_x{}_\flat, \Lambda_y{}_\flat)$ with $\Omega_J = 2\hbar^{-1}d\Lambda_y\wedge d\Lambda_x -2\hbar^{-1}d \Lambda_y{}_\flat\wedge d \Lambda_x{}_\flat$. This smooth torus limit is a classical version of turning off the adjoint mass parameter $m_{\rm adj}$ of $\CN=2^*$ gauge theory to obtain $\CN=4$ gauge theory.

\subsubsection{Shear coordinates}
\label{sec:shearclassMCG}

Shear coordinate transformation rules can also be derived from the loop transformations \eqref{MCGloop}. However, the mapping class group action in shear coordinates has a beautiful interpretation of its own. Historically, this is what allowed the first quantizations of Teichm\"uller space to be carried out in \cite{FockChekhov, Kash-Teich}.

The basic idea is that a mapping class group element $\varphi$ maps any given ideal triangulation $\delta$ of a punctured surface $C$ to a new ideal triangulation $\delta'$. Any two triangulations (such as $\delta$ and $\delta'=\varphi(\delta)$) can also be related by a sequence of elementary ``diagonal flips.''  In turn, each elementary flip induces a simple, well-defined action on the space $Y = \CM_{\rm flat}(G_\C,C)$. This leads to a fully combinatorial decomposition of the action of any $\varphi \in \bm \Gamma(C)$ on $Y$.

\begin{figure}[htb]
\centering
\includegraphics[width=4in]{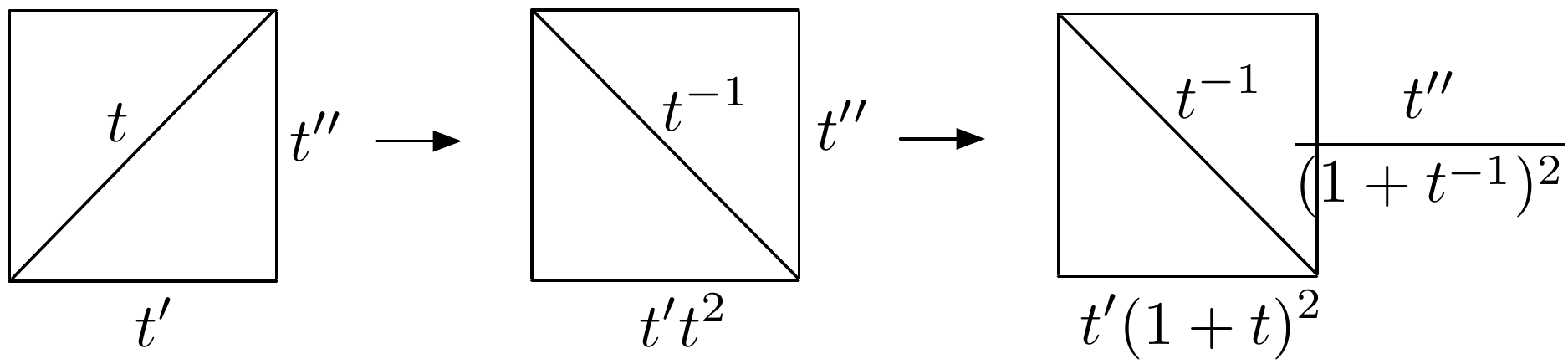}
\caption{The basic flip for a punctured torus.}
\label{fig:flip}
\end{figure}

In the case of the punctured torus $C=T^2\bs\{p\}$, every mapping class group generator $(S,R,L,R^{-1},L^{-1})$ can be realized by a single diagonal flip. The basic flip transforms shear coordinates in two steps, a monomial transformation followed by a more nontrivial symplectomorphism (\cf\ \cite{FG-qdl-cluster}). We illustrate this in Figure \ref{fig:flip}. The actual mapping class group action is then obtained by skewing or twisting the flipped torus on the right to look like the original one on the left. For example, to get the $S$ move we do
\be \includegraphics[width=4.3in]{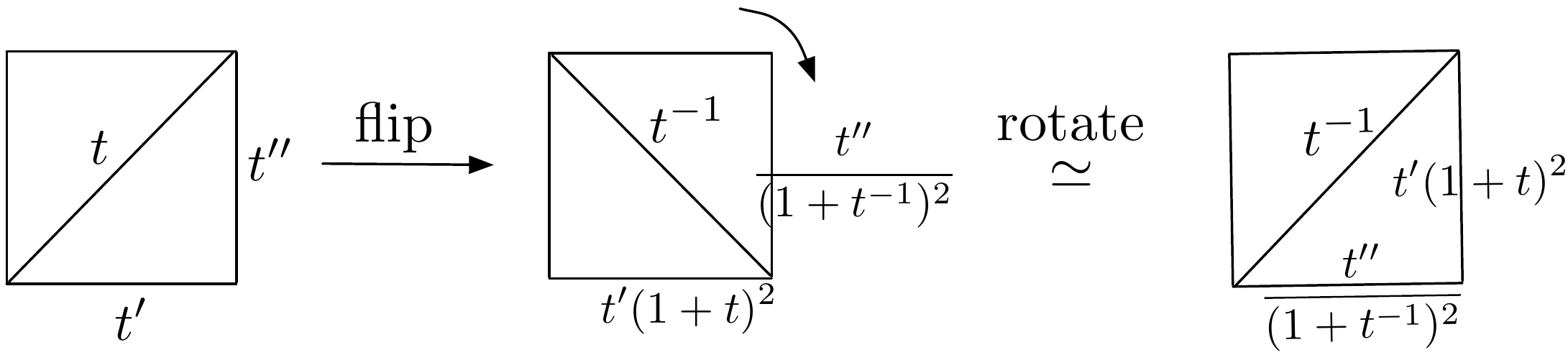} \notag \ee
and find $(t,t',t'')\mapsto(t^{-1},\,t''(1+t^{-1})^{-2},\,t'(1+t)^2)$. The actions of other generators of $\bm\Gamma(C) = PSL(2,\Z)$ are tabulated below. It is most convenient, both now and in the eventual quantization, to take square roots in order to simplify these actions.
\be\label{MCGshear}
\begin{array}{c@{\;\;\;}c@{\;\;}|c@{\quad}c@{\quad}c}
\varphi & &  \sqrt t\,\mapsto & \sqrt{t'}\,\mapsto & \sqrt{t''}\,\mapsto  \\\hline
S & \raisebox{-.2cm}{\includegraphics[height=.3in]{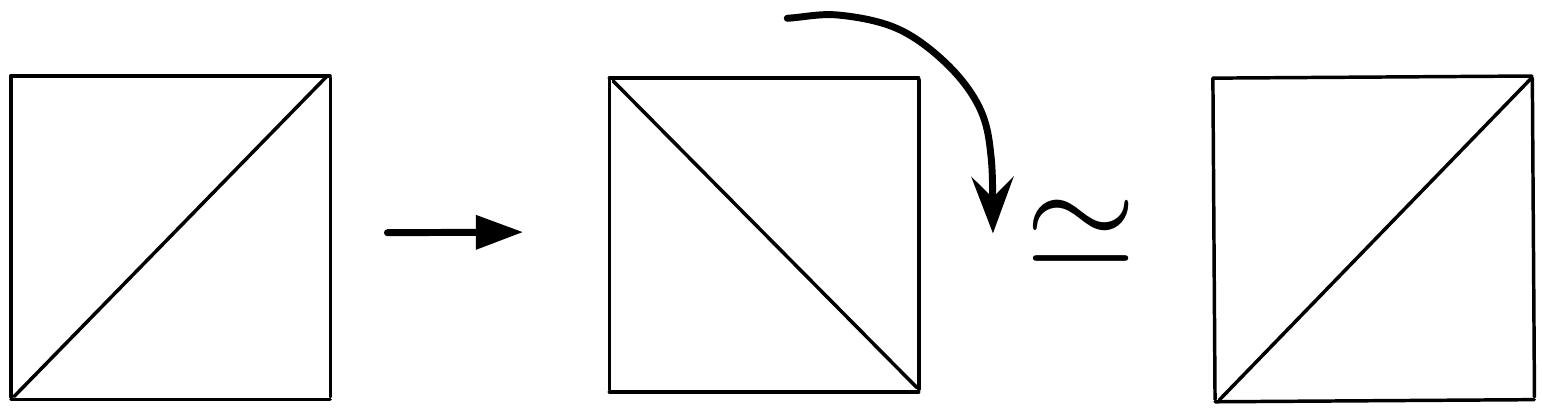}} &  \ds \frac1{\sqrt t} & \ds \frac{\sqrt{t''}}{1+t^{-1}} & \sqrt{t'}(1+t) \\[.35cm]
R = T & \raisebox{-.2cm}{\includegraphics[height=.3in]{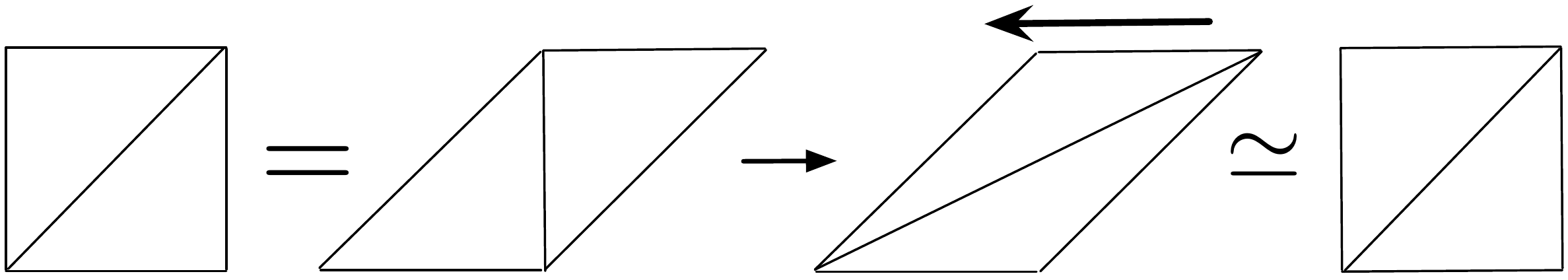}}\; &  \ds \frac{1}{\sqrt {t''}} & \ds \frac{\sqrt{t'}}{1+t''{}^{-1}} & \sqrt t(1+t'') \\[.4cm]
R^{-1} & \!\raisebox{-.2cm}{\includegraphics[height=.3in]{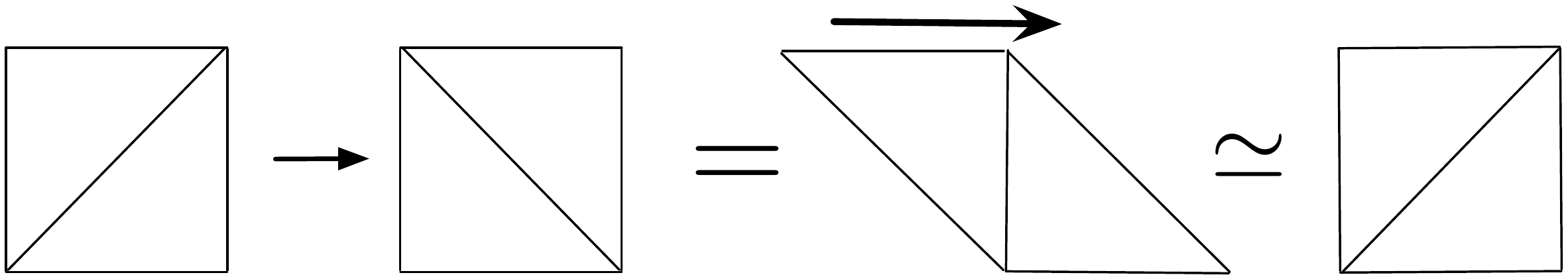}} & \ds \frac{\sqrt{t''}}{1+t^{-1}} & \sqrt{t'}(1+t) & \ds\frac{1}{\sqrt{t}} \\
L = T^t & \raisebox{-.52cm}{\includegraphics[height=.48in]{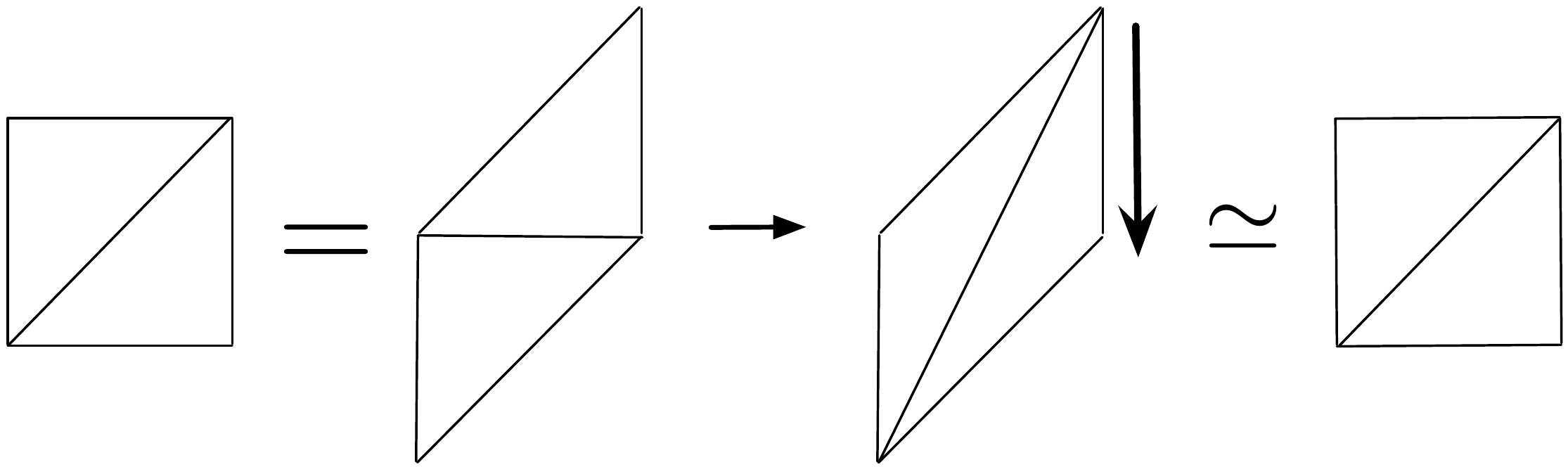}} & \ds \frac{1}{\sqrt{t'}} & \ds \frac{\sqrt{t}}{1+t'{}^{-1}} & \sqrt{t''}(1+t') \\
L^{-1} & \!\raisebox{-.52cm}{\includegraphics[height=.48in]{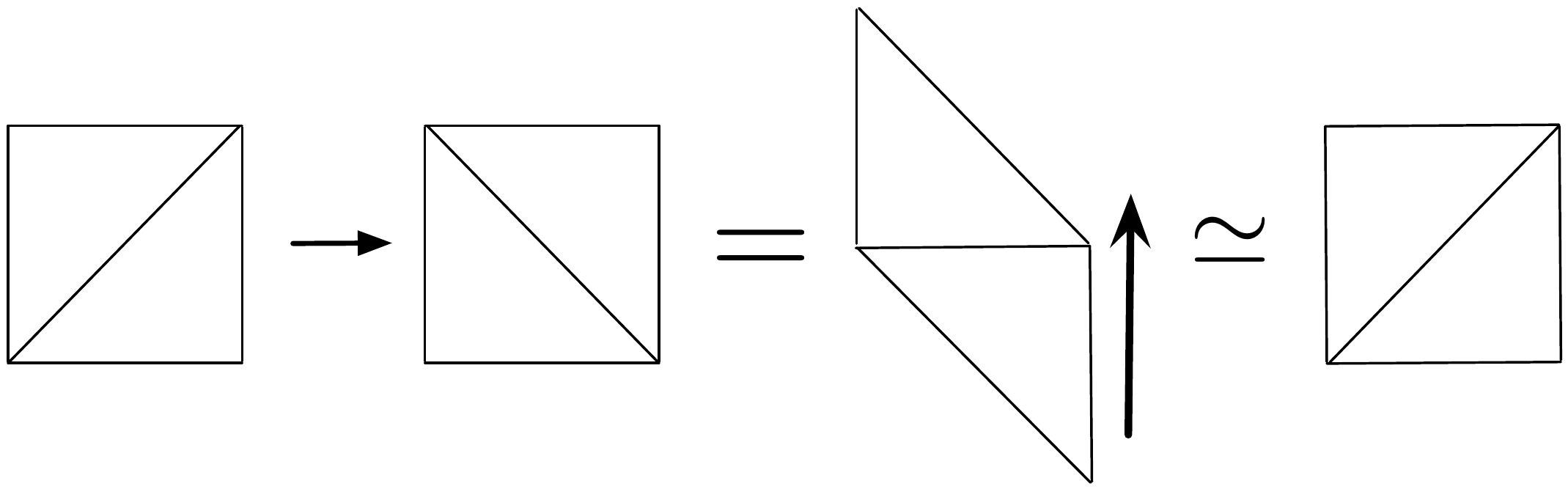}} & \sqrt{t'}(1+t) & \ds \frac{1}{\sqrt{t}} & \ds \frac{\sqrt{t''}}{1+t^{-1}}
\end{array}
\ee

The fact that mapping class group transformations correspond to sequences of diagonal flips in shear coordinates has a nice 3-dimensional interpretation. Namely, in order to implement a diagonal flip in 3-dimensions, one can glue a tetrahedron onto the punctured surface $C$. Geometrically, this well-known construction leads to 3d triangulations of mapping cylinders and mapping tori (formed by closing up the mapping cylinders). However, the precise relation between flat connections on 3d tetrahedra, Chern-Simons theory, and Teichm\"uller theory in this context has so far remained murky. Classical and quantum aspects of this relation are addressed in \cite{DGGV-hybrid}.


\subsection{Mapping tori and eigenbranes}
\label{sec:tori}

Mapping cylinders $M_\varphi$ and the corresponding Lagrangian submanifolds $\Delta_\varphi$
that we described in the previous section have an elegant interpretation in the sigma-model of $Y$.
Indeed, as we already mentioned briefly in \eqref{actiononbranes}, from the sigma-model
point of view $\bm \Gamma(C)$ is the group of autoequivalences of the category of branes on $Y$.
In other words, each $\varphi \in \bm \Gamma(C)$ corresponds to a functor {\em a la} Hecke that acts on
branes in the B-model of $(Y,J)$ and in the A-models of $(Y,\omega_I)$ and $(Y,\omega_K)$:
\be
\varphi \,:\, \adj \CB \mapsto \CB \,.
\label{actonbranes}
\ee
In this section, we illustrate how this interpretation can be used in practice, in particular
for studying the classical moduli spaces and $A$-polynomials of {\em mapping tori},
\be
M \; = \; C \times_\varphi S^1
\; \,\defeq \; C \times I \, / \, (x,0) \sim (\varphi (x),1)
\label{MTdef}
\ee
constructed by gluing the ``top'' and ``bottom'' boundaries of the mapping cylinders $M_\varphi = C\times_\varphi I$,
\cf\ Figure \ref{fig:MCgeneric}. As a result of this gluing procedure,
the moduli space of flat $G_{\C}$ connections on a 3-manifold \eqref{MTdef}
is given simply by the intersection of $\Delta_{\varphi}$
with the diagonal $\Delta_{{\bf 1}} \cong Y \subset Y \times Y$,
\be
\CM_{{\rm flat}} (G_{\C} , M) \; = \; \Delta_{{\bf 1}} \cap \Delta_{\varphi} \,.
\label{mflatgamma}
\ee
In particular, for punctured-torus bundles this becomes the zero locus of the $A$-polynomial,
and \eqref{actonbranes} allows to compute it by studying the mapping class group/braid group action on branes.
We follow the discussion in \cite{surf-op-braids}, where the braid group action on branes was used to
reproduce the $A$-polynomial for torus knots. Here, for balance, we consider the figure-8 knot
and study the mapping class group action.


\subsubsection*{Example: figure-8 knot}

The complement of the figure-8 knot, $M = S^3 \setminus K$, can be represented
as a punctured-torus bundle with the monodromy
\be
RL =
\begin{pmatrix}
2 & 1 \cr 1 & 1
\end{pmatrix}
~:~
\begin{cases}
x \mapsto -y-xz \\
y \mapsto z \\
z \mapsto -x + yz + xz^2
\end{cases}
\label{figeightd}
\ee
where we used \eqref{MCGloop} to find its action on the space
of classical solutions \eqref{toruscubic} or, equivalently, the correspondence $\Delta_{\varphi} \subset Y \times Y$.
Hence, the moduli space of classical solutions ({\it i.e.} flat connections on $M$) is given by \eqref{mflatgamma}:
\be
\Delta_{{\bf 1}} \cap \Delta_{\varphi}
~:~ \quad
\begin{array}{rcl}
x & = & -y-xz \\
y & = & z \\
z & = & -x + yz + xz^2\,.
\label{figeightdel}
\end{array}
\ee
Note that these three equations are not independent.
Indeed, the first two equations imply $x = -\frac{y}{1+y}$
and the third equation then follows automatically.

In order to compare this with the zero locus of the $A$-polynomial,
we have to remember that $(x,y,z)$ must belong to the cubic surface \eqref{toruscubic}.
Therefore, combining \eqref{toruscubic} with \eqref{figeightdel} we obtain a single equation
\be
\Tr g_v = (y^4 + 3 y^3 + y^2 - 4 y - 2) / (y+1)^2
\ee
which is equivalent to the zero locus of the $A$-polynomial $A(\ell,m)=0$,
\be
\Tr g_v = \ell + \ell^{-1} = (1 - m^2 - 2m^4 - m^6 + m^8) / m^4
\label{fig8apol}
\ee
provided that we identify $y = m^2 - 1$.

In order to see how \eqref{figeightd} acts on branes,
let us consider {\it e.g.} a zero-bane $\CB_0$ supported at a point on $Y$:
\be
\CB_0 ~: \quad (x,y,z) \; = \; (x_0,y_0,z_0) \,,
\ee
where $x_0$, $y_0$, and $z_0$ obey the cubic equation \eqref{toruscubic}.
Clearly, this brane is of type $(B,B,B)$. In particular, it is a good brane in the $B$-model of $(Y,J)$,
in which the action of $\varphi$ is holomorphic. Indeed, the first transformation in \eqref{figeightd}
maps this zero-brane into a new brane:
\be
L(\CB_0) ~: \quad y=z_0 \,, \quad z+xy = -y_0 \,.
\ee
Then, applying the $R$-transformation gives:
\be
RL(\CB_0) ~: \quad y=z_0 \,, \quad x - yz - x y^2 = -y_0 \,.
\ee
Notice, that the functors $L$ and $R$ increase the degree of the defining equations. From the vantage point of the zero-brane, the $A$-polynomial equation \eqref{fig8apol} is equivalent to the condition that $\CB_0$ is an eigenbrane of the ``Hecke functor'' $\varphi$, \ie\
\be
\varphi (\CB_0) \; = \; \CB_0 \,.
\ee


\subsection{$A$-polynomial and twisted superpotential}
\label{sec:3dgauge}

Now let us give an interpretation to the classical $A$-polynomial and,
more generally, to the moduli spaces $\Delta_{M} = \CM_{{\rm flat}} (G_{\C} , M)$
in an ``effective'' 3-dimensional gauge theory with $\CN=2$ supersymmetry.
We continue working in the framework of \cite{DGH},
where 3-dimensional $\CN=2$ effective field theory ${\bf T}_M$ is constructed by compactifying
the six-dimensional fivebrane theory on a 3-manifold $M$ and subject
to the $\Omega$-deformation in the remaining three dimensions.
In the special case when $M$ is a mapping cylinder $M_\varphi = C\times_\varphi I$,
the effective theory ${\bf T}_{M_\varphi}$ can be thought of
as a three-dimensional theory on the duality wall\footnote{In this case,
the 4d $\CN=2$ gauge theory is determined by $C$, whereas the duality wall is determined by $\varphi$.}
within a 4d $\CN=2$ gauge theory \cite{DGG-defects, HLP-wall, Yamazaki-3d}).
In fact, for the purposes of the present section --- based on the classical aspects of the geometry ---
we will mainly be interested in the physics of this $\CN=2$ theory with $\Omega$-background parameters $\epsilon_1\to 0$ and $\epsilon_2$ fixed, which in Chern-Simons theory corresponds to the limit $\hbar \to 0$.%
\footnote{A similar limit of 4-dimensional $\Omega$ backgrounds was considered in \cite{NShatashvili, NRS}.} %
Alternatively, as in \cite{DGH}, we can simply turn \emph{off} the $\Omega$-background but put the theory ${\bf T}_M$ on a partially compactified spacetime $\R^2\times S^1_R$, where the radius of $S^1_R$ is $R = \epsilon_2^{-1}$.

As explained in \cite{DGH}, the partial topological twist along $M$
ensures that the supersymmetric vacua of the effective $\CN=2$ theory ${\bf T}_M$
are in one-to-one correspondence with flat $G_{\C}$ connections on $M$.
In the case at hand, supersymmetric vacua are simply the critical points of the twisted superpotential $\CW$.
Therefore, we claim that the moduli space of flat $G_{\C}$ connections
on a 3-manifold $M$ (with or without boundary) is a ``graph" of functions $\frac{\partial \CW}{\partial \Lambda_i}$,
\be
\CM_{{\rm flat}} (G_{\C} , M) \; = \;
\Big\{ \sigma_i, \Lambda_j,  \CT_k \equiv \frac{\partial \CW (\sigma_i, \Lambda_j)}{\partial \Lambda_k} ~\Big|~ \frac{\partial \CW (\sigma_i, \Lambda_j)}{\partial \sigma_i} = 0 \Big\}
\label{moduliSUSY}
\ee
where $\Lambda_i$ is our collective notation for the parameters of the three-dimensional $\CN=2$
effective field theory (that include mass parameters, FI terms, {\it etc.}).
Indeed, the partition function $Z^{CS} (M;\hbar)$ of analytically continued Chern-Simons theory on a 3-manifold $M$
is a wavefunction associated to a classical state \eqref{defDelta} (see \cite{gukov-2003,Dimofte-QRS} for details).
In particular, in the semi-classical limit we have:
\be
Z^{CS} (M;\hbar) \,\overset{\hbar\to0}{\sim}\, \exp\Big(\frac1\hbar \CW + \CO(\log\hbar)\Big)\,,
\label{zcsw}
\ee
where $\CW = \int \theta$ is an integral of a Liouville 1-form $\theta = d^{-1} \Omega$.
(Notice, that $d^{-1} \Omega$ makes sense because $\Delta_M$ is Lagrangian with respect to $\Omega$,
so that $\Omega \big|_{\Delta_M}$ is locally exact.)
On the other hand, using the identification \eqref{Ksquare} we can view \eqref{zcsw}
as a partition function of the three-dimensional $\CN=2$ effective theory, ${\bf T}_M$,
whose leading term is known to be the twisted superpotential $\CW$.
Writing
$\Omega = \sum_i d \CT_i \wedge d \Lambda_i$ as in \eqref{FNholom}
quickly leads to the proposed expression \eqref{moduliSUSY},
where $\Lambda_i$ and $\CT_i$ should be interpreted as (canonically conjugate) coordinates on $\CP_{\pd M}$,
with all other moduli integrated out.
In particular, it means that the superpotential $\CW$ has to be extremized with respect to the complex fields $\sigma_i$,
the scalar components of the twisted chiral superfields $\Sigma_i$ (which are dual to 3d $\CN=2$ vector multiplets).

\FIGURE{
\;\;\includegraphics[width=1.5in]{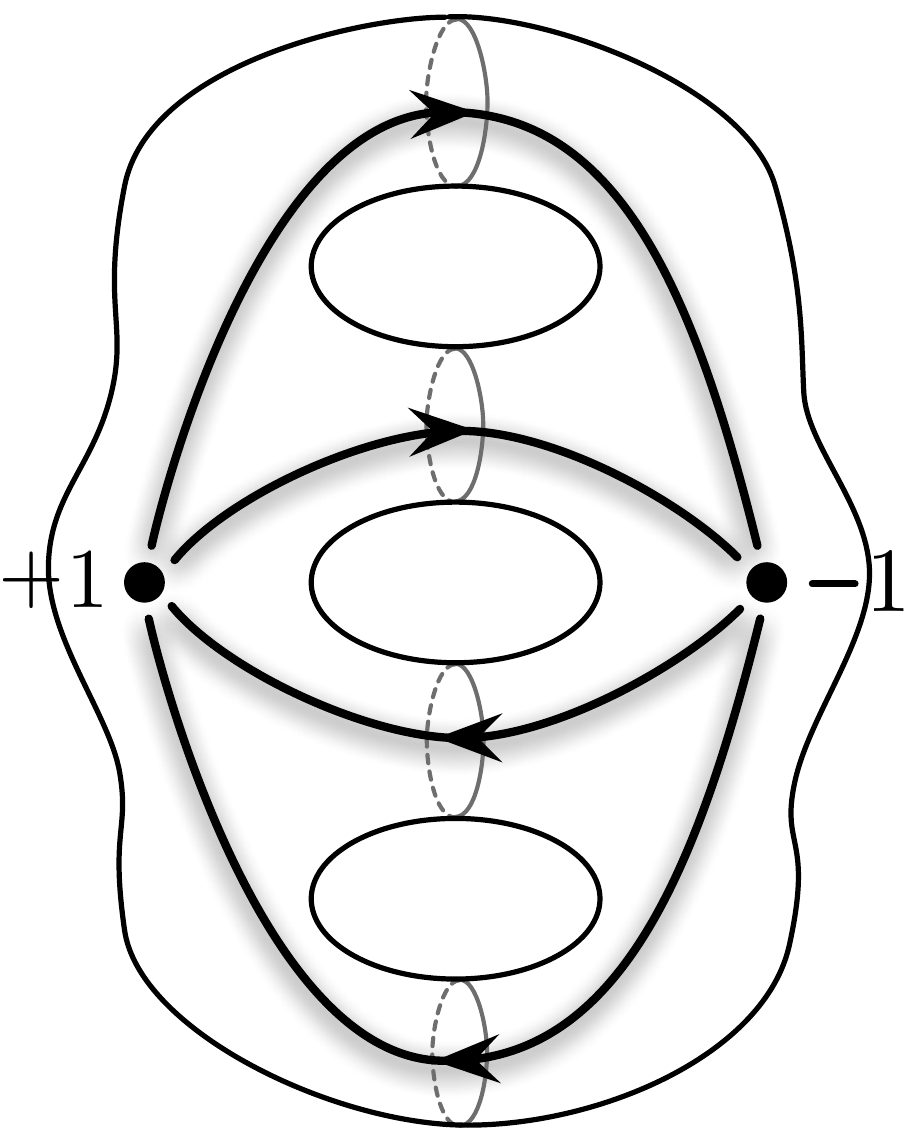}
\caption{For the figure-8 knot, the algebraic curve $A(\ell,m)=0$ has genus 3. The corresponding $\CN=2$ three-dimensional field theory is a $U(1) \times U(1)$ quiver gauge theory with four bifundamental matter multiplets and Chern-Simons terms at level $k_{1,2} = \pm 1$. \\}\label{fig:quiver}
}

For example, if $M$ is a punctured-torus bundle (such as, say, the figure-8 knot complement),
then the only ``external'' parameter is the mass parameter $m_{\rm adj}$ of the four-dimensional $\CN=2^*$ gauge theory
obtained by compactifying the fivebrane theory on $C=T^2\bs \{p\}$.
All other parameters in the superpotential $\CW$ are, in fact, dynamical fields and, therefore, should be integrated out.
In practice, this means extremizing the superpotential $\CW$ with respect to all dynamical fields,
which leads to the ``effective'' twisted superpotential:
\be
\CW_{{\rm eff}} (v) \; \,\defeq \; \CW (v, \Lambda_i) \vert_{\partial_i \CW = 0}
\ee
where we used the identification \eqref{madjvrelation} to express it as a function
of the holonomy eigenvalue $v$ rather than the mass parameter $m_{\rm adj}$.
Then, the $A$-polynomial of the punctured-torus bundle $M$ is simply a graph of the function~$\partial_{v} \CW_{{\rm eff}}$:
\be
\boxed{\phantom{\oint} A (\ell , m) = 0 \qquad \Leftrightarrow \qquad  u = \partial_{v} \CW_{{\rm eff}} (v) \phantom{\oint} }
\label{aviaw}
\ee
where $\ell = e^v$ and $m = e^u$, \cf\ \eqref{gvvholonomy}.

In the present paper, we will focus on a simple class of mapping tori and mapping cylinders $M$,
for which the corresponding three-dimensional $\CN=2$ effective field theory ${\bf T}_M$
is abelian.
The construction of such theories can be modeled on a prototypical example of
a three-dimensional $\CN=2$ SQED with $N_f$ chiral multiplets of charge $q_i$:
\begin{equation}
{\bf 3D~Theory}: \quad \CN=2~~ {\rm SQED~with~} N_f {\rm ~chiral~multiplets~of~charge~} q_i
\nonumber
\end{equation}
(An important example of such theory, relevant to the $S$ transformation \eqref{DSclass},
is the self-mirror\footnote{In the sense of \cite{IS}.} theory usually denoted $T[G]$, for $G=SU(2)$.)
A standard one-loop calculation leads to the twisted superpotential (see \eg\ \cite{DDDS,Witten-phases,HananyHori,HoriVafa}):
\be
\CW = 2\pi i \zeta \sigma + \sum_{i=1}^{N_f}  (q_i\sigma - \wt m_i) \left( \log \frac{q_i\sigma - \wt m_i}{\mu} - 1 \right)
\label{oneloopW}
\ee
where $\wt m_i$ are twisted masses
and $\zeta$ is the complex combination of the FI parameter and the theta parameter.
In order to find the $\epsilon_2$-corrections to $\CW$, we consider the theory on a circle of radius $R = \epsilon_2^{-1}$. Then, each chiral multiplet gives rise to a tower of Kaluza-Klein states with masses
\be
\wt m_i^{(n)} = \wt m_i + 2\pi \frac{n}{R} \,.
\ee
and the twisted superpotential becomes
\be
\CW = 2\pi i \zeta \sigma + \sum_{i=1}^{N_f} \sum_n  (q_i\sigma - \wt m_i^{(n)}) \left( \log \frac{q_i\sigma - \wt m_i^{(n)}}{\mu} - 1 \right)\,.
\label{wkk}
\ee
In the present example of $\CN=2$ SQED, the only dynamical field is $\sigma$. Therefore, extremizing the twisted superpotential \eqref{wkk} with respect to $\sigma$, we find the following equation
\be
0 \; = \; \frac{\partial \CW}{\partial \sigma} \; = \;
2\pi i \zeta + \sum_{i=1}^{N_f} \sum_n \log \Big( \frac{ q_i\sigma - \wt m_i^{(n)}}{\mu} \Big) \,,
\ee
which takes a particularly nice form after exponentiating:
\be
\prod_{i=1}^{N_f} \,\sin \bigg( \frac{R( q_i\sigma - \wt m_i)}{2}\bigg)  \; = \; e^{- 2 \pi i \zeta} \,. \label{sconstraint}
\ee
Integrating \eqref{sconstraint}, we can also re-express the full superpotential as
\be \CW = 2\pi i\zeta\sigma + \sum_{i=1}^{N_f}\bigg(-\frac{iR}{4}(q_i\sigma - \wt m_i)^2+\frac{i}{R}{\rm Li}_2\big(e^{i R(q_i\sigma - \tilde m_i)}\big)\bigg)\,.
\label{WLi2}
\ee

In particular, in the prominent example of the $T[SU(2)]$ theory, we have $\CN_f=4$ and $q_i=(+1,+1,-1,-1)$. The Lagrangian equations $\Delta_S$ in \eqref{DSclass} can be obtained as
\be \Delta_S\;:\quad \Big\{\CT = -iR\,\frac{\pd\CW}{\pd \Lambda}\,,\quad \CT_\flat = iR\,\frac{\pd\CW}{\pd \Lambda_\flat}\,\Big\}\,, \label{DS3d}\ee
provided that we use a familiar identification of the gauge theory parameters with Fenchel-Nielsen coordinates (\cf\ \cite{HLP-wall}),
\be \textstyle \zeta = \frac{\Lambda}{i\pi}\,, \label{ltmmmm} \ee
\be \textstyle \wt m_1 = \frac{i}{R}\big(\Lambda_\flat+\frac{v}{2}\big)\,,\quad
 \wt m_2 = \frac{i}{R}\big(\Lambda_\flat-\frac{v}{2}\big)\,,\quad
 \wt m_3 = \frac{i}{R}\big(\Lambda_\flat+\frac{v}{2}\big)\,,\quad
 \wt m_4 = \frac{i}{R}\big(\Lambda_\flat-\frac{v}{2}\big)\,, \notag
\ee
and impose the extremization \eqref{sconstraint}.
(For example, the first of equations \eqref{DS3d} just says $\sqrt{\tau} = e^{-i R\sigma}$, and substituting this into \eqref{sconstraint} yields the second equation for $\Delta_S$.) This makes sense: as we noted above, the mass-deformed $T[SU(2)]$ theory is the effective 3-dimensional theory for the $S$-transformation.

As an example of a different sort (related to mapping tori rather than to mapping cylinders) let us consider the figure-8 knot complement, $M = {\bf S}^3 \setminus K$, already discussed in Section \ref{sec:tori}. In particular, we already used the fact that $M$ can be represented as a punctured-torus bundle over $S^1$ with monodromy $\varphi = RL = T S T^{-1} S^{-1}$, \cf\ \eqref{figeightd}. Therefore, in this case the effective three-dimensional $\CN=2$ theory ${\bf T}_M$ obtained by compactifying the 6d five-brane theory on $M$ can be constructed by taking a cyclic combination of the basic building blocks, associated to $S$ and $T$ transformations of the punctured torus. The $S$ transformation (and its inverse) corresponds to the basic building block represented by the mass-deformed theory $T[SU(2)]$, whereas the $T^k$ transformation corresponds to a deformation of the Lagrangian by a Chern-Simons term at level $k$ \cite{Witten-SL2, Kapustin-3dloc}. Assembling these blocks in a cyclic order, dictated by the geometry of the mapping torus $M$, means identifying flavor symmetries and integrating out chiral multiplets, much like in four-dimensional generalized quiver gauge theories \cite{AGT}. Hence, the monodromy $\varphi = T S T^{-1} S^{-1}$ translates to the $\CN=2$ quiver gauge theory ${\bf T}_{M}$ shown on Figure~\ref{fig:quiver}.

More generally, the basic building blocks of 3-manifolds and the corresponding 3d theories ${\bf T}_M$ do not need to be limited to mapping cylinders. For example, one can triangulate a 3-manifold $M$ by (ideal) tetrahedra. The partition function of the analytically continued $SL(2,\C)$ Chern-Simons theory on $M$ can be constructed then from the triangulation data \cite{DGLZ, Dimofte-QRS}. The basic ingredient in this approach is a quantum dilogarithm function $\Phi_{\hbar} (z)$ associated to every ideal tetrahedron in a triangulation of $M$. At the level of the $\CN=2$ supersymmetric theory ${\bf T}_M$ it means that each ideal tetrahedron contributes to the twisted superpotential a term of the form:
\be
\boxed{\phantom{\oint}\text{ideal tetrahedron} \quad \longleftrightarrow \quad \delta \CW = Li_2 \phantom{\oint}}
\ee
For example, according to this dictionary, the quiver theory shown on Figure~\ref{fig:quiver} would most naturally be associated to a triangulation of the figure-8 knot complement into eight tetrahedra. Further aspects of this dictionary between the geometry of $M$ and the space of supersymmetric vacua of the corresponding 3d theory ${\bf T}_M$ will be presented in \cite{DGGV-hybrid}.


\section{S-duality actions}
\label{sec:quant}

We turn now to the quantization of the Hitchin moduli space $Y = \CM_{\rm flat}(G_\C,C)$, and the quantum $\CN=2$ and $\CN=4$ S-dualities that act on it. As presented in Section \ref{sec:class}, $Y$ can be thought of as a complexified phase space associated to $C$, with holomorphic symplectic form $\Omega_J$. Quantization of $(Y,\Omega_J)$ then produces a Hilbert space $\CH_\hbar(C)$, and an algebra of holomorphic operators $\hat \CA_\hbar$ acting on it%
\footnote{We again emphasize two subtle but important points. Being precise, the space $\CH_\hbar(C)$ is obtained by first quantizing the real slice $\Teich(C)\subset Y$ with respect to $\omega_I$, and then analytically continuing wavefunctions. This analytic continuation typically breaks the periodicity of $\C^*$ coordinates such as $(t,t',t'')$ on $Y$, resulting in wavefunctions such as $Z(T)$ that depend on a logarithm $T = \log(t)$, with nontrivial behavior as $T\to T+2\pi i$.}:
\be Y = \CP(C) \; \leadsto \; \CH_\hbar(C)\,,\qquad \{\text{functions on $Y$}\}\;\leadsto\; \hat \CA_\hbar(C)\,.\ee
For example, logarithmic shear coordinates $(T,T',T'')$ on $\CP(C)$ for a punctured torus become operators $(\hat T,\hat T',\hat T'')$ with commutation relations
\be [\hat T,\hat T'] = [\hat T',\hat T''] = [\hat T'',\hat T] = 2\hbar\,. \label{Tcommintro} \ee
For a mapping cylinder such as $M_\varphi$ or $M_\xi$, one obtains a wavefunction  $Z_M \in \CH_\hbar(C)\otimes \CH_\hbar(\ol C) = \CH_\hbar(C) \otimes \CH_\hbar(C)^*$, which is annihilated by a system of difference operators. These operators are a quantization of the equations that cut out the Lagrangian $\Delta_M$, so we denote them schematically as $\wh \Delta_M$, with
\be \Delta_M \;\leadsto\; \wh\Delta_M\,,\qquad \wh\Delta_M\,Z_M = 0\,.\ee
Similarly, a mapping torus $M$ as discussed in Section \ref{sec:tori} can lead to a wavefunction $Z_M \in \CH_\hbar(T^2)$, annihilated by the quantum $\hat A$--polynomial of $M$.

The quantization of individual operator algebras is relatively straightforward, and discussed in Section \ref{sec:qalg}. On the other hand, finding the quantum operators $\wh \Delta_M$ that annihilate the wavefunction of a mapping cylinder at first seems a bit difficult. Indeed, if $M$ were a general 3-dimensional cobordism from the surface $C$ to itself, this would be a very nontrivial problem. One approach to solving this problem uses 3d triangulations and quantum gluing \cite{DGGV-hybrid}. Presently, however, we are saved by the fact that mapping cylinders $M_\varphi$ or $M_\xi$ have a very special structure. Topologically, they are trivial manifolds, and their Lagrangians $\Delta_M$ are graphs (``correspondences'') in $Y\times \ol Y$. This turns out to be equivalent to the statement that mapping class group actions $\varphi$ or coordinate transformations $\xi$ can be implemented as \emph{automorphisms} (or, respectively, isomorphisms) of the quantum algebra $\hat \CA_\hbar$.

Given an automorphism of $\hat \CA_\hbar$, say a mapping class group action $\varphi$, it's a short skip to difference equations $\wh\Delta_\varphi$. We observe that a Hilbert space $\CH_\hbar(C)\otimes \CH_\hbar(\ol C)$ has two copies of the full operator algebra, $\hat \CA_\hbar\otimes \hat \CA_\hbar^\flat$, acting on it, where the ``bottom'' copy $\hat \CA_\hbar^\flat$ is the conjugate of $\hat \CA_\hbar^\flat$ in the sense that all operators have been replaced by their adjoints and obey opposite commutation relations. Then, given \emph{any} operator $\wh\CO\in \hat \CA_\hbar$ and its image $\varphi(\wh\CO)$, the wavefunction $Z_\varphi$ will be annihilated by the difference $\wh\CO - [\varphi(\wh\CO)]_\flat \in \hat \CA_\hbar\otimes \hat \CA_\hbar^\flat$. Here $[\;\;]_\flat$ means that we replace $\varphi(\wh\CO)$ by its adjoint. Thus, the set of operators $\wh\Delta_\varphi$ that annihilate $Z_\varphi$ form a left ideal, generated as
\be \wh\Delta_\varphi = \Big\langle \wh\CO - [\varphi(\wh\CO)]_\flat\,\Big|\,\wh\CO \in \hat \CA_\hbar\,\Big\rangle. \label{qDelta} \ee
It is easy to see that in the classical, commuting limit $\hbar \to 0$, \eqref{qDelta} reduces to the ideal \eqref{correspondence} whose zero locus is the Lagrangian correspondence $\Delta_\varphi$.

As a simple example, consider the $S$--move in shear coordinates on the punctured torus, from \eqref{MCGshear}. The quantization of this transformation is  known \cite{FockChekhov} to be
\be \big(\sqrt{\hat t},\,\sqrt{\hat t'}
\,\big) \mapsto \bigg(\frac1{\sqrt {\hat t}},\, \frac{iq^{-\frac14}\ell^{\frac12}}{\sqrt{\hat t\hat t'}}\frac{1}{1+q^{\frac12}\hat t^{-1}}
\,\bigg)\,.\ee
Then the difference equations annihilating $Z_{S}(T,T_\flat)$ are
\be \wh\Delta_{S}\,:\quad \sqrt{\hat t} - \frac{1}{\sqrt{\hat t_\flat'}} \simeq 0\,,\quad\;\; \sqrt{\hat t'} - \frac{iq^{-\frac14}\ell^{\frac12}}{\sqrt{\hat t_\flat\hat t'_\flat}}\frac{1}{1+q^{\frac12}\hat t^{-1}_\flat} \simeq  0\,. \label{qSeqsintro} \ee
(From now on we introduce the notation $\simeq 0$ to mean ``annihilates a physical wavefunction.'') While the operators $\hat t$ and $\hat t'$ $q$--commute according to the symplectic structure \eqref{shearholom}, the adjoints $\hat t_\flat$ and $\hat t'_\flat$ $q^{-1}$--commute,
\be \hat t\hat t' = q^2\hat t'\hat t\,,\qquad \hat t_\flat\hat t'_\flat = q^{-2}\hat t'_\flat\hat t_\flat\,,\qquad (q=e^\hbar)\,.\ee
They act on $Z_S(T,T_\flat)$ as $\hat t=t=e^T,\,\hat t'=e^{-2\hbar \pd_T},\,\hat t_\flat = t_\flat = e^{T_\flat},$ and $\hat t_\flat' =e^{+2\hbar\pd_{T_\flat}}$. More interesting examples will be explored in Sections \ref{sec:qcyl} and \ref{sec:eg}. In particular, we will consider mapping kernels $Z_\varphi(\Lambda,\Lambda_\flat)$ in Fenchel-Nielsen coordinates, which are relevant for Liouville theory and gauge theory.

Physically, equations \eqref{qSeqsintro} say that the S-kernel $Z_S$ intertwines the action of a Wilson loop and a 't Hooft loop. This is a little more obvious when we rewrite the ideal $\wh\Delta_\varphi$ in loop coordinates:
\be \wh\Delta_{S}\,:\quad \hat x-\hat y_\flat \simeq 0\,,\quad\;\; \hat y-\hat x_\flat \simeq 0\,. \label{qSloopintro} \ee
The quantized loop operators $\hat x,\,\hat y,\,\hat z$ are identified, respectively, with a spin-1/2 Wilson loop $W_{1/2}$, a 't Hooft loop $H_{1/2}$, and a mixed Wilson--'t Hooft loop $W^{(1,1)}_{1/2}$ of charge (1,1) in $\CN=2^*$ gauge theory. They act on the Hilbert space of $\CN=2^*$ theory on $\Omega$-deformed $S^3\times \R$, $\CH_{\epsilon_{1,2}}^{\CN=2^*}(S^3) \cong \CH_\hbar(C)$. Thus, yet another schematic way to write \eqref{qSloopintro} is (\cf\ \cite{AGGTV, Tesch-loop})
\be W_{1/2}K_S = K_SH_{1/2}\,,\qquad H_{1/2}K_S = K_SW_{1/2}\,.\ee

Underlying the quantization of $Y$ and its mapping class group actions (in any coordinate system) is $\CN=4$ S-duality. It acts on the quantization parameter $\hbar$ as
\be {}^L:\;\hbar\to {}^L\hbar = -\frac{4\pi^2}{\hbar}\,,\qquad\text{or}\qquad \epsilon_1\leftrightarrow\epsilon_2\,.\ee
When we described Wilson loops in Section \ref{sec:params}, we found that in the classical limit $\hbar = 2\pi i\epsilon_1/\epsilon_2\to 0$ only the loops wrapping one of the great circles $S^1_{\epsilon_2}$ of $S^3$ were visible, with (\eg)
\be \langle W_{1/2}\rangle = \langle\hat x\rangle = \cosh\Big(\frac{2\pi i}{\epsilon_2}a\Big)\,. \label{KSinter} \ee
\FIGURE{
\;\;\;\includegraphics[width=2.2in]{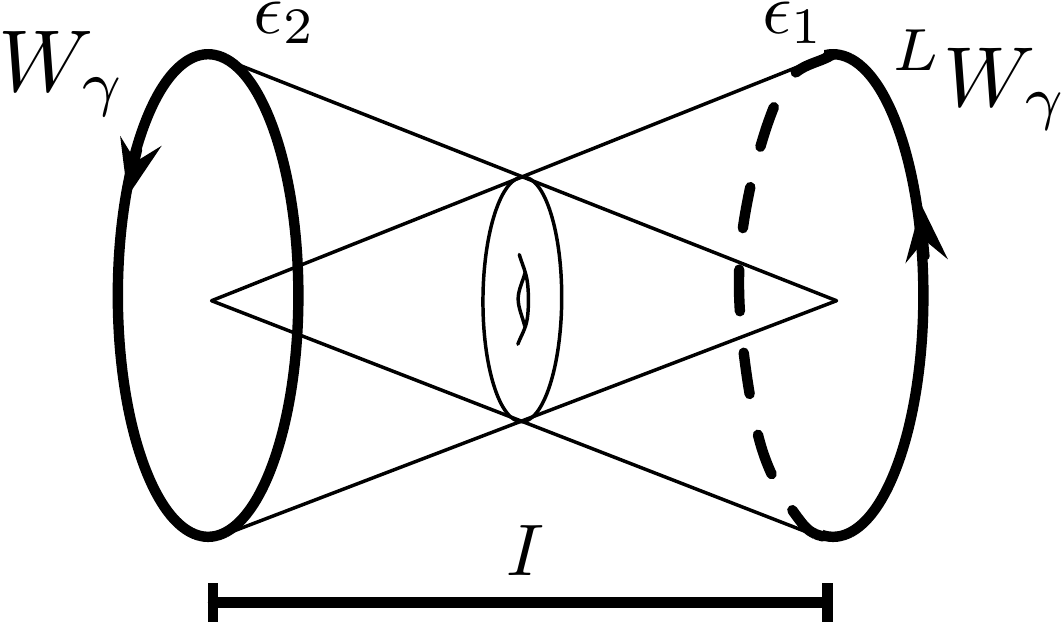}\smallskip
\caption{$\CN=4$ dual loop operators on $S^3$.\bigskip}
\label{fig:S3dual}
}
\noindent Now, upon quantization at finite $\hbar$, we can find the ``missing'' loops that wrap $S^1_{\epsilon_1}$ (Figure \ref{fig:S3dual}). They are just the $\CN=4$ duals to the original loops, and have expectation values (\eg)
\be \langle {}^LW_{1/2}\rangle = \langle{}^L\hat x\rangle = \cosh\Big(\frac{2\pi i}{\epsilon_1}a\Big)\,.\ee

Algebraically, $\CN=4$ S-duality appears naturally in the quantization of the operator algebras $\hat \CA_\hbar$ \cite{Dimofte-QRS}. For example, the full operator algebra in shear coordinates on a punctured torus is generated by the logarithmic operators $\hat T,\hat T',\hat T''$ (with a central constraint $\hat T+\hat T'+\hat T'' = v+i\pi+\hbar/2$, \cf\ \ref{logvshear}). Working instead in exponentiated coordinates, one finds that $\hat \CA_\hbar$ actually factors%
\footnote{In order to make this factorization mathematically precise, one must take an appropriate closure of the tensor product on the RHS.} %
into two commuting subalgebras \cite{Fad-modular, FockChekhov}
\be \hat \CA_\hbar \,=\, \C(\hbar)[\hat T,\hat T',\hat T''] \,\cong\, \C(q)[\hat t,\hat t',\hat t'']\otimes \C({}^Lq)[{}^L\hat t,{}^L\hat t',{}^L\hat t''] \,\eqdef \, \hat \CA_q \otimes \hat \CA_{{}^Lq}\,, \label{qalgfact} \ee
where $q=e^\hbar$, ${}^Lq = \exp({}^L\hbar)$, ${}^L\hat t = \exp({}^L\hat T) = \exp\big(\frac{2\pi i}{\hbar}\hat T)$, and similarly for ${}^L\hat t'$ and ${}^L\hat t''$. Note that $\hat t\hat t' = q^2\hat t'\hat t$, while ${}^L\hat t{}^L\hat t' = {}^Lq^2\,{}^L\hat t'{}^L\hat t$, and that all the ${}^L\hat t$'s commute with $\hat t$'s. $\CN=4$ S-duality simply exchanges the two factors on the right side of \eqref{qalgfact}, while preserving $\hat \CA_\hbar$. An S-duality-invariant wavefunction such as $Z_\varphi$ for a mapping cylinder is annihilated not only by operators $\wh\Delta_\varphi$ but also by their duals ${}^L\wh\Delta_\varphi$. Or, saying that a bit more physically, a kernel such as $K_S$ in \eqref{KSinter} intertwines the action of dual loop operators ${}^LW_{1/2}$ and ${}^LH_{1/2}$ as well as the original $W_{1/2}$ and $H_{1/2}$. The two sets of operators have (almost) no mutual interactions.

A final interesting aspect of quantization involves the return to the classical limit $\hbar\to 0$. The fact that a system $\wh\Delta_M$ reduces to the defining classical equations of the Lagrangian $\Delta_M$ as $\hbar \to 0$ implies that the leading asymptotics of a wavefunction $Z_M$ should be determined by $\Delta_M$. In fact, one finds that
\be Z_M \,\overset{\hbar\to0}{\sim}\, \exp\Big(\frac1\hbar \CW_M+ \CO(\log\hbar)\Big)\,,\ee
where $\CW_M$ is a ``potential function'' for the Lagrangian $\Delta_M$, calculated by the WKB approximation. For example, in the case of a mapping class kernel $Z_\varphi(T,T_\flat)$ in shear coordinates, this means that $\Delta_\varphi$ is cut out by equations $T' = -2\pd_T\CW_\varphi$ and $T'_\flat = 2\pd_T\CW_\varphi$. This observation has nontrivial consequences when $Z_M$ is interpreted as a physical partition function in various contexts. Indeed, in Section \ref{sec:3dgauge} the identification of $\CW_M$ with the effective superpotential $\CW$ of effective 3d gauge theory allowed us to obtain Lagrangian equations from the superpotential, as in \eqref{moduliSUSY}.

\subsection{Operators and $\CN=4$ S-duality}
\label{sec:qalg}

As in Section \ref{sec:class}, we specialize to a punctured surface $C=T^2\bs\{p\}$. We begin here by describing the quantization of the operator algebra $\hat \CA_\hbar$ in this case. In shear coordinates, both the quantization and the mapping class group action on $\hat \CA_\hbar$ were found in \cite{FockChekhov, Kash-Teich}, and the action of $\CN=4$ S-duality is a simple extension of the general ``modular'' structure discussed by \cite{Fad-modular}.%
\footnote{Strictly speaking, \cite{FockChekhov, Kash-Teich} only used shear coordinates to quantize the real slice $\Teich(C)\subset Y$. However, as far as operator algebras are concerned, everything can be analytically continued to algebras of holomorphic operators. Alternatively, in terms of brane quantization \cite{GW-branes}, the algebra $\hat\CA_\hbar$ is realized as ${\rm Hom}(\CB_{cc},\CB_{cc})$ for a canonical coisotropic brane $\CB_{cc}$ wrapping all of $Y$; thus $\hat\CA_\hbar$ cannot depend on which particular real slice of $(Y,\Omega_J)$ one chooses to quantize.} %
Quantization in Fenchel-Nielsen coordinates also appears straightforward, but the actions of $\CN=2$ and $\CN=4$ S-duality are much more subtle. To understand them, we will construct isomorphisms between algebras $\hat\CA_\hbar$ in all three of our coordinate systems.

\subsubsection{Shear algebra: quantum Teichm\"uller space}

We promote the logarithmic shear coordinates $(T,T',T'')$ of Section \ref{sec:shear} to operators $(\hat T,\hat T',\hat T'')$. According to the symplectic structure \eqref{shearholom}, they should obey
\be [\hat T,\hat T']=[\hat T',\hat T'']=[\hat T'',\hat T]=2\hbar\,. \label{Tcomm} \ee
We also impose a central constraint
\be \hat T+\hat T'+\hat T'' = v+i\pi + \hbar/2\,.
\label{qTv} \ee
Note that the factor of 2 on the right side of \eqref{Tcomm} comes from the fact that any two edges in the triangulated punctured torus $C$ share exactly two sides. We then find $\hat \CA_\hbar = \C(\hbar)[\hat T,\hat T',\hat T'']$ modulo \eqref{qTv}.

In exponentiated coordinates, we obtain operators $\hat t = e^{\hat T}$, $\hat t'=e^{\hat T'}$, $\hat t''=e^{\hat T''}$ that obey
\be \hat t\hat t'=q^2\hat t'\hat t\quad\text{(+ cyclic)}\,,\qquad
\hat t\hat t'\hat t'' = -q^{\frac32}\ell\,,\ee
with $\boxed{q=e^\hbar}$.
One might venture to guess that we could also define $\hat \CA_\hbar \sim \C(q)[\hat t,\hat t',\hat t'']$. However, this is not quite true. In order to obtain the full operator algebra, we must also introduce ($\CN=4$) S-dual exponential operators \cite{Fad-modular, Dimofte-QRS}
\be {}^L\hat t = \exp{}^L\hat T\, \defeq \exp\bigg( \frac{2\pi i}{\hbar} \hat T\bigg)\,, \label{defLT} \ee
and similarly for ${}^L\hat t'$ and ${}^L\hat t''$. It is easy to see from the commutation relations
\be [{}^L\hat T,{}^L\hat T'] = 2\,{}^L\hbar\qquad \text{(+ cyclic)} \ee
that ${}^L\hat t{}^L\hat t' = {}^Lq^2\,{}^L\hat t'{}^L\hat t$ and that $({}^L\hat t,{}^L\hat t',{}^L\hat t'')$ commute with $(\hat t,\hat t',\hat t'')$. Moreover, the constraint \eqref{qTv} is invariant under $\CN=4$ S-duality, in the sense that
\be {}^L\hat T+{}^L\hat T'+{}^L\hat T'' = {}^Lv+i\pi + {}^L\hbar/2\,,\label{LqTv} \ee
with ${}^Lv = \frac{2\pi i}{\hbar}v$. (Note that the ``quantum correction'' by $\hbar/2$ in \eqref{qTv} was necessary to achieve this invariance.) Thus, ${}^L\hat t{}^L\hat t'{}^L\hat t'' = -{}^Lq^{\frac32}{}^L\ell$. Altogether, we find that
\be \hat \CA_\hbar = \hat \CA_q \otimes \hat \CA_{{}^Lq}\,, \label{qalgfact2} \ee
as anticipated in \eqref{qalgfact}, with
\be \begin{array}{cl}
\multirow{2}{*}{
\raisebox{-.35cm}{
$\CN=4$ S-duality\;\raisebox{-.4cm}{\includegraphics[height=1cm]{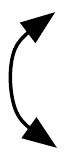}}}
}
& \hat \CA_q=\C(q)[\hat t,\hat t',\hat t'']\big/({\small \hat t\hat t'\hat t'' = -q^{\frac32}\ell)} \\[.2cm]
& \hat \CA_{{}^Lq}=\C({}^Lq)[{}^L\hat t,{}^L\hat t',{}^L\hat t'']\big/({\footnotesize {}^L\hat t{}^L\hat t'{}^L\hat t'' = -{}^Lq^{\frac32}{}^L\ell)}\,.
\end{array}
\label{Salg}
\ee

\subsubsection{Fenchel-Nielsen algebra}

Now consider Fenchel-Nielsen coordinates. Since the symplectic form $\Omega_J$ \eqref{FNholom} is diagonal, we can promote $(\Lambda,\CT)\,\defeq(\Lambda_x,\CT_x)$ to operators $(\hat \Lambda,\hat \CT)$ satisfying
\be [\hat \CT,\hat \Lambda]=\hbar\,,
\ee
and set $\hat \CA_\hbar  = \C(\hbar)[\hat \CT,\hat L]$. In exponentiated coordinates, it would be natural to define
\be \hat \tau = e^{\hat \CT}\,,\quad \hat\lambda =e^{\hat \Lambda}\,,\qquad {}^L\hat \tau = e^{{}^L\hat \CT}\,,\quad {}^L\hat\lambda =e^{{}^L\hat \Lambda}\,,\ee
with ${}^L\hat \CT = \frac{2\pi i}{\hbar}\hat \CT$, ${}^L\hat\Lambda=\frac{2\pi i}{\hbar}\hat\Lambda$ so that $\hat\tau\hat\lambda = q\hat\lambda\hat\tau$ and ${}^L\hat\tau{}^L\hat\lambda = {}^Lq\,{}^L\hat\lambda{}^L\hat\tau$. Then we expect that $\hat \CA_\hbar  = \hat \CA_q\otimes\hat \CA_{{}^Lq}$, with two mutually commuting subalgebras $\hat\CA_q = \C(q)[\hat\tau,\hat\lambda]$ and $\hat\CA_{{}^Lq} = \C(q)[{}^L\hat\tau,{}^L\hat\lambda]$ that are exchanged by $\CN=4$ S-duality.

The relations \eqref{shearFN} relating shear coordinates to Fenchel-Nielsen coordinates can be fully quantized to produce an isomorphism between the respective operator algebras. One (partial) derivation of this quantization uses mapping cylinders $M_\xi$ and is given in Appendix \ref{app:map}. The result is most easily expressed by setting
\be \sqrt{\hat t} = i\frac{1}{\hat a-\hat a^{-1}}(\hat b-\hat b^{-1})\,,\qquad \sqrt{\hat t'} = i\frac{1}{\hat c-\hat c^{-1}}(\hat a-\hat a^{-1})\,,\qquad \sqrt{\hat t''} = i\sqrt{\ell}\frac{1}{\hat b-\hat b^{-1}}(\hat c-\hat c^{-1})\,, \label{qshearFN} \ee
with
\be \hat a \,\defeq \hat \lambda\,,\qquad \hat b \,\defeq e^{-\frac{\hat \CT}{2}}=\frac{1}{\sqrt{\hat \tau}}\,,\qquad \hat c \,\defeq e^{\frac{\hat \CT}{2}-\hat \Lambda}=q^{-1/4}\hat \lambda^{-1}\sqrt{\hat \tau}\,. \ee
These new operators $\hat a,\,\hat b,\,\hat c$ are a more symmetric way of writing the Fenchel-Nielsen algebra $\hat \CA_q$. They satisfy commutation relations
\be \hat a\hat b = q^{\frac12}\hat b\hat a\,,\qquad  \hat b\hat c = q^{\frac12}\hat c\hat b\,,\qquad  \hat c\hat a = q^{\frac12}\hat a\hat c\,, \label{abcalg} \ee
and a central constraint $\hat a\hat b\hat c = q^{\frac14}$. It is obvious from \eqref{qshearFN} that the square roots of shear coordinates obey the constraint $\sqrt{\hat t''}\sqrt{\hat t'}\sqrt{\hat t}=i\sqrt\ell$ (as expected from \eqref{qTv}). However, it is not at all obvious that the $\hat t$'s have the proper $q$-commutation relations, or equivalently that $\sqrt{\hat t}\sqrt{\hat t'}\sqrt{\hat t''} = iq^{\frac12}\sqrt\ell$. Both of these facts follow from a marvelous identity in the abstract $(\hat a,\hat b,\hat c)$ algebra:
\be \frac{1}{\hat a-\hat a^{-1}}(\hat b-\hat b^{-1})\frac{1}{\hat c-\hat c^{-1}}(\hat a-\hat a^{-1})\frac{1}{\hat b-\hat b^{-1}}(\hat c-\hat c^{-1}) = q^{\frac12}\,. \label{abcid} \ee
It is relatively straightforward to prove that \eqref{abcid} holds, given \eqref{abcalg} and the condition $\hat a\hat b\hat c = q^{\frac14}$.

In addition to \eqref{qshearFN}, we also find dual relations
\be \sqrt{{}^L\hat t} = \frac{i}{{}^L\hat a-{}^L\hat a^{-1}}({}^L\hat b-{}^L\hat b^{-1})\,,\quad \sqrt{{}^L\hat t'} = \frac{i}{{}^L\hat c-{}^L\hat c^{-1}}({}^L\hat a-{}^L\hat a^{-1})\,,\quad \sqrt{{}^L\hat t''} = \frac{i\sqrt{{}^L\ell}}{{}^L\hat b-{}^L\hat b^{-1}}({}^L\hat c- {}^L\hat c^{-1})\,, \label{LqshearFN} \ee
where ${}^L\hat a = e^{{}^L\hat\Lambda}$, ${}^L\hat b = e^{-{}^L\hat \CT/2}$, and $\hat c = e^{{}^L\hat\CT/2-{}^L\hat\Lambda}$.
It is the presence of these dual relations that ensures that our definition of $\CN=4$ S-duals in Fenchel-Nielsen coordinates $\big({}^L\hat\CT = \frac{2\pi i}{\hbar}\hat \CT,\,{}^L\hat\Lambda = \frac{2\pi i}{\hbar}\hat\Lambda\big)$ is compatible with the definition \eqref{defLT} in shear coordinates. This is a mathematically nontrivial statement. For example, comparing \eqref{qshearFN} and \eqref{LqshearFN} leads to new operator identities of the form $\Big[\frac{i}{\hat a-\hat a^{-1}}(\hat b-\hat b^{-1})\Big]^{\frac{2\pi i}{\hbar}} = \frac{i}{{}^L\hat a-{}^L\hat a^{-1}}({}^L\hat b-{}^L\hat b^{-1})$ when acting on an appropriate Hilbert space (\cf\ similar identities in \cite{volkov-2003}).

\subsubsection{Loop algebra}
\label{qloop}

Finally, for the connection to Wilson and 't Hooft operators, we should quantize the algebra $\hat \CA_\hbar$ in loop coordinates. Aside from $\CN=4$ S-duality, this quantization is already known mathematically (\cf\ \cite{FockChekhov, ChekhovPenner, Kash-kernel} and earlier work \cite{Turaev-skein}). Physically, we obtain the algebras of Wilson and 't Hooft loops that were studied in \cite{GMNIII, Tesch-loop}.

We start by quantizing the relations between loop and (exponentiated) shear coordinates \eqref{loopshear}. To do so, we normal-order each of the terms appearing on the right-hand sides; for example the binomial $\sqrt{tt'}$ becomes
\be \sqrt{tt'} \,=\, e^{\frac{ T+ T'}{2}} \;\;\rightarrow\;\; e^{\frac{\hat T+\hat T'}{2}} \,=\, q^{-\frac14}\sqrt{\hat t}\sqrt{\hat t'}\,. \ee
Then we find
\begin{subequations} \label{qloopshear}
\begin{align}
-\hat x &= q^{-\frac14}\hat t{}^{\frac12}\hat t'{}^{\frac12}+q^{-\frac14}\hat t{}^{-\frac12}\hat t'{}^{-\frac12}+q^{\frac14}\hat t{}^{-\frac12}\hat t'{}^{\frac12}\,,\\
-\hat y &= q^{-\frac14}\hat t''{}^{\frac12}\hat t{}^{\frac12}+q^{-\frac14}\hat t''{}^{-\frac12}\hat t{}^{-\frac12}+q^{\frac14}\hat t''{}^{-\frac12}\hat t{}^{\frac12}\,,\\
-\hat z &= q^{-\frac14}\hat t'{}^{\frac12}\hat t''{}^{\frac12}+q^{-\frac14}\hat t'{}^{-\frac12}\hat t''{}^{-\frac12}+q^{\frac14}\hat t'{}^{-\frac12}\hat t''{}^{\frac12}\,.
\end{align}
\end{subequations}
This leads to interesting $q$-deformed commutation relations in the loop algebra:
\be q^{\frac14}\hat x\hat y-q^{-\frac14}\hat y\hat x = -(q^{\frac12}-q^{-\frac12})\hat z\,, \label{loopcomm} \ee
and similarly for the cyclic permutations $\hat x\to\hat y\to\hat z\to\hat x$. In other words, for Wilson and 't Hooft loops we have $q^{\frac14}W_{1/2}H_{1/2}-q^{-\frac14}H_{1/2}W_{1/2}=-(q^{\frac12}-q^{-\frac12})W_{1/2}^{(1,1)}$ \cite{GMNIII, Tesch-loop}. Moreover, the loop operators satisfy a quantized version of the Markov cubic constraint \eqref{toruscubic},
\be q^{\frac12}\hat x^2+q^{-\frac12}\hat y^2+q^{\frac12}\hat z^2+q^{\frac14}\hat x\hat y\hat z = q^{\frac12}\ell+\frac{1}{q^{\frac12}\ell}+q^{\frac12}+q^{-\frac12}\,. \label{qtoruscubic} \ee

We can similarly define $\CN=4$ dual loop coordinates $({}^L\hat x,{}^L\hat y,{}^L\hat z)$ by dualizing (adding ${}^L$'s to) the right-hand sides of (\ref{qloopshear}a-c). Then it is certainly true that $\hat \CA_\hbar = \C(q,{}^Lq)[\hat x,\hat y,\hat z,{}^L\hat x,{}^L\hat y,{}^L\hat z]$, modulo the quantized cubic \eqref{qtoruscubic} and its dual. However, the loop algebra does not \emph{quite} split into two mutually commuting copies $\hat \CA_q\otimes \hat \CA_{{}^Lq}$ in the obvious way. The main problem is the presence of square roots in \eqref{qloopshear}; for while (say) $\hat t$ and ${}^L\hat t'$ commute, their square roots anticommute: $\sqrt{\hat t}\sqrt{{}^L\hat t'}=-\sqrt{{}^L\hat t'}\sqrt{\hat t}$. For the loop operators, this implies that
\be \hspace{.6in} \hat x\,{}^L\hat x = {}^L\hat x\,\hat x\,,\qquad\text{but}\qquad  \hat x\,{}^L\hat y = -\,{}^L\hat y\,\hat x\,,\qquad  \hat x\,{}^L\hat z = -\,{}^L\hat z\,\hat x\qquad\text{(+ cyclic)}\,.
\label{xLyphase}
\ee
Thus, in $\CN=2^*$ gauge theory on $S^3$, a spin-1/2 Wilson loop on one great circle of $S^3$ and a 't Hooft loop on the other should anticommute. This is completely natural, since these operators are mutually nonlocal. When an electric quark is brought into the presence of a magnetic monopole, there is a nontrivial Poynting vector corresponding to half a unit of angular momentum, hence an extra phase $e^{i\pi}=-1$ in \eqref{xLyphase}.

By combining equations \eqref{qloopshear} and \eqref{qshearFN}, we also obtain (after some algebra) the quantized relation between loop and Fenchel-Nielsen operators:
\begin{subequations} \label{qloopFN}
\begin{align}
-\hat x &= \hat \lambda+\hat \lambda^{-1}\,,\\
-\hat y &= \frac{\ell^{-\frac12}\hat \lambda-\ell^{\frac12}\hat \lambda^{-1}}{\hat \lambda-\hat \lambda^{-1}}\sqrt{\hat \tau}+\frac{\ell^{\frac12}\hat \lambda-\ell^{-\frac12}\hat \lambda^{-1}}{\hat \lambda-\hat \lambda^{-1}}\frac{1}{\sqrt{\hat \tau}}\,, \label{qloopFNy} \\
-\hat z &= \frac{\ell^{-\frac12}\hat \lambda-\ell^{\frac12}\hat \lambda^{-1}}{\hat \lambda-\hat \lambda^{-1}}e^{\frac{\hat\CT}{2}-\hat \Lambda}+\frac{\ell^{\frac12}\hat \lambda-\ell^{-\frac12}\hat \lambda^{-1}}{\hat \lambda-\hat \lambda^{-1}}e^{-\frac{\hat\CT}{2}+\hat \Lambda}\,.
\end{align}
\end{subequations}
These relations also hold with $(\hat x,\hat y,\hat z,\hat\lambda,\hat \tau)$ replaced by their $\CN=4$ duals $({}^L\hat x,{}^L\hat y,{}^L\hat z,{}^L\hat\lambda,{}^L\hat \tau)$.

Altogether, (\ref{qloopFN}a-c) explicitly show the action of gauge theory loop operators on $\CH^{\CN=2^*}_{\epsilon_{1,2}}(S^3)$. For example, in a polarization such that Fenchel-Nielsen operators act on (Chern-Simons) wavefunctions as $\hat\Lambda = \Lambda$ and $\hat \CT = \hbar \pd_\Lambda$, the dictionary of Section \ref{sec:params} dictates that they act on instanton partition functions $Z(a,m_{\rm adj};\epsilon_1,\epsilon_2)\in \CH^{\CN=2^*}_{\epsilon_{1,2}}(S^3)$ as
\be \hat\CT = \hbar\pd_\Lambda \to -\epsilon_1\pd_a\,,\qquad \hat\Lambda = \Lambda\to -\frac{2\pi i}{\epsilon_2}a\,,\ee
and (\cf\ \cite{Tesch-loop})
\begin{subequations}
\begin{align} W_{1/2} &= 2\cosh(2\pi i a/\epsilon_2)\,, \\
H_{1/2} &= \frac{\sinh(\frac{2\pi i}{\epsilon_2}a-\frac{i\pi}{\epsilon_2}m_{\rm adj})}{\sinh(2\pi i a/\epsilon_2)}e^{-\frac{\epsilon_1}{2}\pd_a}+\frac{\sinh(\frac{2\pi i}{\epsilon_2}a+\frac{i\pi}{\epsilon_2}m_{\rm adj})}{\sinh(2\pi i a/\epsilon_2)}e^{\frac{\epsilon_1}{2}\pd_a}\,,
\end{align}
\end{subequations}
\emph{etc}. Similarly, by replacing $\Lambda \to 2\pi bp_x$ and $v\to 2\pi b(p_v-iQ/2)$ we obtain Verlinde loop operators in Liouville theory. The normalization (or polarization) for wavefunctions being used here corresponds to rescaling conformal blocks by part of the DOZZ 3-point functions \cite{Tesch-loop}. It differs somewhat from the polarization used in (\eg) \cite{AGGTV} but is more natural for the connection to Chern-Simons theory; in particular, it allows all loop operators as in \eqref{qloopFN} to be \emph{algebraic} functions of $\hat\lambda$ and $\sqrt{\hat\tau}$.

\subsection{Kernels and operator equations}
\label{sec:qcyl}

Having described the quantum algebra $\hat \CA_\hbar$ in various isomorphic coordinate systems, we now continue to discuss the mapping class group actions (automorphisms of $\hat \CA_\hbar$) induced by elements $\varphi\in\bm\Gamma(C)$. Here we start with the $\bm \Gamma(C)$ action on the shear algebra. Just as in the classical case, this action is combinatorial, and allows wavefunctions $Z_\varphi(T,T_\flat)$ to be easily computed. We then consider the $\bm\Gamma(C)$ action on the loop algebra. In ``physical'' Fenchel-Nielsen coordinates, both the quantum $\bm\Gamma(C)$ action and the wavefunctions $Z_\varphi(\Lambda,\Lambda_\flat)$ will be discussed in Section \ref{sec:eg}.

\subsubsection{Automorphisms of the shear algebra}

For illustrative purposes, let us start with the element $\varphi = S\;\in\;PSL(2,\Z)\cong \bm\Gamma(C)$. Following \cite{FockChekhov} and \cite{FG-qdl-cluster}, the induced automorphism of $\hat\CA_\hbar$ factors as a composition of a linear symplectic transformation and conjugation by a quantum dilogarithm:
\be S:\;(\hat T,\hat T',\hat T'')\;\overset{\rm lin}{\longmapsto}\; (-\hat T,\hat T'',\hat T'+2\hat T)\;\overset{\rm QDL}{\longmapsto}\;\Phi_\hbar(-\hat T)(-\hat T,\hat T'',\hat T'+2\hat T)\Phi_\hbar(-\hat T)^{-1}\,, \label{qTS} \ee
where the quantum dilogarithm can be defined as
\be \Phi_\hbar(T) = \prod_{r=1}^\infty\frac{1+q^{r-\frac12}e^T}{1+{}^Lq^{\frac12-r}e^{{}^LT}}\qquad (|q|<1) \ee
(for further details, see Appendix \ref{app:qdl}). In particular, in exponentiated operators we find
\begin{subequations} \label{qtS}
\begin{align}
\big(\sqrt{\hat t},\sqrt{\hat t'},\sqrt{\hat t''}\big)
&\;\overset{\rm lin}{\longmapsto}\;
\Big(\frac{1}{\sqrt{\hat t}},\sqrt{\hat t''},\sqrt{q\hat t'}\,\hat t\Big)
\;\overset{\rm QDL}{\longmapsto}\;
\Big(\frac{1}{\sqrt{\hat t''}},\sqrt{\hat t''}\frac{1}{1+q^{\frac12}\hat t^{-1}},\sqrt{\hat t'}(1+q^{\frac12}\hat t)\Big)\,, \\
\big(\sqrt{{}^L\hat t},\sqrt{{}^L\hat t'},\sqrt{{}^L\hat t''}\big)
&\;\overset{\rm lin}{\longmapsto}\;
\Big(\frac{1}{{}^L\sqrt{\hat t}},\sqrt{{}^L\hat t''},\sqrt{{}^Lq{}^L\hat t'}\,{}^L\hat t\Big) \notag \\ &\hspace{.6in}
\;\overset{\rm QDL}{\longmapsto}\;
\Big(\frac{1}{\sqrt{{}^L\hat t''}},\sqrt{{}^L\hat t''}\frac{1}{1+\!{}^Lq^{\frac12}{}^L\hat t^{-1}},\sqrt{{}^L\hat t'}(1+\!{}^Lq^{\frac12}{}^L\hat t)\Big)\,.
\end{align}
\end{subequations}
Observe that before and after the transformations \eqref{qTS} or \eqref{qtS} we still have $\hat T+\hat T'+\hat T''=v+i\pi+\hbar/2$. The mapping class group actions for other generators of $\bm\Gamma(C)$ are just permutations of \eqref{qTS}. They can be encoded in diagonal flips (and skews/rotations), as described in the classical setup of Section \ref{sec:shearclassMCG}. We summarize the results for exponentiated operators here, with the understanding that $\CN=4$ S-dual operators have identical transformations:
\be\label{qMCGshear}
\begin{array}{c@{\;\;}|c@{\quad}c@{\quad}c@{\;}|@{\;\;}c}
\varphi & \sqrt{\hat t}\,\mapsto & \sqrt{\hat t'}\,\mapsto & \sqrt{\hat t''}\,\mapsto & Z_\varphi(T,T_\flat) \\\hline
S & 
 \ds \frac1{\sqrt {\hat t}} & \ds \sqrt{\hat t''}\frac{1}{1+q^{\frac12}\hat t^{-1}} & \sqrt{\hat t'}(1+q^{\frac12}\hat t) &
 \raisebox{-.1cm}{$\frac{\ds e^{-\frac1{2\hbar}(v+c_\hbar)T-\frac{1}{4\hbar}T^2}}{\ds \Phi_\hbar(T)}\delta(T\!+\!T_\flat)$} \\[.4cm]
R = T & 
\ds \frac{1}{\sqrt {\hat t''}} & \ds \sqrt{\hat t'}\frac{1}{1+q^{\frac12}\hat t''{}^{-1}} & \sqrt{\hat t}(1+q^{\frac12}\hat t'') &
e^{-\frac1{4\hbar}(T+v+c_\hbar-T_\flat)^2}\Phi_\hbar(T)^{-1} \\[.4cm]
R^{-1} & 
 \ds \sqrt{\hat t''}\frac{1}{1+q^{\frac12}\hat t^{-1}} & \sqrt{\hat t'}(1+q^{\frac12}\hat t) & \ds\frac{1}{\sqrt{\hat t}} &
 e^{\frac1{4\hbar}(T-v-c_\hbar-T_\flat)^2}\Phi(T_\flat) \\[.4cm]
L = T^t & 
\ds \frac{1}{\sqrt{\hat t'}} & \ds \sqrt{\hat t}\frac{1}{1+q^{\frac12}\hat t'{}^{-1}} & \sqrt{\hat t''}(1+q^{\frac12}\hat t') &
e^{-\frac1{2\hbar}TT_\flat}\,\Phi_\hbar(T)^{-1} \\[.4cm]
L^{-1} & 
\sqrt{\hat t'}(1+q^{\frac12}\hat t) & \ds \frac{1}{\sqrt{\hat t}} & \ds \sqrt{\hat t''}\frac{1}{1+q^{\frac12}\hat t^{-1}} &
e^{\frac1{2\hbar}T T_\flat}\,\Phi_\hbar(T_\flat)\,.
\end{array}
\ee

The final column of \eqref{qMCGshear} displays the wavefunction $Z_\varphi$ that is annihilated by the quantum operators $\wh \Delta_\varphi$ coming from each mapping class group element $\varphi$ (in these formulas, $c_\hbar = i\pi+\hbar/2$).%
\footnote{We produce these wavefunctions up to an overall normalization by an $\hbar$- and $v$-dependent constant. The dependence on $v$ can become important when mapping cylinders are glued together to form mapping tori, as in Section \ref{sec:tori}. A detailed analysis of the resulting wavefunctions for punctured torus bundles, including the $v$-dependence, recently appeared in \cite{Yamazaki-layered}.} %
Recall that the wavefunction $Z_\varphi(T,T_\flat)$ depends on coordinates at the top and bottom of a mapping cylinder $M_\varphi$, and satisfies
\be \big(\hat t{}^{(}{}'{}^{)}{}^{(}{}''{}^{)} - \big[\varphi(\hat t{}^{(}{}'{}^{)}{}^{(}{}''{}^{)})\big]_\flat\big) Z_\varphi(T,T_\flat)= 0\,,\qquad
\big({}^L\hat t{}^{(}{}'{}^{)}{}^{(}{}''{}^{)} - \big[\varphi({}^L\hat t{}^{(}{}'{}^{)}{}^{(}{}''{}^{)})\big]_\flat\big) Z_\varphi(T,T_\flat)=0\,.
\ee
For example, in the case of the S-move, we have four equations
\begin{subequations} \label{shearSops}
\be \sqrt{\hat t}-\frac1{\sqrt{\hat t_\flat}} \,\simeq\, 0\,,\qquad\qquad\qquad \sqrt{{}^L\hat t}-\frac1{\sqrt{{}^L\hat t_\flat}}\,\simeq\, 0\,, \ee
\be \sqrt{\hat t'}-\frac{1}{1+q^{\frac12}\hat t^{-1}_\flat}\sqrt{\hat t''_\flat} \,\simeq\, 0\,,\qquad \sqrt{{}^L\hat t'}-\frac{1}{1+{}^Lq^{\frac12}{}^L\hat t^{-1}_\flat}\sqrt{{}^L\hat t''_\flat} \,\simeq\, 0\,,
\ee
\end{subequations}
that annihilate $Z_S(T,T_\flat)$ and generate the left ideal $\wh\Delta_S$. We have chosen a polarization such that logarithmic operators act as
\begin{subequations}
\be \hat T = T\,,\qquad \hat T'=-2\hbar\pd_T\,,\qquad \hat T'' = v+i\pi+\hbar/2-T+2\hbar\pd_T\,,\ee
\be \hat T_\flat = T_\flat\,,\qquad \hat T'_\flat=2\hbar\pd_{T_\flat}\,,\qquad \hat T''_\flat = v+i\pi+\hbar/2-T_\flat-2\hbar\pd_{T_\flat}\,.\ee
\end{subequations}

The calculation of $Z_\varphi$ for any general $\varphi\in\bm\Gamma(C)$ is completely straightforward. One simply decomposes $\varphi$ into generators, say $\varphi = \varphi_N\cdots\varphi_1$. Then the wavefunction is obtained by gluing together mapping cylinders for each $\varphi_i$\,:\; $Z_\varphi(T,T_\flat) = \int dT_{N-1}... dT_1\, Z_{\varphi_N}(T,T_{N-1})\cdots Z_{\varphi_1}(T_1,T_\flat)$\,.

\subsubsection{Automorphisms of the loop algebra}

The mapping class group action in the loop algebra can be derived directly from the quantized relation to shear coordinates \eqref{qloopshear} and the action in the shear algebra \eqref{qMCGshear}. Alternatively, the transformations for two of the three loop operators are usually very intuitive, just as in the classical case. For example, the $S$-move maps cycles $(\gamma_x,\gamma_y)\mapsto(\gamma_y^{-1},\gamma_x)$, so it switches $\hat x\leftrightarrow\hat y$; whereas the $T^t$-move fixes $\gamma_x$ and sends $\gamma_y\mapsto\gamma_z$, so that $(\hat x,\hat y)\mapsto(\hat x,\hat z)$. The quantized action on the third loop operator can then be obtained by requiring invariance of the quantized cubic \eqref{qtoruscubic}. One way or another, we find:
\be \label{qMCGloop}
\begin{array}{c|c@{\quad}c@{\quad}c}  \varphi & \hat x\mapsto & \hat y\mapsto & \hat z\mapsto  \\[.1cm] \hline
 S = {\small \begin{pmatrix} 0 & -1 \\ 1 & 0\end{pmatrix}} & \hat y & \hat x & -q^{\frac12}\hat z-q^{\frac14}\hat x\hat y \\[.4cm]
R = T = {\small \begin{pmatrix} 1 & 1 \\ 0 & 1\end{pmatrix}} & \hat z & \hat y & -q^{\frac12}x-q^{\frac14}\hat y\hat z \\[.4cm]
R^{-1} = {\small \begin{pmatrix} 1 & -1 \\ 0 & 1\end{pmatrix}} & -q^{\frac12}\hat z-q^{\frac14}\hat x\hat y & \hat y & \hat x\\[.4cm]
L = T^t = {\small \begin{pmatrix} 1 & 0 \\ 1 & 1\end{pmatrix}} & \hat x & \hat z & -q^{\frac12}y-q^{\frac14}\hat z\hat x \\[.4cm]
L^{-1} = {\small \begin{pmatrix} 1 & 0 \\ -1 & 1\end{pmatrix}} & \hat x &-q^{\frac12}z-q^{\frac14}\hat x\hat y & \hat y
\end{array}
\ee

In the complete algebra $\hat \CA_\hbar$ we also have $\CN=4$ S-dual loop operators $({}^L\hat x,{}^L\hat y,{}^L\hat z)$, which obey identical transformations. A mapping-cylinder wavefunction $Z_\varphi$ (in any polarization or representation) is annihilated by a left ideal of operators generated by four elements. For example, for the $S$-move, we would have
\begin{subequations} \label{loopSops}
\be \hat x -\hat y_\flat\,\simeq\,0\,,\qquad {}^L\hat x -{}^L\hat y_\flat\,\simeq\,0\,, \ee
\be \hat y -\hat x_\flat\,\simeq\,0\,,\qquad {}^L\hat y -{}^L\hat x_\flat\,\simeq\,0\,, \ee
\end{subequations}
all annihilating $Z_\varphi$. These are analogous to \eqref{shearSops} in shear coordinates. Note, however, that equations \eqref{loopSops} and their duals only commute up to a phase; for example $(\hat x -\hat y_\flat)({}^L\hat y -{}^L\hat x_\flat)=-({}^L\hat y -{}^L\hat x_\flat)(\hat x -\hat y_\flat)$ (with a phase of $e^{i\pi}=-1$). This is no obstacle to imposing all four equations simultaneously on a wavefunction.

\subsection{Examples}
\label{sec:eg}

We consider several more detailed examples of mapping cylinders $M$, focusing especially on the interplay between their quantum wavefunctions $Z_M$ and the ideals of operators (``quantized Lagrangians'') $\wh\Delta_M$ that annihilate these wavefunctions. None of the actual wavefunctions presented below are entirely new formulas. We hope, however, that our treatment of these examples will be enlightening.

Since we have largely neglected coordinate-transformation cylinders $M_\xi$ above, we begin with a simple but fundamental example of the mapping cylinder that interpolates between shear and Fenchel-Nielsen coordinates. We then consider the mapping cylinders for $T^t$ and $S$ elements of the mapping class group of the punctured torus, writing down wavefunctions and equations in the more physical Fenchel-Nielsen algebra.

\subsubsection{Shear\,--\,Fenchel-Nielsen cylinder}
\label{sec:qshearFN}

\FIGURE{
\includegraphics[width=2in]{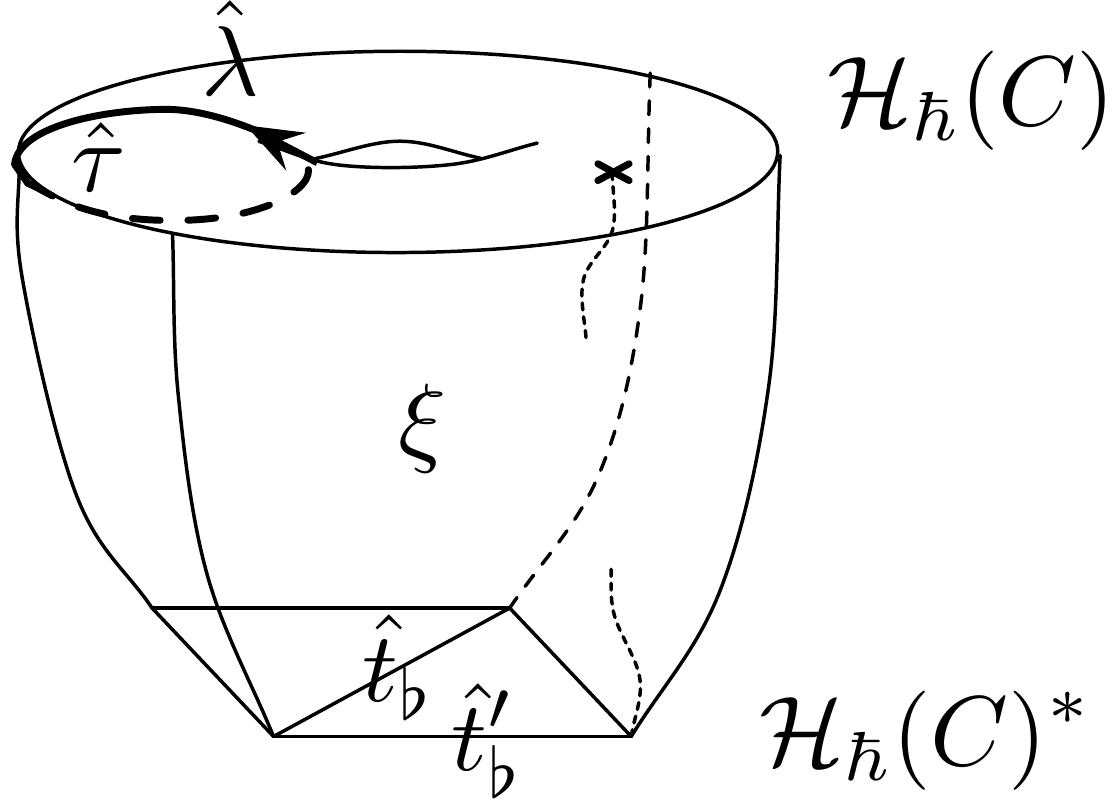}
\caption{Cylinder $M_\xi$ for the shear--F.N. coordinate transformation.}
\label{fig:qshearFN}
}

The mapping cylinder $M_\xi$ that implements a transformation from shear to Fenchel-Nielsen coordinates is shown in Figure \ref{fig:qshearFN} (\cf\ Figure \ref{fig:sFNcyl}a). A Hilbert space $\CH_\hbar(C)$ and its dual $\CH_\hbar(C)^*$ are obtained by quantizing the spaces $Y$ and $\ol{Y}$ of flat connections on the top and bottom boundaries, respectively. Operator algebras $\hat \CA_\hbar$ and $[\hat \CA_\hbar]_\flat$ act on these spaces, and we write them in terms of Fenchel-Nielsen and shear operators, respectively:
\be \hat\CA_\hbar  = \C(\hbar)[\hat \CT,\hat\Lambda]\,,\qquad [\hat \CA_\hbar]_\flat = \C(\hbar)[\hat T_\flat,\hat T_\flat'
,\hat T''_\flat],
\ee
where $[\hat \CT,\hat \Lambda]=\hbar$ but $[\hat T_\flat,\hat T'_\flat]=-2\hbar$.

The quantum wavefunction $Z_\xi(\Lambda,T_\flat;\hbar)$ of $M_\xi$ was first found (as a coordinate-transformation kernel) in \cite{Kash-kernel}, and studied extensively in the context of relating quantum Teichm\"uller, Liouville, and gauge theories \cite{Teschner-TeichLiouv, Tesch-LtoQG, Teschner-TeichMod, Tesch-loop}. In a polarization where
\be \hat \CT = \hbar\,\pd_\Lambda\,,\qquad \hat\Lambda = \Lambda\,,\qquad \hat T_\flat' = 2\hbar\, \pd_{T_\flat}\,,\qquad \hat T_\flat = T_\flat\,,
\ee
we find\footnote{This is rescaled from the kernel of (\eg) \cite{Kash-kernel} by a factor $\exp(\Lambda^2/2\hbar)$.}
\be Z_\xi(\Lambda,T_\flat) = \frac{e^{\frac1\hbar c_\hbar\Lambda+\frac{\Lambda^2}{2\hbar}}}{\sqrt{i\pi\hbar}}\int dS\,e^{-\frac1\hbar\big(S-\frac T2\big)^2+\frac{S^2}{2\hbar}-\frac1\hbar\Lambda S}\,
\frac{\Phi_\hbar(\Lambda-S+c_\hbar)}{\Phi_\hbar(\Lambda+S-c_\hbar)}
\label{sFNwf}
\ee
(with $c_\hbar = i\pi+\hbar/2$).
This wavefunction is annihilated by two difference equations
\be
\frac{1}{\hat r_\flat}+\hat s_\flat+\frac{1}{\hat s_\flat}-\hat \lambda-\frac{1}{\hat \lambda}\,\simeq\, 0\,,\qquad
(\hat s_\flat-q\hat\lambda)\,\hat\tau+q\hat\lambda(1-\hat s_\flat\hat\lambda)\,\simeq 0\,,
\label{sFNops1}
\ee
and their $\CN=4$ S-duals
\be
\frac{1}{{}^L\hat r_\flat}+{}^L\hat s_\flat+\frac{1}{{}^L\hat s_\flat}-{}^L\hat \lambda-\frac{1}{{}^L\hat \lambda}\,\simeq\, 0\,,\qquad
({}^L\hat s_\flat-{}^Lq{}^L\hat\lambda)\,{}^L\hat\tau+{}^Lq{}^L\hat\lambda(1-{}^L\hat s_\flat{}^L\hat\lambda)\,\simeq 0\,,
\label{sFNLops1}
\ee
where
$\hat s_\flat \,\defeq\, \exp\big(\frac{\hat T_\flat+\hat T_\flat'}{2}\big)$,
$\hat r_\flat \,\defeq\, \exp\big(\frac{\hat T_\flat-\hat T_\flat'}{2}\big)$,
and similarly
${}^L\hat s_\flat \,\defeq\, \exp\big(\frac{{}^L\hat T_\flat+{}^L\hat T_\flat'}{2}\big)$ and
${}^L\hat r_\flat \,\defeq\, \exp\big(\frac{{}^L\hat T_\flat-{}^L\hat T_\flat'}{2}\big)$.
Together, equations \eqref{sFNops1}--\eqref{sFNLops1} generate $\wh\Delta_\xi$. With a little bit of work (which we defer to Appendix \ref{app:map}), we can show that equations \eqref{sFNops1} are equivalent to
\be
\sqrt{\hat t_\flat}+\frac{i}{\hat\lambda-\hat\lambda^{-1}}(\hat \tau^{\frac12}-\hat\tau^{-\frac12})\,\simeq\, 0\,,\quad
\sqrt{\hat t_\flat'}-\frac{i}{\frac{q^{-\frac14}}{\hat\lambda}\hat\tau^{\frac12}-\hat\lambda\frac{q^{-\frac14}}{\hat \tau^{\frac12}}}(\hat\lambda-\hat\lambda^{-1})\,\simeq\,0\,,
\label{sFNops}
\ee
which correspond directly to the algebra isomorphism \eqref{qshearFN} for $\hat \CA_q$. Similarly, \eqref{sFNLops1} can be rewritten in a form that corresponds to the isomorphism \eqref{LqshearFN} for $\hat \CA_{{}^Lq}$.

It is easily verified%
\footnote{One should be careful about integration contours when dualizing wavefunctions. In the case of $Z_\xi$, however, there is a unique infinite integration contour that passes between the half-lines of zeroes and the half-lines of poles of the quantum dilogarithms in the integrand. This is the contour that should always be used (\cf\ \cite{volkov-2003}).} %
that the wavefunction \eqref{sFNwf} has $\CN=4$ S-duality, in the sense that
\be Z_\xi(\Lambda,T_\flat;\hbar) = Z_\xi({}^L\Lambda,{}^LT_\flat;{}^L\hbar)\,. \label{ZsFNS} \ee
Given the initial equations \eqref{sFNops1}, the S-duality property \eqref{ZsFNS} immediately implies that the dual equations \eqref{sFNLops1} must hold as well. Conversely, the combined system \eqref{sFNops1}--\eqref{sFNLops1} annihilates a unique%
\footnote{The solution is unique up to multiplication by $\hbar$- and $v$-dependent ``constants'', and multiplication by elliptic functions. The presence of nontrivial elliptic functions can typically be ruled out by requiring some regularity for wavefunction asymptotics.} %
wavefunction $Z_\xi$, and this wavefunction must have $\CN=4$ S-duality.

As discussed in Section \ref{sec:FN}, the Fenchel-Nielsen twist $\tau$ is canonically defined, either at the classical or quantum levels. Our present choice for $\tau$ was used extensively in the connection between quantum Teichm\"uller and Liouville theory, though the more geometric $\tau^{(FN)}$ has also made recent appearances in the physics literature \cite{NRS}. Such choices for $\tau$ correspond to different quantum polarizations for $Z_\xi(\Lambda,T_\flat)$. For example, in order to obtain Lagrangian equations $\Delta_\xi$ for the geometric twist $\tau^{(FN)}$ instead of $\tau$ (in a classical limit) we could modify
\be Z_\xi(\Lambda,T_\flat) \;\longrightarrow\; Z_\xi^{(FN)}(\Lambda,T_\flat) \,=\, e^{-\frac1\hbar v\Lambda}\, \sqrt{\frac{\Phi_\hbar(2\Lambda-v+c_\hbar)}{\Phi_\hbar(2\Lambda+v-c_\hbar)}} \,Z_\xi(\Lambda,T_\flat)\,, \ee
with $c_\hbar =i\pi+\hbar/2$.
On the new wavefunction $Z_\xi^{(FN)}$, the logarithmic $\hat \CT^{(FN)}$ will now act as $\hbar\,\pd_\Lambda$.

\subsubsection{$T$-move}

We now come back to mapping class group transformations $\varphi$, and the corresponding mapping cylinder wavefunctions $Z_\varphi(\Lambda,\Lambda_\flat)$ and quantized Lagrangians $\wh\Delta_\varphi$. When working in Fenchel-Nielsen coordinates, and specifically in a polarization such that
\be \hat \CT = \hbar\,\pd_\Lambda\,,\qquad \hat \Lambda=\Lambda\,,\qquad \hat \CT_\flat = -\hbar\,\pd_{\Lambda_\flat}\,,\qquad\hat \Lambda_\flat = \Lambda_\flat\,,\ee
the wavefunction $Z_\varphi(\Lambda,\Lambda_\flat)$ can be interpreted as a Moore-Seiberg kernel for Liouville conformal blocks, or as a mapping kernel for gauge theory partition functions in $\CH^{\CN=2^*}_{\epsilon_{1,2}}(S^3)$. This was described in the introduction, and in Section \ref{sec:cyl}. For example, using the dictionary of Section \ref{sec:params}, the instanton partition function $Z(a,m_{\rm adj};\epsilon_1,\epsilon_2)$ of $\CN=2^*$ gauge theory should transform under an $\CN=2$ S-duality $\varphi$ as%
\footnote{\label{foot:measure}%
In the literature, a measure factor $\nu(a_\flat) \sim \sinh(2\pi ia_\flat/\epsilon_2)\sinh(2\pi ia_\flat/\epsilon_1)\sim \sinh(\Lambda_\flat)\sinh({}^L\Lambda_\flat)$ often appears in expressions like \eqref{generickernel}. We always include this measure as part of the mapping cylinder wavefunction $Z_\varphi$. This turns out to be crucial for a unified treatment in terms of operator algebra.} %
\be
Z(a,m_{\rm adj};\epsilon_1,\epsilon_2)\,\overset{\varphi}{\longmapsto}\,\int da_\flat\,Z_\varphi\Big(-\frac{2\pi i}{\epsilon_2}a,-\frac{2\pi i}{\epsilon_2}a_\flat\Big)\,Z(a_\flat,m_{\rm adj};\epsilon_1,\epsilon_2)\,. \label{generickernel}
\ee

In Fenchel-Nielsen coordinates, some elements $\varphi\in \bm\Gamma(C)$ have a particularly simple action in $\hat\CA_\hbar$ and on wavefunctions, while others are quite complicated. Just as in the classical case (Section \ref{sec:FNclassMCG}), the $T^t$ transformation is very simple. From \eqref{qMCGloop} and \eqref{qloopFN}, we first find that the wavefunction must be annihilated by
\be \hat x-\hat x_\flat\,\simeq\,0 \qquad\Rightarrow\qquad  \hat\lambda+\hat\lambda^{-1}-\hat\lambda_\flat-\hat\lambda_\flat^{-1}\,\simeq\,0\,,
\ee
which implies that $\hat\lambda-\hat\lambda_\flat\simeq 0$ or $\hat\lambda-\hat\lambda_\flat^{-1}\simeq 0$. (The difference between the two possibilities is the Weyl group action on $Y$, which we ought to quotient out by in the operator algebras.) We will choose $\hat\lambda-\hat\lambda_\flat\simeq 0$. Together with the dual equation ${}^L\hat x-{}^L\hat x_\flat \simeq 0$, this actually implies $\hat \Lambda-\hat \Lambda_\flat\simeq 0$, in logarithmic coordinates.

The second equation in $\wh\Delta_{T^t}$ (and its $\CN=4$ dual) is $\hat y-\hat z_\flat\simeq 0$, which looks rather more complicated:
\be
\frac{\ell^{-\frac12}\hat \lambda-\ell^{\frac12}\hat \lambda^{-1}}{\hat \lambda-\hat \lambda^{-1}}\sqrt{\hat \tau}+\frac{\ell^{\frac12}\hat \lambda-\ell^{-\frac12}\hat \lambda^{-1}}{\hat \lambda-\hat \lambda^{-1}}\frac{1}{\sqrt{\hat \tau}}
-
\sqrt{\hat \tau_\flat}\frac{q^{-\frac14}}{\hat\lambda_\flat}\frac{\ell^{-\frac12}\hat \lambda_\flat-\ell^{\frac12}\hat \lambda_\flat^{-1}}{\hat \lambda_\flat-\hat \lambda_\flat^{-1}}-\frac{q^{-\frac14}}{\sqrt{\hat \tau_\flat}}\hat\lambda_\flat\frac{\ell^{\frac12}\hat \lambda_\flat-\ell^{-\frac12}\hat \lambda_\flat^{-1}}{\hat \lambda_\flat-\hat \lambda_\flat^{-1}}\,\simeq\,0 \label{Tyeq1}
\ee
However, after using $\hat\Lambda-\hat\Lambda_\flat\simeq 0$, we find
\be \big(\ell^{-\frac12}\hat \lambda-\ell^{\frac12}\hat \lambda^{-1})\bigg(\sqrt{\hat \tau}-\frac{q^{-\frac14}}{\hat\lambda}\sqrt{\hat \tau_\flat}\bigg)+\big(\ell^{\frac12}\hat \lambda-\ell^{-\frac12}\hat \lambda^{-1}\big)\bigg(\frac{1}{\sqrt{\hat\tau}}-\hat\lambda\frac{q^{-\frac14}}{\sqrt{\hat \tau_\flat}}\bigg)\,\simeq\,0\,,
\ee
whose only consistent ``solution'' is $q^{\frac14}\hat\lambda\sqrt{\hat\tau} - \sqrt{\hat \tau_\flat}\,\simeq\,0$, or in fact $2\hat\Lambda+\hat\CT-\hat \CT_\flat\,\simeq\,0$. Thus, we find that we can write
\be \wh\Delta_{T^t}\;=\;\{\hat\Lambda-\hat\Lambda_\flat\,,\; \hat\CT  - \hat\CT_\flat+2\hat\Lambda_\flat\}\,.
\ee
This is obviously a direct quantization of \eqref{DTclass}. Geometrically, a $T^t$ move on the punctured torus cuts open the Fenchel-Nielsen pair of pants, and glues it back together with one less full twist, decreasing $\hat\CT$ by $2\hat \Lambda$ while preserving $\hat\Lambda$.

The unique solution to $\wh\Delta_{T^t}\,Z_{T^t} = 0$ is
\be Z_{T^t}(\Lambda,\Lambda_\flat) = e^{-\frac1\hbar\Lambda^2}\,\delta(\Lambda-\Lambda_\flat)\,, \label{ZT}
\ee
up to multiplication by $\hbar$- and $v$--dependent constants.
The exponential in \eqref{ZT} can immediately be identified with the conformal weight $\Delta_x = \frac{Q^2}{4}+p_x^2$ of a primary field with Liouville momentum $p_x = \Lambda/(2\pi b)$; thus we have recovered the standard, diagonal $T^{(t)}$--transformation of Liouville conformal blocks, as multiplication by $e^{2\pi i\Delta_x}$.

The wavefunction \eqref{ZT} could also have been obtained by starting with a wavefunction in shear coordinates from Table \eqref{qMCGshear}, and ``sandwiching'' it between two coordinate-transformation cylinders as in Figure \ref{fig:sFNcyl}(b). This leads to the representation
\be Z_{T^t}(\Lambda,\Lambda_\flat) = \int dT\,dT_\flat\,Z_\xi(\Lambda,T_\flat)\,Z_{T^t}(T_\flat,T)\,Z_{\xi^{-1}}(T,\Lambda_\flat)\,, \label{ZT2}
\ee
sometimes quoted in the literature. Deriving \eqref{ZT} directly from \eqref{ZT2} involves several nontrivial quantum dilogarithm identities (\cf\ Appendix \ref{app:qdl}) and is \emph{much} less trivial than the manipulation of difference operators that we have just performed.

\subsubsection{$S$-move}

The operators $\wh\Delta_S$ corresponding to the element $\varphi=S\;\in\bm\Gamma(C)$ do not simplify quite as nicely as those for $\varphi=T^t$. Translating the loop equations $\hat x-\hat y_\flat \,\simeq\, 0$ and $\hat y-\hat x_\flat\,\simeq\,0$ into Fenchel-Nielsen operators, we find
\begin{subequations} \label{ZSeqs}
\begin{align}
&\hat \lambda+\hat \lambda{}^{-1}-\sqrt{\adj{\hat\tau}}\frac{\ell^{-\frac12}\adj{\hat \lambda}-\ell^{\frac12}\adj{\hat \lambda}^{-1}}{\adj{\hat \lambda}-\adj{\hat \lambda}^{-1}}-\frac{1}{\sqrt{\adj{\hat\tau}}}\frac{\ell^{\frac12}\adj{\hat \lambda}-\ell^{-\frac12}\adj{\hat \lambda}^{-1}}{\adj{\hat \lambda}-\adj{\hat \lambda}^{-1}} \simeq 0\,, \\
&\hat\lambda_\flat+\hat\lambda_\flat^{-1}-\frac{\ell^{-\frac12}{\hat \lambda}-\ell^{\frac12}{\hat \lambda}^{-1}}{{\hat \lambda}-{\hat \lambda}^{-1}}\sqrt{\hat\tau}-\frac{\ell^{\frac12}{\hat \lambda}-\ell^{-\frac12}{\hat \lambda}^{-1}}{{\hat \lambda}-{\hat \lambda}^{-1}}\frac{1}{\sqrt{\hat\tau}} \simeq 0\,.
\end{align}
\end{subequations}
These, along with their $\CN=4$ S-duals, annihilate the $\CN=2$ S-duality kernel, or wavefunction, $Z_S(\Lambda,\Lambda_\flat)$.

The unique solution for $Z_S(\Lambda,\Lambda_\flat)$, up to an $\hbar$- and $v$--dependent constant, appears in \cite{Tesch-Liouv, Tesch-LtoQG}. In our present conventions, it is given by
\be Z_S(\Lambda,\Lambda_\flat)= \frac{\sqrt{i}\ell^{-\frac14}}{\sqrt{2}\pi\hbar}\frac{e^{\frac{1}{\hbar}\adj \Lambda(v+2\pi i+\hbar)-\frac{i\pi}{2\hbar}v-\frac{v^2}{4\hbar}+C_\hbar}}{\Phi_\hbar(v+c_\hbar)}\nu(\adj \Lambda)\int dR\,  e^{-\frac2\hbar R\Lambda} \prod_{\delta=\pm} \frac{\Phi_\hbar\big(\adj \Lambda+v/2+c_\hbar+\delta\hspace{.02cm}R\big)}{\Phi_\hbar\big(\adj \Lambda-v/2-c_\hbar+\delta\hspace{.02cm}R\big)}\,,
\label{ZS}
\ee
with $\nu(\Lambda_\flat) = \frac{1}{2\sqrt{-2\pi i\hbar}}(\lambda-\lambda^{-1})({}^L\lambda-{}^L\lambda^{-1})$ (this is the measure that was mentioned in Footnote \ref{foot:measure}) and $c_\hbar =i\pi+\hbar/2$, $C_\hbar = \frac{\pi^2-\hbar^2/4}{6\hbar}$. It is fairly straightforward to show that \eqref{ZS} does indeed satisfy \eqref{ZSeqs}. Moreover, since $Z_S(\Lambda,\Lambda_\flat)$ is manifestly $\CN=4$ S-dual,
\be Z_S(\Lambda,\Lambda_\flat;\hbar) = Z_S({}^L\Lambda,{}^L\Lambda_\flat;{}^L\hbar)\,,\ee
it must satisfy the S-dual analogues of \eqref{ZSeqs}. A first-principles derivation of \eqref{ZS} using an ideal triangulation of the mapping cylinder $M_S$ will appear in \cite{DGGV-hybrid}.
We observe that in the classical limit $\hbar\to 0$, the leading asymptotic $Z_S \sim \exp\big(\frac1\hbar\CW+...\big)$ of \eqref{ZS} reproduces the effective superpotential of 3-dimensional $T[SU(2)]$ theory \eqref{WLi2} (\cf\ \cite{HLP-wall}).

Just as in the classical construction of Section \ref{sec:FNclassMCG}, the special limit $v\to 0$ (or alternatively $\ell\to 1$) removes the puncture from the torus $C$. In terms of gauge theory, this effectively turns off the adjoint mass parameter $m_{\rm adj}$, thus promoting 4-dimensional $\CN=2^*$ gauge theory to $\CN=4$.%
\footnote{According to \eqref{madjvrelation}, we are actually sending $m_{\rm adj}\to -(\epsilon_1+\epsilon_2)/2$. In the presence of an $\Omega$ deformation, there are several ``$\CN=4$ limits'' that reproduce different features of physical $\CN=4$ gauge theory (\cf\ \cite{OkudaPestun}). The current limit is relevant for obtaining the correct S-duality kernel.} %
The operators $\wh\Delta_S$ then simplify to become
\be
\hat\lambda+\hat \lambda^{-1}-\hat \tau_\flat^{\frac12}-\hat\tau_\flat^{-\frac12}\,\simeq\,0\,,\qquad \hat\lambda_\flat+\hat \lambda_\flat^{-1}-\hat \tau^{\frac12}-\hat\tau^{-\frac12}\,\simeq\,0\,.
\label{ZSeqs4}
\ee
For $v=0$, the (now quantized) Fenchel-Nielsen twist $\hat\CT$ is just twice the holonomy eigenvalue for the cycle $\gamma_y$ dual to $\gamma_x$ on $C$. Therefore, we can switch from $(\hat\CT,\hat\Lambda)$ to two canonically conjugate eigenvalues $(\hat\Lambda_y,\hat\Lambda_x)=(\hat\CT/2,\hat\Lambda)$, with
\be [\hat \Lambda_y,\hat\Lambda_x]=\hbar/2\,.\ee
Equations \eqref{ZSeqs4} and their $\CN=4$ duals can be ``solved'' to give (\cf\ \eqref{classSlimit})
\be \wh\Delta_S^{\CN=4}\;:\quad \{\hat\Lambda_x-\hat\Lambda_y{}_\flat\simeq0,\,\hat\Lambda_y+\hat\Lambda_x{}_\flat\simeq0\}\quad\text{or}\quad \{\hat\Lambda_x+\hat\Lambda_y{}_\flat\simeq0,\,\hat\Lambda_y-\hat\Lambda_x{}_\flat\simeq0\}\,,
\ee
and it is easy to see that $\wh\Delta_S^{\CN=4}$ then annihilates any linear combination of the functions $e^{\pm\frac{2}{\hbar}\Lambda_x\,\Lambda_x{}_\flat}$. The particular combination with the right symmetries for Liouville theory and gauge theory is
\be Z_S^{\CN=4}(\Lambda_x,\Lambda_x{}_\flat) = \cosh\Big(\frac2\hbar\Lambda_x\Lambda_x{}_\flat\Big)\,. \label{ZS4} \ee
This result can also be obtained directly by specializing \eqref{ZS} to $v=0$, though this requires somewhat more work (\cf\ \cite{HLP-wall}).

Amusingly, a version of \eqref{ZS4} has been well known for a long time in Chern-Simons theory. As a 3-manifold, the mapping cylinder $M_S$ is just the complement of the Hopf link in the 3-sphere. Its (compact!) $SU(2)$ Chern-Simons partition function can be written as $Z_S^{SU(2)}=(\text{const})\times\sinh\Big(\frac2\hbar\Lambda_x\Lambda_x{}_\flat\Big)$ where $k=\frac{2\pi i}\hbar\in\Z$ and $\frac2\hbar\Lambda_x,\,\frac2\hbar\Lambda_y\,\in\Z$ are integers parametrizing the colors (or representations) on the two loops of the Hopf link \cite{Wit-Jones}. The naive analytic continuation of this partition function is clearly annihilated by $\wh\Delta_S^{\CN=4}$, though it appears to have slightly different symmetries from the noncompact wavefunction \eqref{ZS4}.


\section{Flat connections and S-duality}
\label{sec:mirrors}

In the previous discussion we mainly emphasized the role of $(B,A,A)$ branes,
of which $\CB_{\rm Teich}$ is a prime example.
We also have seen branes of type $(A,B,A)$; for instance, correspondences $\Delta_{\varphi}$
associated with mapping cylinders $M_{\varphi}$ define $(A,B,A)$ branes holomorphic in complex structure $J$.
More generally, flat $G_{\C}$ connections on any 3-manifold $M$ with boundary $C$
define an $(A,B,A)$ brane in $Y = \CM_{{\rm flat}} (G_{\C},C)$.

In the present section, we shall study the $\CN=4$ S-duality transformation of such branes that come from flat connections.
Previously, we were able to ignore the fact that $\CN=4$ S-duality exchanges the gauge group $G$ with its Langlands or GNO dual group ${}^LG$, as in $\CN=4$ super-Yang-Mills theory. Now, however, the distinction between $G$ and ${}^LG$ becomes important.
In particular, for $G_{\C} = SL(2,\C)$ the dual group is ${}^LG_\C = SO(3,\C) = PGL(2,\C)$.

To keep things simple, we focus on $(A,B,A)$ branes in the target space
\begin{align}
Y & \, =\; \CM_{\rm flat}(SL(2,\C),T^2) \notag \\ &\,= \; ( \C^* \times \C^* ) / \Z_2 \,.
\label{sl2ctorus}
\end{align}
which has many important applications, ranging from the special case $\Tr g_v = 2$ of the mapping cylinders in Sections \ref{sec:cyl}--\ref{sec:qcyl}, to punctured torus bundles as in Sections \ref{sec:tori}--\ref{sec:3dgauge} and to general knot complements.
In all of these examples, the boundary Riemann surface $C=T^2$ is a torus and the moduli space
$Y = \CM_{{\rm flat}} (G_{\C},C)$ can be conveniently parametrized by the $\C^*$-valued
eigenvalues of the holonomies around A- and B-cycles of $C$:
\be
m = e^u\,, \qquad \ell = e^v\,,
\label{mluv}
\ee
defined modulo the action of the Weyl group, $\CW = \Z_2$, which maps $m \mapsto m^{-1}$ and $\ell \mapsto \ell^{-1}$.

In the sigma-model of $Y$,
any 3-manifold $M$ with a torus boundary $\partial M = C$ defines an $(A,B,A)$ brane $\CB$
supported on the subspace $\CM_{{\rm flat}} (G_{\C},M)$, {\it i.e.} on the zero-locus of the $A$-polynomial,
\be
\CB ~:~~~ A(\ell , m) = 0 \,.
\ee
The $\CN=4$ S-duality maps $Y$ into $\wt Y = \CM_{{\rm flat}} ({}^LG_{\C},C)$ and,
similarly, transforms the brane $\CB$ into a dual brane $\wt \CB$,
which is supported on $\CM_{{\rm flat}} ({}^LG_{\C},M)$
and clearly is also of type $(A,B,A)$ with respect to the hyper-K\"ahler structure on $\wt Y$.
Since in our examples the dual group is ${}^LG_\C = SO(3,\C)$ we shall denote by $A_{SO(3)} (\wt \ell, \wt m)$
the polynomial whose zero locus defines $\CM_{{\rm flat}} ({}^LG_{\C},M)$, so that
\be
\wt \CB ~:~~~ A_{SO(3)} (\wt \ell, \wt m) = 0 \,.
\ee

In order to work out $\wt \CB$ and $A_{SO(3)}$ in concrete examples,
it is convenient to approach the problem from the viewpoint of
the two-dimensional sigma-model with target space $Y$.
In the sigma-model, the S-duality that exchanges $G$ and ${}^LG$
is realized as a composition of the mirror transform and a hyper-K\"ahler rotation \cite{Kapustin-Witten}:
\be
\text{S-duality} \; = \; \left( \text{mirror symmetry} \right) \circ \left(
\begin{smallmatrix}
J&\to&K \\
K&\to&-J
\end{smallmatrix}
\right) \,.
\label{Svsmirror}
\ee
Indeed, $Y = \CM_{{\rm flat}} (G_{\C},C)$ and $\wt Y = \CM_{{\rm flat}} ({}^LG_{\C},C)$
make a famous pair of mirror manifolds~\cite{Hausel-Thaddeus}. Moreover, since $SO(3) \cong SU(2) / \Z_2$
in our examples the mirror variety $\wt Y$ is simply a quotient of the original space $Y$
by the ``group of sign changes'' $\Xi = \Z_2 \times \Z_2$,
\be
\wt Y \; \cong \; Y / \, \Xi \,,
\label{yyquotient}
\ee
whose action on holomorphic coordinates
is generated by $(\ell,m) \mapsto (\ell, -m)$ and  $(\ell,m) \mapsto (- \ell, m)$
(see \cite{mirrorbranes} for further discussion in a closely related context).
As a result, the holomorphic coordinates $\wt \ell$ and $\wt m$ that parametrize the mirror variety $\wt Y$
can be expressed as invariant combinations of $\ell$ and $m$:
\be
\wt \ell = \ell^2 \,, \qquad \wt m = m^2 \,.
\label{lmsquares}
\ee
This is a familiar relation between holomorphic coordinates on the moduli spaces of
$SL(2,\C)$ and $PSL(2,\C)$ flat connections (see {\it e.g.} \cite{Champ-hypA})
and we shall return to it later, in our discussion of $A_{SO(3)} (\wt \ell, \wt m)$.

The eigenvalues $\ell$ and $m$ are clearly $J$--holomorphic coordinates for $Y$. Alternatively, given a complex structure $\tau$ on $C=T^2$, we could define $I$--holomorphic coordinates $b$ and $f$ such that $b$ parametrizes the base ${\bf B}\simeq \C/\Z_2$ of the Hitchin fibration, and $f$ parametrizes a generic fiber ${\bf F}\simeq T^2$. Let us take $\tau=i$ for simplicity. Then, setting
\be u = x_0+ix_2\,,\qquad v = -x_1+ix_3 \label{uvreal} \ee
for real $x_i$ as in \cite{mirrorbranes}, we have
\be b = x_0+ix_1\,,\qquad f = x_2+i x_3\,. \label{bfreal} \ee
Notice that if we write a flat $G_\C$ connection as $\CA = A+i\phi$, then $f$ parametrizes the holonomies of the hermitian connection $A$, whereas $b$ parametrizes the holonomies of the Higgs field $\phi$ (a 1-form on $C$). Generically, the fiber ${\bf F}$ is nonsingular, but it degenerates to a ``pillowcase'' $T^2/\Z_2$ at the origin of ${\bf B}$.

In real coordinates $x_i$, the three K\"ahler forms of $Y$ are given by
\begin{eqnarray}
\omega_I & = & (2\hbar^{-1})\big(dx_0 \wedge dx_1 + dx_2 \wedge dx_3\big)\,, \nonumber \\
\omega_J & = & (2\hbar^{-1})\big(dx_0 \wedge dx_2 - dx_1 \wedge dx_3\big)\,, \label{ijkonr4} \\
\omega_K & = & (2\hbar^{-1})\big(dx_0 \wedge dx_3 + dx_1 \wedge dx_2\big)\,. \nonumber
\end{eqnarray}
Note, moreover, that the $J$--holomorphic symplectic form is $\Omega_J=\omega_I-i\omega_K = (2\hbar^{-1})dv\wedge du$. Similarly, we have $\Omega_I = (2i\hbar^{-1})df\wedge db\,.$

The Hitchin fibration of $Y$ coincides with its ``SYZ'' fibration. Therefore, according to the SYZ picture \cite{SYZ}, mirror symmetry is simply a T-duality along the generic fibers ${\bf F} = T^2$. Furthermore, a hyper-K\"ahler rotation $J\leftrightarrow K$ is given by the transformation $(x_2,x_3)\mapsto(x_3,-x_2)$. Both of these transformations preserve the base $\bf B$. Indeed, from \eqref{lmsquares}, we would expect to identify the natural coordinates on $\bf B$ and $\bf \wt B$ as
\be \wt x_0 = 2x_0\,,\qquad \wt x_1=2x_1\,. \label{x01squares} \ee

\subsection{Simple S-dualities}
\label{sec:simpleS}

As our first example of S-duality action on flat connections,
let us consider the complement of a $(p,q)$ torus knot.
The corresponding $A$-polynomial has a very simple form\footnote{If both $p>2$ and $q>2$, then \eqref{Aforpqknots} is only one irreducible component of the full nonabelian $A$-polynomial $A(\ell,m) = (\ell^2 m^{2pq}-1)$. This is not a problem; we simply take the brane $\CB$ to be defined by this irreducible component.}:
\be
A(\ell,m) = \ell m^{pq} + 1
\label{Aforpqknots}
\ee
so that $\CB$ is middle-dimensional $(A,B,A)$ brane supported on a linear variety:
\be
\CB ~:~~~
\begin{cases}
x_1 + pq x_0 = 0 \\
x_3 - pq x_2 = \pi\,. \label{Bpqknots}
\end{cases}
\ee
Under mirror symmetry ({\it i.e.} T-duality along $x_2$ and $x_3$) nothing happens to the first equation.
The second equation, on the other hand, defines a 1-dimensional submanifold in the torus ${\bf F} = T^2$
parametrized by $x_2$ and $x_3$.
Focusing on these two directions for a moment, we can think of the 1-dimensional submanifold
defined by the equation $x_3 - pq x_2 = \pi$
as a D1-brane wrapped on a homology cycle $(1,pq)$ in $H_1 ({\bf F},\Z)$.
T-duality along both directions of ${\bf F} = T^2$ maps it to a D1-brane
supported\footnote{A simple way to see this is to perform the T-duality in two steps,
dualizing one circle at a time. Then, the first duality along the A-cycle of ${\bf F} = T^2$
maps the D1-brane on $(1,pq)$ cycle into a D0-D2 system with charges $(1,pq)$.
The second T-duality along the B-cycle on $T^2$ maps it back to a D1-brane
supported on a different cycle, namely $(pq,-1)$.}
on the homology cycle $(pq,-1)$ obtained by acting with a matrix
$S =
(\begin{smallmatrix}
0&1\\ -1&0
\end{smallmatrix}
)$ on $(1,pq)$.
Therefore, we conclude that mirror symmetry transforms the second equation in \eqref{Bpqknots}
to the new equation $pq \wt x_3 + \wt x_2 =0$.

Now we are ready to describe the dual $(A,B,A)$ brane on $\wt Y$.
We should not forget, however, to perform the hyper-K\"ahler rotation in \eqref{Svsmirror}
which acts as $\wt x_2 \mapsto \wt x_3$ and $\wt x_3 \mapsto - \wt x_2$ and,
therefore, transforms $pq \wt x_3 + \wt x_2 =0$ into $\wt x_3 - pq \wt x_2 =0$.
Combining this result with the first equation in \eqref{Bpqknots},
we finally obtain the dual $(A,B,A)$ brane on $\wt Y$,
\be
\wt \CB ~:~~~
\begin{cases}
\wt x_1 + pq \wt x_0 = 0 \\
\wt x_3 - pq \wt x_2 = 0\,, \label{mirrorBpqknots}
\end{cases}
\ee
where we identified $\wt x_0 = 2 x_0$ and $\wt x_1 = 2 x_1$ in order to account for \eqref{lmsquares}.
Indeed, if we introduce holomorphic coordinates on $\wt Y$ as in \eqref{uvreal}:
\be
\wt u = \wt x_0 + i \wt x_2\,, \qquad \wt v = \wt x_1 - i \wt x_3\,,
\label{uvrealY}
\ee
then $\wt \ell = e^{\tilde v}$ and $\wt m = e^{\tilde u}$ are automatically
consistent with \eqref{lmsquares}.
In terms of the holomorphic variables $\wt \ell$ and $\wt m$, equation \eqref{mirrorBpqknots}
represents the zero locus of the polynomial
\be
A_{SO(3)} (\wt \ell, \wt m) = \wt \ell \wt m^{pq} - 1\,.
\ee

Comparing it with the original $A$-polynomial \eqref{Aforpqknots} in the $SL(2,\C)$ theory,
we see that $A(\ell,m)$ and $A_{SO(3)} (\wt \ell, \wt m)$ obey%
\protect\footnote{Note that there is a curious
alternative to computing the polynomial $A_{SO(3)} (\tilde \ell, \tilde m)$ based on the embedding
of $SO(3,\C)$ into $SL(3,\C)$. Indeed, $SO(3,\C)$ can be realized as the group of fixed points
in $SL(3,\C)$ under the involution $\tau : g  \mapsto  J ( g^t )^{-1} J$, where
$\displaystyle J = \left(\begin{smallmatrix}
0 &  0 & 1 \\
0 & -1 & 0 \\
1 &  0 & 0
\end{smallmatrix}\right)$.
}
\be
A_{SO(3)} (\wt \ell, \wt m) \; = \; A (-\ell, m) \cdot A (\ell, m)\,,
\label{aaa}
\ee
with the identification of variables \eqref{lmsquares}.
In fact, this relation holds for any knot complement in $S^3$, for which the $A$-polynomial is already a function of $\tilde m = m^2$. More generally, for any 3-manifold with torus boundary, we have $A_{SO(3)} (\wt \ell, \wt m) =  A(-\ell,-m)A(-\ell,m)A(\ell,-m)A(\ell,m)$.
Obviously, this prescription is consistent with the fact that for $G_{\C} = SL(2,\C)$ and ${}^LG_{\C} = SO(3,\C)$
the corresponding moduli spaces are related by the quotient \eqref{yyquotient},
which leads to the identification of variables \eqref{lmsquares}.
For example, the $SO(3,\C)$ version of the $A$-polynomial for the figure-$8$ knot \eqref{fig8apol} looks like:
$$
A_{SO(3)} (\wt \ell, \wt m) = \wt \ell + \wt \ell^{-1}
- (\wt m + \wt m^{-1})^4 + 2(\wt m + \wt m^{-1})^3 + 7(\wt m + \wt m^{-1})^2 - 8(\wt m + \wt m^{-1}) - 14 \,.
$$

\subsection{A-polynomial and amoebas}
\label{sec:amoeba}

\FIGURE{
\includegraphics[height=3.5in,width=1.5in]{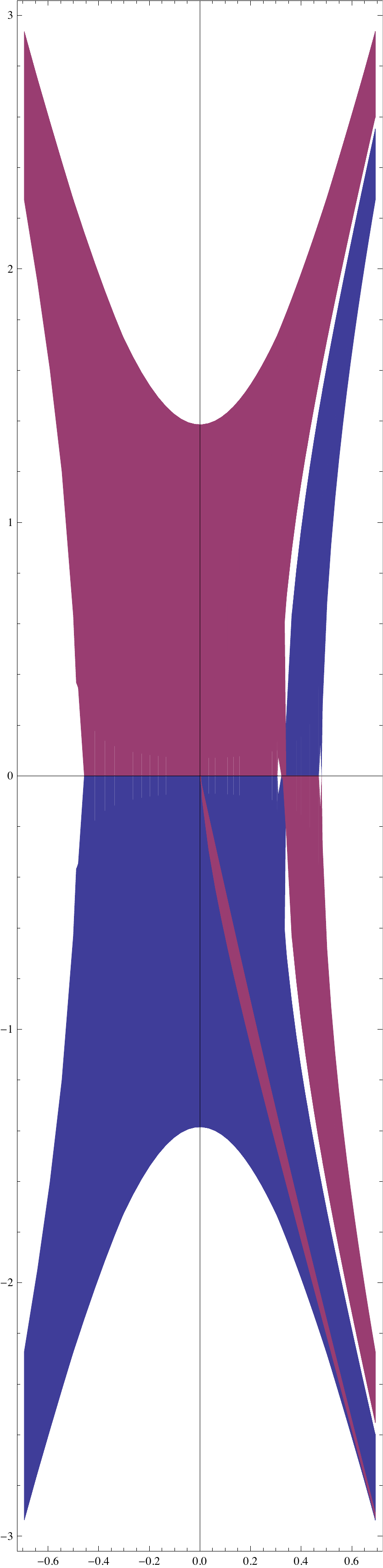}
\caption{Amoeba for the figure-$8$ knot.}
\label{amoeba}
}

Now, it is clear how to generalize $\CN=4$ S-duality to arbitrary 3-manifolds $M$
with torus boundary $C = T^2$ and, in fact, also to arbitrary groups $G$.
The moduli spaces of complex flat connections define mirror manifolds
$Y = \CM_{{\rm flat}} (G_{\C},C)$ and $\wt Y = \CM_{{\rm flat}} ({}^LG_{\C},C)$.
Similarly, complex flat connections that can be extended to $M$ define
the support of dual $(A,B,A)$ branes $\CB$ and $\wt \CB$ in $Y$ and $\wt Y$, respectively.
As we proceed to explain, a criterion generalizing the T-duality argument of the previous subsection is
that mirror branes $\CB$ and $\wt \CB$ have the same ``amoeba.''

Recall that for a polynomial $A(\ell,m)$ defining a curve in $\C^* \times \C^*$,
the corresponding amoeba is the image of $A(\ell,m)=0$ under the map
\be
{\rm Log} ~:~ (\ell,m) \mapsto (\log |\ell| , \log |m|)\;= \;(\Re(v),\Re(u)) \,.
\label{logmap}
\ee
When a polynomial $A(\ell,m)$ is invariant under the ``Weyl reflection''  $(\ell,m) \mapsto (\ell^{-1},m^{-1})$ --- a property enjoyed by all $A$-polynomials  ---
its zero locus defines a curve in $( \C^* \times \C^* ) / \Z_2$, and a corresponding amoeba in $(\R\times \R)/\Z_2$.
For example, the amoeba for the $A$-polynomial of the figure-8 knot is shown
on Figure \ref{amoeba}.

In the present context, $( \C^* \times \C^* ) / \Z_2$ or, more generally, $\frac{(\C^*)^r \times (\C^*)^r}{\CW}$
is the moduli space of flat connections on $C=T^2$,
and the map \eqref{logmap} is precisely the projection to the base of the SYZ fibration \cite{SYZ}\,,
\be
\begin{array}{ccccc}
Y = \CM_{{\rm flat}} (G_{\C},C) & \; & \; & \; & \CM_{{\rm flat}} ({}^LG_{\C},C) = \widetilde{Y} \\
\; & \searrow & \; & \swarrow & \; \\
\; & \; & {\bf B} & \; & \;
\end{array}
\label{mhmhmirror}
\ee
whose generic fibers are dual abelian varieties ${\bf F}$ and $\wt {\bf F}$.
Since mirror symmetry between $Y$ and $\wt Y$
can be understood as a T-duality along the fibers of \eqref{mhmhmirror},
the projections of any pair of mirror branes $\CB$ and~$\wt \CB$,
 to the base ${\bf B}$ must be \emph{identical}.
In other words, $\CB$ and~$\wt \CB$ must have the same ``amoeba.''

Note that in our case the branes $\CB$ and~$\wt \CB$ that parametrize flat $G_{\C}$ and ${}^LG_{\C}$
connections on $M$ are almost uniquely determined by a small open fragment of the amoeba's boundary.
The ambiguity in determining the branes includes multiplicative factors which, in the rank-1 case,
account for the signs in \eqref{aaa}. Indeed, it is easy to see that the polynomials
$A (\ell, m)$ and $A (-\ell, m)$ have the same amoeba, so that $\CB$ and~$\wt \CB$ defined
by the zero loci of the polynomials $A (\ell, m)$ and $A_{SO(3)} (\wt \ell, \wt m)$
have the same set of ``fingerprints'' in the base ${\bf B}$.

\subsection{Dual loop coordinates}
\label{sec:SL}

To finish this classical discussion, we can reconnect with $\CN=4$ S-duality as it was described in Sections \ref{sec:class}--\ref{sec:quant}. There, we suppressed the dualization of the gauge group $G\to {}^LG$. Or, more precisely, for $G=SL(2,\C)$, we ignored the quotient in ${}^LG = SO(3,\C)=SL(2,\C)/\Z_2$ and tacitly related partition functions and operators in Chern-Simons theory with gauge group ${}^LG$ to the same objects in Chern-Simons theory with gauge group $G$. Semiclassically, we would have said that the S-dual of an $A$-polynomial
\be A(\ell,m)=0 \ee
is the curve $A({}^L\ell,{}^L\ell)= 0$, where
\be {}^L\ell = \exp{}^Lv=\exp\big(\frac{2\pi i}{\hbar}v\big)\,,\qquad{}^Lm=\exp{}^Lu=\exp\big(\frac{2\pi i}{\hbar}u\big)\,. \label{LdefA} \ee
In this, we missed the ``doubling'' effect of the $\Z_2\times \Z_2$ quotient on $\wt Y$. Taking it into account, we should correct our statement to
\be S\,:A(\ell,m)\quad\mapsto\quad A_{SO(3)}({}^L\wt\ell,{}^L\wt m) = A(-{}^L\ell,{}^L m)A({}^L\ell,{}^L m)\,, \label{SdualA} \ee
with ${}^L\wt\ell={}^L\ell^2$ and ${}^L\wt m = {}^L m^2$.

The appearance of nonperturbative dual coordinates ${}^L\ell,{}^L m$ in \eqref{SdualA} follows from the proper identification of objects in Chern-Simons theory with gauge group $G$ with objects in Chern-Simons theory with gauge group ${}^LG$. To understand this, observe (\cf\ \cite{mirrorbranes}) that under mirror symmetry plus a hyper-K\"ahler rotation the K\"ahler forms \eqref{ijkonr4} are dualized to
\begin{eqnarray}
\wt\omega_I & = & \frac{2}{\hbar}\bigg(dx_0 \wedge dx_1 + \Big(\frac{\hbar}{2\pi i}\Big)^2 d{}^Lx_2 \wedge d{}^Lx_3\bigg)\,, \nonumber \\
\wt\omega_J & = & \frac{2}{\hbar}\bigg(\frac{\hbar}{2\pi i}dx_0 \wedge d{}^Lx_2 - \frac{\hbar}{2\pi i}dx_1 \wedge d{}^Lx_3\bigg)\,, \label{Lijkonr4} \\
\wt\omega_K & = & \frac{2}{\hbar}\bigg(\frac{\hbar}{2\pi i}dx_0 \wedge d{}^Lx_3 + \frac{\hbar}{2\pi i}dx_1 \wedge d{}^Lx_2\bigg)\,, \nonumber
\end{eqnarray}
so that, in particular, the volume of the generic fiber $\bf \wt F$ is inverted. Here ${}^Lx_{2,3}$ are simply the new coordinates on $\bf \wt F$, \emph{a priori} unrelated to $x_{2,3}$. The forms $\wt\omega_{I,J,K}$ look identical to the original K\"ahler forms $\omega_{I,J,K}$ at dualized coupling $\hbar\to {}^L\hbar = -\frac{4\pi^2}\hbar$, provided that we \emph{rescale} the base coordinates $x_0$ and $x_1$ as
\be x_{0,1} = \frac{\hbar}{2\pi i}{}^Lx_{0,1}\,. \ee
This additional rescaling, necessary for bringing $\wt\omega_{I,J,K}$ back to canonical form, is responsible for the full identification \eqref{LdefA}. Once we rescale $x_{0,1}$, it is natural to also identify the fiber coordinates as ${}^Lx_{2,3}=\frac{2\pi i}{\hbar}x_{2,3}$, and \eqref{LdefA} results.

\subsection{Dual D-modules}
\label{sec:D}

Upon quantization, a $J$--holomorphic Lagrangian brane (\emph{a.k.a.} a mid-dimensional $(A,B,A)$ brane) $\CB\subset Y$ is promoted to a ``holonomic $\CD$-module'' \cite{Kashiwara-book}. Practically, this means that the $J$--holomorphic equations that define $\CB$ classically become differential or difference operators in an appropriate algebra. We encountered several instances of this phenomenon in Section \ref{sec:quant}, where a Lagrangian in $Y\times Y$, cut out by equations $\Delta$, led to a system of operators $\wh\Delta$.

In the case of a knot complement $M$, the $A$-polynomial $A(\ell,m)=0$ is quantized to an operator \cite{gukov-2003}
\be \hat A(\hat\ell,\hat m;q) \simeq 0\,,\ee
which annihilates the Chern-Simons wavefunction of $M$. Here, operators $\hat \ell = e^{\hat v}$ and $\hat m = e^{\hat u}$ satisfy
\be \hat \ell\hat m = q^{\frac12}\hat m\hat \ell\,,\qquad\text{or}\qquad [\hat v,\hat u]=\hbar/2\,,\ee
consistent with the classical symplectic structure $\Omega_J = (2\hbar^{-1})dv\wedge du$.

The classical $\CN=4$ S-duality of previous sections can be extended to the quantum operators $\hat A(\hat \ell,\hat m;q)$. The first examples of this were explored in \cite{Dimofte-QRS}. For simplicity, let us drop the hats on $\hat\ell$ and $\hat m$. Then, mirror symmetry plus hyper-K\"ahler rotation on $Y$ maps
\be S\;:\;\; \hat A(\ell,m;q)\quad\mapsto\quad
 \hat A_{SO(3)}(\tilde\ell,\tilde m;q)\,, \label{qaaa}
\ee
where $\hat A_{SO(3)}$ is the quantum $SO(3,\C)\simeq PSL(2,\C)$ $A$-polynomial of $M$. For a knot complement in $S^3$, it is given by
\be \hat A_{SO(3)}(\tilde\ell,\tilde m;q) = \hat A'(-\ell,m;q)\cdot\hat A(\ell,m;q)\,,\ee
with $\tilde \ell = \ell^2$, $\tilde m = m^2$, where $\hat A'(\tilde\ell,\tilde m;q)$ is the unique\footnote{Up to multiplication by functions of $m$.} polynomial operator such that 1) the product $\hat A'(-\ell,m;q)\cdot\hat A(\ell,m;q)$ only contains even powers of $\ell$ and $m$, and 2) its classical $q\to 1$ limit is $\hat A(-\ell,m;1)=A(\ell,m)$. Obviously, \eqref{qaaa} is a quantum generalization of \eqref{aaa}.

For example, for an $(a,b)$ torus knot, the quantum $\hat A$-polynomial and its S-dual are given by%
\footnote{These expressions are written in a polarization that differs from that of \cite{Dimofte-QRS} by $\ell\to q^{-1/2}\ell$. This makes the $\CN=4$ S-duality more manifest. Again, if both $a>2$ and $b>2$ the $\hat A$-polynomial here is only a single factor of the full quantum operator.} \cite{garoufalidis-2004, Hikami-torusAhat, Dimofte-QRS}
\begin{subequations}
\begin{align} \hat A^{a,b}(\ell,m;q) &\,=\, q^{\frac{ab}{4}-\frac12}m^{ab}\ell+1\,,\\
\hat A^{a,b}_{SO(3)}(\tilde\ell,\tilde m;q) &\,=\, q^{ab-1}\tilde m^{ab}\tilde \ell-1\;=\,(q^{\frac{ab}{4}-\frac12}m^{ab}\ell-1)(q^{\frac{ab}{4}-\frac12}m^{ab}\ell+1)\,.
\end{align}
\end{subequations}
Similarly, for the figure-eight knot we have
\be \hat A^{\bm{4_1}}(\ell,m;q) =\big(q^{-1}m^{2}-m^{-2}\big)\ell -\big(m^2-m^{-2}\big)\big(m^4+m^{-4}-m^2-m^{-2}-q-q^{-1}\big)+\big(qm^2-m^{-2}\big)\ell^{-1}\,, \notag \ee
and its rather more complicated S-dual is readily obtained by multiplying $\hat A(\ell,m;q)$ with the operator
\begin{align}   {\hat A^{\bm{4_1}}}{}'(-\ell,m;q) \,=\,& q^2(m^2-qm^{-2})(m^2-q^2m^{-2})(m^4+q^4m^{-4}-qm^2-q^3m^{-2}-q-q^3)\ell \notag \\
&+ (m^2-qm^{-2})(qm^2-m^{-2})(m^4+q^4m^{-4}-qm^2-q^3m^{-2}-q-q^3) \notag \\
&\qquad \times(q^4m^4+m^{-4}-q^3m^2-qm^{-2}-q-q^3) \notag \\
&+q^3(q^2m^2-m^{-2})(qm^2-m^{-2})(q^4m^4+m^{-4}-q^3m^2-qm^{-2}-q-q^3)\ell^{-1}\,. \notag
\end{align}
The intrepid reader is invited to check that $\hat A_{SO(3)}^{\bm{4_1}}={\hat A^{\bm{4_1}}}{}'\cdot\hat A^{\bm{4_1}}$ has no odd powers of $\ell$.

If we want to identify (objects in) $G=SL(2,\C)$ and ${}^LG=SO(3,\C)$ Chern-Simons theories as in Section \ref{sec:SL}, we find that $\CN=4$ S-duality maps $\hat A(\ell,m;q)\;\mapsto\;\hat A_{SO(3)}({}^L\tilde\ell,{}^L\tilde m;{}^L q)$, where ${}^L\ell = \exp {}^L \hat v = \exp\big(\frac{2\pi i}{\hbar}\hat v\big)$ and ${}^Lm = \exp{}^L\hat u = \exp\big(\frac{2\pi i}\hbar \hat u\big)$. However, in Section \ref{sec:quant} (see also \cite{Dimofte-QRS}), we argued that Chern-Simons wavefunctions in a single $\CN=4$ duality frame are actually annihilated by \emph{both} operators $\hat A(\ell,m;q)$ and $\hat A({}^L\ell,{}^Lm;{}^Lq)$. Therefore, the complete picture of S-duality actions appears to be
\be
\hspace{.3in}\begin{array}{rcl@{\qquad}c} \hat A(\ell,m;q) && \hat A({}^L\ell,{}^Lm;q) &\text{annihilate $Z^{CS}_{SL(2,\C)}$} \\
& \hspace{-.3cm}S\;\;\,\raisebox{.1cm}{$\nearrow$}\hspace{-.58cm}\raisebox{-.1cm}{$\swarrow$}\hspace{-.39cm}\raisebox{.1cm}{$\nwarrow$}\hspace{-.19cm}\raisebox{-.1cm}{$\searrow$} && \\
\hat A_{SO(3)}(\ell,m;q) && \hat A_{SO(3)}({}^L\ell,{}^Lm;q) &\text{annihilate $Z^{CS}_{SO(3,\C)}$}
\end{array} \label{fullqS}
\ee
From the point of view of analytically continued wavefunctions, it is generally impossible to detect the difference between being annihilated by operators $\hat A(\ell,m;q)$ and $\hat A_{SO(3)}(\ell,m;q)=\hat A'(-\ell,m;q)\cdot \hat A(\ell,m;q)$ (and similarly for the ${}^L$--versions), because the leftmost factor $\hat A'$ in $\hat A_{SO(3)}$ is invertible. This was the case in all the examples of Section \ref{sec:quant}. Then \eqref{fullqS} reduces to the simpler picture
\be \hat A(\ell,m;q)\;\overset{S}{\longleftrightarrow}\;\hat A({}^L\ell,{}^Lm;q)\,,\ee
which is what we found in \eqref{Salg}, \eqref{shearSops}, \eqref{loopSops}, \eqref{sFNLops1}, and many other cases.

For example, the wavefunction of the $(a,b)$ torus knot complement
\be Z^{a,b}(u;\hbar) = \exp\bigg(\frac{ab}\hbar u^2 +\frac{1}{\hbar}(2\pi i+\hbar)u\bigg) \ee
is clearly S-invariant, in the usual sense that $Z^{a,b}(u;\hbar)=Z^{a,b}({}^Lu;{}^L\hbar)$. It is annihilated by the operator $\hat A^{a,b}(\ell,m;q)= q^{\frac{ab}{4}-\frac12}m^{ab}\ell+1$ as well as its honest S-dual $\hat A^{a,b}_{SO(3)}({}^L\ell,{}^Lm;{}^Lq) = ({}^Lq^{\frac{ab}{4}-\frac12}\,{}^Lm^{ab}\,{}^L\ell-1)({}^Lq^{\frac{ab}{4}-\frac12}\,{}^Lm^{ab}\,{}^L\ell+1)$. But in fact, the rightmost factor of $\hat A^{a,b}_{SO(3)}({}^L\ell,{}^Lm;{}^Lq)$, namely $\hat A^{a,b}({}^L\ell,{}^Lm;{}^Lq)={}^Lq^{\frac{ab}{4}-\frac12}\,{}^Lm^{ab}\,{}^L\ell+1$ is sufficient to kill $Z^{a,b}(u;\hbar)$. The factor on the left plays no important role, aside from assuring that $\hat A^{a,b}_{SO(3)}({}^L\ell,{}^Lm;{}^Lq)$ is invariant under the $\Z_2\times \Z_2$ group of sign changes.

\acknowledgments{We wish thank G. Mikhalkin, G. Moore, A. Neitzke, R. van der Veen, D. Zagier, and especially D. Gaiotto and J. Teschner for many enlightening and helpful discussions. We would also like to thank the Aspen Center for Physics for their hospitality during the 2010 Summer Program, where some of the ideas presented here originated. The work of TD is supported in part by NSF Grant PHY-0969448. The work of SG is supported in part by DOE Grant DE-FG03-92-ER40701 and in part by NSF Grant PHY-0757647. Opinions and conclusions expressed here are those of the authors and do not necessarily reflect the views of funding agencies.}


\appendix


\section{Fenchel-Nielsen and Markov coordinates}
\label{app:FNMarkov}

Here we explain the relation between Fenchel-Nielsen (or Darboux) coordinates and the holonomy eigenvalues (\ie\ Markov coordinates) of $G_\R=SL(2,\R)$ or $G_\C=SL(2,\C)$ flat connections on a punctured torus
\be C = T^2\bs \{p\}. \ee
A similar treatment can be found in \cite{Okai-pants, Goldman-Fricke}.
Let us first consider $G_\R=SL(2,\R)$ flat connections on $C$, and  assume further that all holonomies around nontrivial cycles of $C$ are hyperbolic.%
\footnote{Recall that an $SL(2,\R)$ matrix is called hyperbolic if its trace is $\ge 2$, or, equivalently, if its eigenvalues are real.} %
This condition picks out the component of $\CM_{\rm flat}(G_\R;C)$ corresponding to a (cover%
\footnote{In order to get classical Teichm\"uller space, we should really be using $PSL(2,\R)$ flat connections, but all relevant constructions can be lifted to $SL(2,\R)$.} %
of) the Teichm\"uller space $\Teich(C)$. Thus, by standard uniformization arguments, we can describe our flat $SL(2,\R)$ connections in terms of hyperbolic metrics on $C$.  In particular, given any nontrivial closed curve $\gamma$ on $C$, the hyperbolic length $2\Lambda_\gamma$ of the minimal geodesic homotopic to $\gamma$ is related to the $SL(2,\R)$ holonomy matrix $g_\gamma$ around $\gamma$ as
\be |\Tr g_\gamma| = 2\cosh(\Lambda_\gamma) = 2\cosh\big({\rm length}(\gamma)/2\big) \,.\ee
Equivalently, the eigenvalues of $g_\gamma$ are $\pm\exp(\pm \Lambda_\gamma)$.

\begin{figure}[htb]
\centering
\includegraphics[width=4.6in]{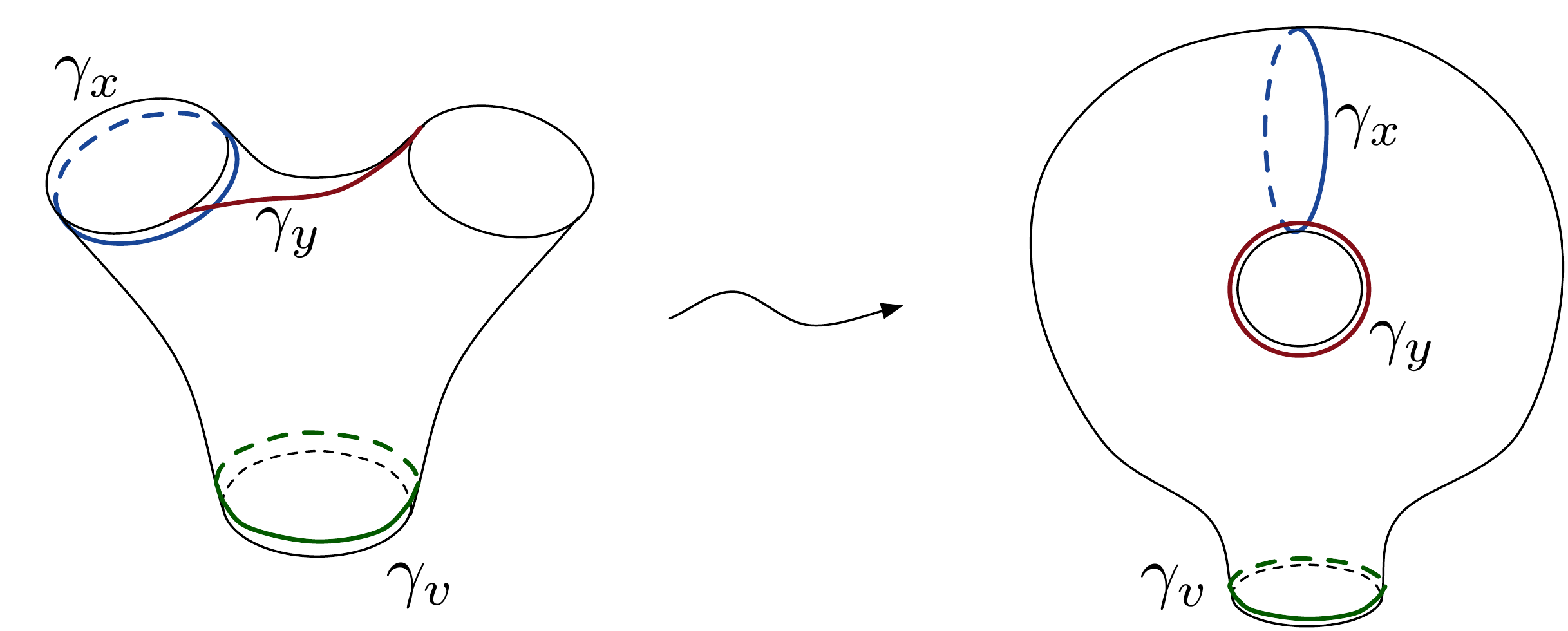}
\caption{Gluing a hyperbolic pair of pants to form $C=T^2\bs\{p\}$.}
\label{fig:apptorus}
\end{figure}

We have $\dim_\R \Teich(C)=2$, so two real coordinates suffice to describe Teichm\"uller space. One possibility is to take the lengths of the two closed cycles $\gamma_x,\gamma_y$ of $C$ drawn in Figure \ref{fig:apptorus}. Letting $g_x$, $g_y$ be the corresponding holonomy matrices, we immediately obtain a relation to Markov coordinates,
\be -x = \Tr g_x = 2\cosh(\Lambda_x)\,,\qquad -y = \Tr g_y = 2\cosh(\Lambda_y) \,. \label{Marlengths} \ee
The holonomy matrix $V$ around the puncture (or hole) $p$ must also have a trace that equals
\be \Tr g_v = -2\cosh \Lambda_v\,, \label{Vminus} \ee
where $\Lambda_v$ is half the hyperbolic length of the geodesic around the puncture. In the notation of Section \ref{sec:MHit}, it is clear that $\boxed{\Lambda_v=v+i\pi}$\,.)

\begin{figure}[htb]
\centering
\includegraphics[width=5in]{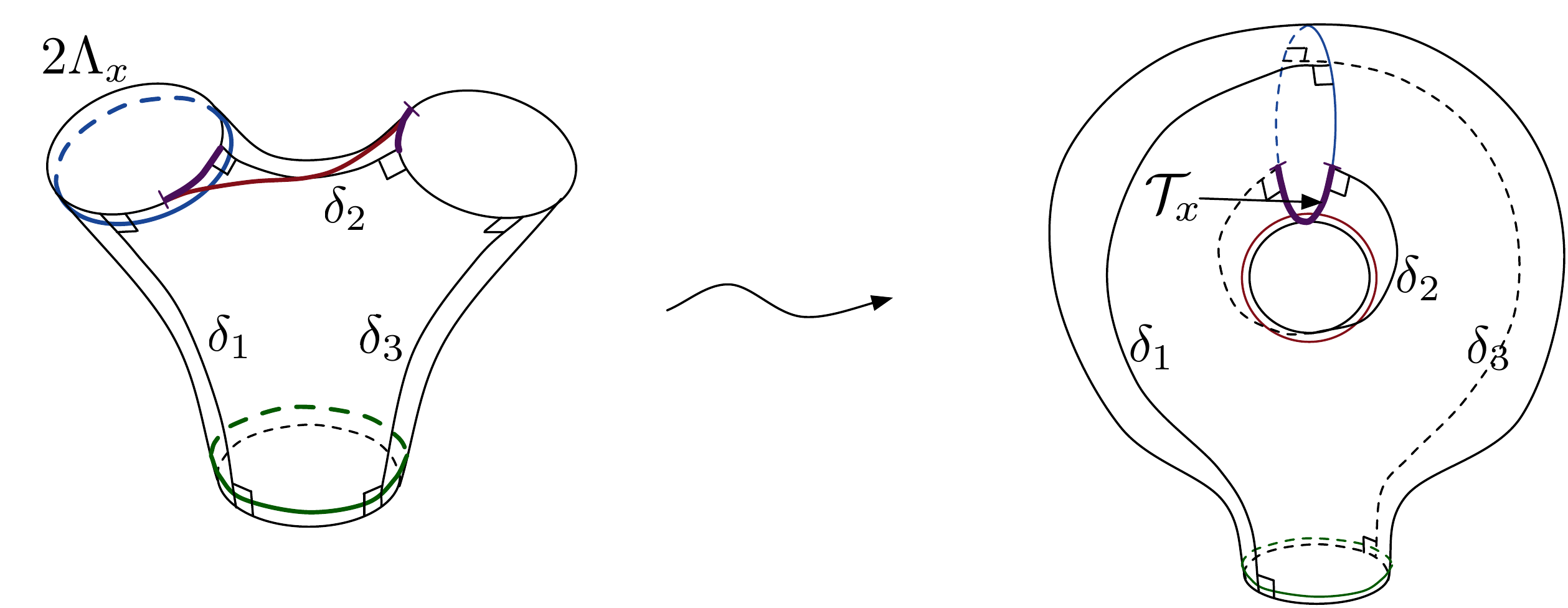}
\caption{Defining the Fenchel-Nielsen twist.}
\label{fig:appFN}
\end{figure}

The hyperbolic Fenchel-Nielsen coordinates on $C$ \cite{Fenchel-Nielsen} (see also \cite{Wolpert-spectra, Wolpert-deformation}) induced by the pants decomposition of Figure \ref{fig:apptorus} are $(\Lambda_x,\CT_x)$, where $\Lambda_x$ is half the length of $\gamma_x$ as above, and $\CT_x$ is the so-called twist dual to $\Lambda_x$. In order to define the twist, one draws minimal open geodesics $\delta_i$ between the three boundaries of the pair of pants that is glued to form $C$. This is shown on the left of Figure \ref{fig:appFN}. Note that these open geodesics must intersect the boundary geodesics at right angles in the hyperbolic metric. Upon gluing, one can twist the two top boundaries of the pair of pants relative to one another by an arbitrary real amount, and this twist is parametrized by the hyperbolic length $\CT_x$ indicated on the right of Figure \ref{fig:appFN}. In terms of the twist angle $\theta_x$, we have
\be \CT_x = 2\Lambda_x\theta_x\,.\ee
Note that neither $\theta_x$ nor $\CT_x$ are periodic: twisting by a full $2\pi$ degrees corresponds to a Dehn twist along $\gamma_x$ (and element of the mapping class group $\bm\Gamma(C)$, and defines a distinct point in Teichm\"uller space $\Teich(C)$.

\begin{figure}[htb]
\centering
\includegraphics[width=6in]{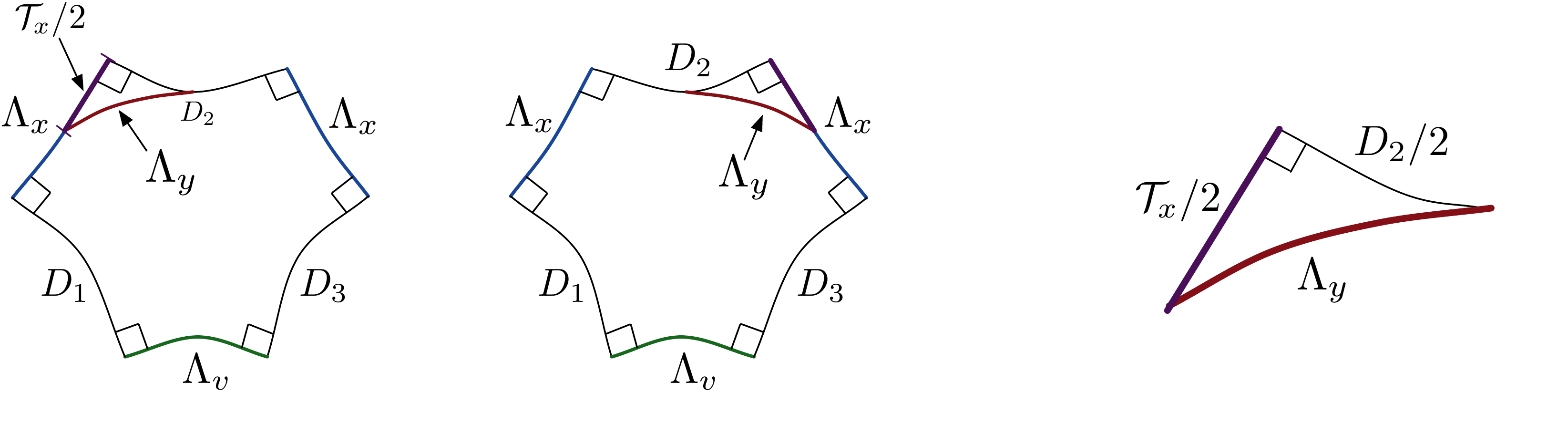}
\caption{Cutting pants into right-angle hexagons.}
\label{fig:hexFN}
\end{figure}

The Fenchel-Nielsen twist $\CT_x$ can be related to $\Lambda_y$ (and in turn the Markov coordinate $y$) via a short exercise in classical hyperbolic geometry. One begins by cutting the pair of pants on the left of Figure \ref{fig:appFN} on the three geodesics $\delta_i$, obtaining two hyperbolic right-angle hexagons (left of Figure \ref{fig:hexFN}). Let $D_1,D_2,D_3$ be the lengths of the geodesics $\delta_1,\delta_2,\delta_3$, respectively.

Since the hyperbolic structure on a right-angle hexagon is fully determined by any three edge lengths, and the two hexagons here share lengths $D_1,D_2,D_3$, they must be isometric. Therefore, the remaining three edge lengths in each hexagon must be $\Lambda_x,\,\Lambda_x$, and $\Lambda_v$. There is a hyperbolic law of cosines for right-angle hexagons (\cf\ \cite{Fenchel-book}), which can be applied here to give
\be \cosh (\Lambda_v) = -\cosh^2(\Lambda_x)+\sinh^2(\Lambda_x)\cosh(D_2) \label{hexcos} \ee
\be \Rightarrow\qquad \cosh(D_2) = \frac{\cosh(\Lambda_v)+\cosh^2(\Lambda_x)}{\sinh^2(\Lambda_x)}\,, \label{solveD2} \ee
solving for the length $D_2$.

Symmetry considerations further require that the geodesic $\gamma_y$ intersects $\delta_2$ at its midpoint. We obtain a small hyperbolic right triangle, as expanded on the right of Figure \ref{fig:hexFN}. The hyperbolic law of cosines for triangles%
\footnote{The hyperbolic laws of sines and cosines for triangles are sufficient to prove the law of cosines for right-angle hexagons \eqref{hexcos}. This may be a fun exercise for the reader.} gives
\be \cosh(\Lambda_y) = \cosh(\CT_x/2)\cosh(D_2/2)\,, \label{triangcos} \ee
which, when combined with \eqref{solveD2}, finally yields
\begin{align} -y &= 2\cosh\Lambda_y \notag \\
 &= 2 \cosh\frac{\CT_x}{2}\sqrt{\frac{\cosh (\Lambda_v)+2\sinh^2(\Lambda_x)-1}{2\sinh^2(\Lambda_x)}} \notag \\
 &= 2\cosh\frac{\CT_x}{2}\frac{\sqrt{\lambda_x^2+\lambda_x^{-2}-\ell-\ell^{-1}}}{\lambda_x-\lambda_x^{-1}}\,,
\label{MarFNyapp}
\end{align}
which is formula \eqref{loopFNgeom} of Section \ref{sec:FN}.

Having obtained \eqref{Marlengths} and \eqref{MarFNyapp} to relate the Fenchel-Nielsen twist coordinate $(\Lambda_x,\CT_x)$ to holonomy matrices $g_x$ and $g_y$ on $C$ in the case of $G_\R=SL(2,\R)$ and hyperbolic holonomies, we can now extend to $G_\C=SL(2,\C)$. The coordinates $x=-\Tr g_x$ and $y=-\Tr g_y$ make perfect sense for $G_\C$ matrices; they are just complexifications (in complex structure $J$) of the real coordinates $x$ and $y$ above. Then, as in (\eg) \cite{NRS}, it is simplest to just \emph{define} complexified Fenchel-Nielsen or Darboux coordinates $(\Lambda_x,\CT_x)$ (modulo $4\pi i$) via the equations $-x = 2\cosh(\Lambda_x)$ and \eqref{MarFNyapp}. A more intrinsic, geometric definition of complex Fenchel-Nielsen coordinates for $\CM_{\rm flat}(G_\C;C)$ is given in \cite{Kourouniotis-cxFN, Tan-cxFN}, and further analyzed by Goldman \cite{Goldman-cxFN}.

In the $G_\R=SL(2,\R)$ case, we could allow $(\Lambda_x,\CT_x)$ to take on pure imaginary values (as well as real values) in order to cover all of $\CM_{\rm flat}(G_\R;C)$ --- that is, components where holonomies may not be hyperbolic. In the $G_\C=SL(2,\C)$ case, we automatically hit all of $\CM_{\rm flat}(G_\C;C)$ by allowing $(\Lambda_x,\CT_x)$ to take arbitrary complex values, so no extra stipulations are necessary.


\section{Quantum dilogarithms}
\label{app:qdl}

The quantum dilogarithm function has a long history dating back to Barnes \cite{Barnes-QDL}, though it came to particular prominence in the last decades with the discovery of its quantum pentagon identities \cite{Fad-Kash, Fad-modular}. Unfortunately, there is no universally established convention for denoting this function.

In this paper, we use notations that are most common in the study of complex Chern-Simons theory (\cf\ \cite{hikami-2006, DGLZ, Dimofte-QRS}), defining the ``noncompact'' quantum dilogarithm as
\be \Phi_\hbar(p) = \prod_{r=1}^\infty \frac{1+q^{r-\frac12}e^p}{1+{}^Lq^{\frac12-r}e^{{}^Lp}} = \exp\bigg(\frac14 \int_{\R+i\epsilon}\frac{dx}{x}\, \frac{e^{-ipx}}{\sinh(\pi x)\sinh \big(\frac{\hbar}{2i}x)}\bigg)\,, \label{QDLdef} \ee
where as usual ${}^Lp = \frac{2\pi i}{\hbar} p$ and ${}^Lq=\exp({}^L\hbar) = \exp\big(\!-\!\frac{4\pi^2}{\hbar}\big)$. This function was called $\Phi_{\hbar/2}(p)$ in \cite{Dimofte-QRS}, and is related to the function $\Phi(z;\tau)$ of \cite{DGLZ} as $\Phi_{\hbar}(p) = \Phi\Big({\displaystyle \frac{p}{2\pi i};\frac{\tau}{2\pi i}}\Big)$. The infinite product and the integral in \eqref{QDLdef} each converge for a particular range of $p$ and $\hbar$, but can be analytically continued to define a meromorphic function of $p\in \C$, for $\hbar$ in the cut plane $\C\bs[0,-i\infty)$; details can be found in \cite{volkov-2003}, \cite{DGLZ} and many references therein.

The function $\Phi_\hbar(p)$ has poles at $p=m(i\pi)+n(\hbar/2)$ and zeroes at $p = -m(i\pi)-n(\hbar/2)$ for $m,n$ odd positive integers. Some simple functional identities that it satisfies are
\be \Phi_\hbar(p-\hbar/2) = (1+e^p)\Phi_\hbar(p+\hbar/2)\,,\qquad
\Phi_\hbar(p-i\pi) = (1+e^{{}^Lp})\Phi_\hbar(p+i\pi)\,,\ee
as well as
\be \Phi_\hbar(p)\Phi_\hbar(-p) = e^{-\frac{p^2}{2\hbar}-C_\hbar}\,, \ee
with $\displaystyle C_\hbar \,\defeq \frac{\pi^2-\hbar^2/4}{6\hbar}\,.$ It obeys an important integral identity \cite{FKV, PonsotTeschner} (a version of the quantum pentagon relation)
\be
\frac{1}{\sqrt{2\pi \hbar}}\int d\zeta\, e^{-\frac1\hbar\zeta z}\frac{\Phi_\hbar(\zeta+y+c_\hbar)}{\Phi_\hbar(\zeta-c_\hbar)}=e^{-C_\hbar}\frac{\Phi_\hbar(y+c_\hbar)\Phi_\hbar(z+c_\hbar)}{\Phi_\hbar(y+z+c_\hbar)}\,, \label{doubleFT}
\ee
where $c_\hbar\,\defeq i\pi+\hbar/2$\,. Using the fact that $\Phi_\hbar(y)\to 1$ as $y\to-\infty$ (for $\Im\,\hbar>0$), one can easily derive from \eqref{doubleFT} the ordinary Fourier transforms
\begin{subequations}\label{QDLFT}
\begin{align} \frac{1}{\sqrt{2\pi\hbar}}\int d\zeta\, \Phi_\hbar(\zeta)e^{\frac1\hbar \zeta z} &=\Phi_\hbar(-z+c_\hbar)e^{\frac{z^2}{2\hbar}-C_\hbar}\,, \\
\frac{1}{\sqrt{-2\pi \hbar}}\int d\zeta\,\frac{1}{\Phi_\hbar(\zeta)}e^{\frac1\hbar \zeta z} &= \frac{1}{\Phi_\hbar(z-c_\hbar)}e^{-\frac{z^2}{2\hbar}+C_\hbar}\,.
\end{align}
\end{subequations}
Moreover, setting $y=-2c_\hbar$ in \eqref{doubleFT} yields, heuristically,
\be \frac{\Phi_\hbar(-c_\hbar)\Phi_\hbar(z+c_\hbar)}{\Phi_\hbar(z-c_\hbar)}\;\text{``}=\text{''}\;\sqrt{-2\pi\hbar}\,\delta(z)\, \label{Phidelta}\ee
(note that $\Phi_\hbar(p)$ has a single pole at $p=0$, so the left side of \eqref{Phidelta} vanishes unless $z=0$).
Capitalizing on the symmetries of the right-hand side of \eqref{doubleFT} under $y\leftrightarrow z$ and $y\to -y-z$, this then leads to the more rigorous relations
\begin{subequations}
\begin{align} \int d\zeta\,e^{\frac{2\pi i+\hbar}{\hbar}\zeta}\frac{\Phi_\hbar(\zeta+z+c_\hbar)}{\Phi_\hbar(\zeta-c_\hbar)}&=2\pi i\hbar\,\delta(z)\,, \label{Phidelta2}\\
 \int d\zeta\,e^{-\frac{1}{\hbar}\zeta z}\frac{\Phi_\hbar(\zeta-z+c_\hbar)}{\Phi_\hbar(\zeta-c_\hbar)}&=2\pi i\hbar\,\delta(z)\,. \label{Phidelta3}
\end{align}
\end{subequations}

In Liouville theory, a much more common choice for noncompact quantum dilogarithm is
\be e_b(x) = \Phi_\hbar(p)\,, \ee
with $p=2\pi b x$ alongside the usual identification $\hbar =2\pi ib^2$. One also encounters the closely related and somewhat more ``balanced'' function
\be s_b(x) = e^{-\frac{i\pi}{2}x^2+\frac12C_\hbar}e_b(x) = e^{\frac{p^2}{4\hbar}+\frac12C_\hbar}\Phi_\hbar(p)\,\qquad(p=2\pi bx)\,,\ee
where now we can write $C_\hbar  = -\frac{i\pi}{12}(b^2+b^{-2})=-\frac{i\pi}{12}(Q^2-2)$ with $Q = b+b^{-1}$. The commonly used Liouville parameter $c_b=iQ/2$ is related to $c_\hbar$ above as $c_\hbar=2\pi b\,c_b$. Some of the functional identities for $s_b$ are
\be \frac{s_b(x-ib/2)}{s_b(x+ib/2)} = 2\cosh(\pi b x)\,,\qquad \frac{s_b(x-ib^{-1}\!/2)}{s_b(x+ib^{-1}\!/2)} = 2\cosh(\pi b^{-1} x)\,, \ee
and
\be s_b(x)s_b(-x) = 1\,. \ee


\section{Quantizing the shear\,--\,Fenchel-Nielsen transformation}
\label{app:map}

In Section \ref{sec:qalg}, we argued that the classical transformation from shear to Fenchel-Nielsen coordinates on $Y = \CM_{\rm flat}(SL(2,\C),\,T^2\bs\{p\})$ can be quantized, providing an isomorphism of the operator algebra $\hat\CA_\hbar$. In particular, the map of operators is described by any two of the three equations \eqref{qshearFN}
\be \sqrt{\hat t} = i\frac{1}{\hat a-\hat a^{-1}}(\hat b-\hat b^{-1})\,,\qquad \sqrt{\hat t'} = i\frac{1}{\hat c-\hat c^{-1}}(\hat a-\hat a^{-1})\,,\qquad \sqrt{\hat t''} = i\sqrt{\ell}\frac{1}{\hat b-\hat b^{-1}}(\hat c-\hat c^{-1})\,, \label{app:qshearFN} \ee
and their $\CN=4$ duals \eqref{LqshearFN}, where
\be \hat a = \hat \lambda\,,\qquad \hat b = e^{-\frac{\hat \CT}{2}}=\frac{1}{\sqrt{\hat \tau}}\,,\qquad \hat c = e^{\frac{\hat \CT}{2}-\hat \Lambda}=q^{-1/4}\hat \lambda^{-1}\sqrt{\hat \tau}\,. \label{app:abc} \ee
Here we want to show that relations \eqref{app:qshearFN} are consistent with (and can be intuited from) the integral kernel $Z_\xi(\Lambda,T_\flat)$ that implements on wavefunctions the change of coordinates from shear to Fenchel-Nielsen,
\be Z(T)\;\mapsto\; Z(\Lambda) = \int dT_\flat\, Z_\xi(\Lambda,T_\flat)\,Z(T_\flat)\,.\ee

As in Figure \ref{fig:sFNcyl}(a) or Figure \ref{fig:qshearFN}, the kernel $Z_\xi(\Lambda, T_\flat)$ can also be interpreted as the Chern-Simons wavefunction of a mapping cylinder $M_\xi$. In a normalization that is natural for Chern-Simons theory as well as Liouville theory, this wavefunction was given in \eqref{sFNwf}, following \cite{Kash-kernel, Teschner-TeichLiouv, Teschner-TeichMod}. By converting \eqref{sFNwf} to a polarization in which operators $\hat R_\flat = \frac12(\hat T_\flat-\hat T_\flat')$ and $\hat S_\flat = \frac12(\hat T_\flat+\hat T_\flat')$ act as
\be \hat R = -2\hbar \pd_S\,,\qquad \hat S = S\,,\ee
expression \eqref{sFNwf} acquires the simpler form
\be Z_\xi(\Lambda,S_\flat) = e^{\frac{\Lambda^2}{2\hbar}-\frac1\hbar (S_\flat-c_\hbar)\Lambda}\frac{\Phi_\hbar(\Lambda-S_\flat+c_\hbar)}{\Phi_\hbar(\Lambda+S_\flat-c_\hbar)}\,,
\label{app:sFNsf}
\ee
where as usual $c_\hbar = i\pi+\hbar/2$.
One easily checks that \eqref{app:sFNsf} is annihilated by the two difference equations
\be
\frac{1}{\hat r_\flat}+\hat s_\flat+\frac{1}{\hat s_\flat}-\hat \lambda-\frac{1}{\hat \lambda}\,\simeq\, 0\,,\qquad
(\hat s_\flat-q\hat\lambda)\,\hat\tau+q\hat\lambda(1-\hat s_\flat\hat\lambda)\,\simeq 0\,,
\label{app:sFNops}
\ee
where $\hat s_\flat = \exp\hat S_\flat$ and $\hat r_\flat = \exp\hat R_\flat$ obey $q$-commutation relations $\hat r_\flat \hat s_\flat = q^{-1}\hat s_\flat\hat r_\flat$, as do $\hat \tau = e^{\hbar\,\pd_\Lambda}$ and $\hat \lambda=e^\Lambda$, with $\hat\tau\hat\lambda=q\hat\lambda\hat\tau$.

In order to rewrite \eqref{app:sFNops} in a form more obviously consistent with \eqref{app:qshearFN}, we first solve the second equation in \eqref{app:sFNops} for $\hat s_\flat^{-1}$:
\be \frac1{\hat s_\flat}\,\simeq\, \frac{1}{1-\hat\tau}\frac{q^{-1}}{\hat\lambda}(q\hat\lambda^2-\hat\tau)\,. \label{app:sflat} \ee
Here by ``$\simeq$'' we mean equality modulo $Z_\xi(\Lambda,\hat S_\flat)$, in the sense that $\hat A\simeq\hat B$ if and only if $(\hat A-\hat B)Z_\xi = 0$. Then we find that
\begin{align} \allowdisplaybreaks
 q^{\frac12}\frac{1}{\hat t_\flat} &= \frac1{\hat s_\flat}\frac1{\hat r_\flat} \notag \\
&\overset{\eqref{app:sFNops}}{\simeq} \frac{1}{\hat s_\flat}\bigg(-\hat s_\flat-\frac1{\hat s_\flat}+\hat \lambda+\frac{1}{\hat\lambda}\bigg) \notag \\
&= -1+\bigg(\hat\lambda+\frac1{\hat \lambda}\bigg)\frac{1}{\hat s_\flat}-\frac{1}{\hat s_\flat^2} \notag \\
&\overset{\eqref{app:sflat}}{\simeq}
 -1 + \bigg(\hat\lambda+\frac1{\hat \lambda}\bigg)\frac{1}{1-\hat\tau}\frac{q^{-1}}{\hat\lambda}(q\hat\lambda^2-\hat\tau) - \frac{1}{1-\hat\tau}\frac{q^{-1}}{\hat\lambda}(q\hat\lambda^2-\hat\tau)\frac{1}{1-\hat\tau}\frac{q^{-1}}{\hat\lambda}(q\hat\lambda^2-\hat\tau)\,. \notag
\end{align}
After commuting all $\hat\tau$'s to the left and all $\hat\lambda$'s to the right, and much simplification, this becomes
\begin{align} q^{\frac12}\frac1{\hat t_\flat} &\simeq \frac{q^{-1}}{1-q^{-1}\hat\tau}(q\hat\lambda^2-\hat\tau) + \frac{1}{1-q\hat\tau}(\hat\lambda^2-q\hat\tau)\frac1{\hat\lambda^2} \notag \\
 &= q^{\frac12} \bigg[\frac{i}{\hat\tau^{\frac12}-\hat\tau^{-\frac12}}\bigg(\hat\lambda-\frac1{\hat\lambda}\bigg)\bigg]^2\,,
\end{align}
which upon taking a square root is equivalent to
\be \sqrt{\hat t_\flat}-\frac{i}{\hat \lambda-\hat\lambda^{-1}}\big(\hat\tau^{\frac12}-\hat\tau^{-\frac12}\big) \simeq 0\,. \ee
This final equation is precisely what would follow from the first operator algebra transformation in \eqref{app:qshearFN}, following the rule \eqref{qDelta} described at the beginning of Section \ref{sec:quant}. In a similar way, it is possible to show that
\be \hat t_\flat' = q^{\frac12}\hat s_\flat \frac{1}{\hat r_\flat} \simeq \bigg[\frac{i}{q^{-\frac14}\hat\lambda^{-1}\hat\tau^{\frac12}-q^{-\frac14}\hat\lambda\hat\tau^{-\frac12}}\big(\hat\lambda-\hat\lambda^{-1}\big)\bigg]^2\,,\ee
consistent with the second algebra transformation in \eqref{app:qshearFN}. The last relation, involving $\hat t_\flat''$, then follows automatically.

As described in Section \ref{sec:qshearFN}, the kernel $Z_\xi(\Lambda,T_\flat)$ is annihilated not just by the ideal \eqref{app:sFNops} but by the $\CN=4$ dual equations \eqref{sFNLops1}. Repeating our derivations above leads immediately to the conclusion that
\be \sqrt{{}^L\hat t_\flat}-\frac{i}{{}^L\hat \lambda-{}^L\hat\lambda^{-1}}\big({}^L\hat\tau^{\frac12}-{}^L\hat\tau^{-\frac12}\big) \simeq 0\,,\qquad
\sqrt{{}^L\hat t_\flat'} - \frac{i}{{}^Lq^{-\frac14}{}^L\hat\lambda^{-1}{}^L\hat\tau^{\frac12}-{}^Lq^{-\frac14}{}^L\hat\lambda{}^L\hat\tau^{-\frac12}}\big({}^L\hat\lambda-{}^L\hat\lambda^{-1}\big)\,,
\ee
\emph{etc.} This was the basis for our claim in \eqref{LqshearFN} that the $\CN=4$ dual transformation in the operator algebra can acquire such a (nontrivially) simple form.

Finally, we note that just as the kernel $Z_\xi(\Lambda,T_\flat)$ provides a change of variables at the ``top'' of a mapping cylinder, there also exists a kernel $Z_{\xi^{-1}}(T,\Lambda_\flat)$ that implements a change of variables at the ``bottom'' (\cf\ Figure \ref{fig:sFNcyl}(b)),
\be Z(T_\flat)\;\mapsto\;Z(\Lambda_\flat) = \int dT\,Z(T)\,Z_{\xi^{-1}}(T,\Lambda_\flat)\,.\ee
In a polarization such that operators $\hat S = \frac{1}{2}(\hat T+\hat T')$ and $\hat R=\frac12(\hat T-\hat T')$ act as $\hat S = S$ and $\hat R = \hbar\pd_S$ (and similarly $\hat \CT_\flat = -\hbar\pd_{\Lambda_\flat}$ and $\hat \Lambda_\flat = \Lambda_\flat$), we can write
\be Z_{\xi^{-1}}(S,\Lambda_\flat) = e^{-\frac{1}{2\hbar}\Lambda_\flat^2-\frac1\hbar(c_\hbar-S)\Lambda_\flat}\frac{\Phi_\hbar(\Lambda_\flat+S+c_\hbar)}{\Phi_\hbar(\Lambda_\flat-S-c_\hbar)}\,\nu(\Lambda_\flat)\,, \label{Zxiinv}
\ee
where $\nu(\Lambda_\flat) = \frac{1}{2\sqrt{-2\pi i\hbar}}\big(\lambda_\flat-\lambda_\flat^{-1}\big)\big({}^L\lambda_\flat-{}^L\lambda_\flat^{-1}\big)$ is the appropriate measure for integrating over $\Lambda_\flat$. With this crucial measure factor included, the difference equations that annihilate \eqref{Zxiinv} can be manipulated into the form
\be \sqrt{\hat t}-\frac{i}{\hat \lambda_\flat-\hat\lambda_\flat^{-1}}\big(\hat\tau_\flat^{\frac12}-\hat\tau_\flat^{-\frac12}\big) \simeq 0\,,\qquad
\sqrt{\hat t''}-\frac{i}{q^{-\frac14}\hat\lambda_\flat^{-1}\hat\tau_\flat^{\frac12}-q^{-\frac14}\hat\lambda_\flat\hat\tau_\flat^{-\frac12}}\big(\hat\lambda_\flat-\hat\lambda_\flat^{-1}\big)\,,
\ee
again consistent with the algebra transformation \eqref{app:qshearFN}. This provides very strong evidence in support of \eqref{app:qshearFN}.


\bibliographystyle{JHEP_TD}
\bibliography{CSsdual}

\end{document}